\documentclass[secnumarabic, graphics,floatfix, nofootinbib,tightenlines,nobibnotes, aps, 12pt]{revtex4}
\usepackage{graphicx}
\usepackage[english]{babel}
\usepackage[utf8]{inputenc}
\usepackage{amsmath,amssymb}
\usepackage{amsfonts}
\usepackage{multirow}
\usepackage{booktabs}
\usepackage[section]{placeins}
\usepackage{float}%
\usepackage[colorlinks]{hyperref} 
\usepackage{graphicx,epstopdf}

\usepackage[utf8]{inputenc}
\usepackage[english]{babel}
\usepackage{ragged2e}

\usepackage{tikz}
\newsavebox{\tempbox}

\begin{document}

\newcommand{\bq}{\begin{equation}}
\newcommand{\eq}{\end{equation}}
\newcommand{\bqn}{\begin{eqnarray}}
\newcommand{\eqn}{\end{eqnarray}}
\newcommand{\nb}{\nonumber}
\newcommand{\lb}{\label}
\newcommand{\PRL}{Phys. Rev. Lett.}
\newcommand{\PL}{Phys. Lett.}
\newcommand{\PR}{Phys. Rev.}
\newcommand{\CQG}{Class. Quantum Grav.}

\title{Detailed Model of the Growth of Fluffy Dust Aggregates in a Protoplanetary Disk: Effects of Nebular Conditions}

\author{C. Xiang$^{1}$}
\author{L.S. Matthews$^{1}$}
\author{A. Carballido$^{1}$}
\author{T.W. Hyde$^{1}$}
\affiliation{$^{1}$ Center for Astrophysics, Space Physics, and Engineering Research, Baylor University, Waco, TX 76798-7316, USA;}

\begin{abstract}
Coagulation of dust aggregates plays an important role in the formation of planets and is of key importance to the evolution of protoplanetary disks (PPDs). Characteristics of dust, such as the diversity of particle size, porosity, charge, and the manner in which dust couples to turbulent gas, affect the collision outcome and the rate of dust growth. Here we present a numerical model of the evolution of the dust population within a PPD which incorporates all of these effects. The probability that any two particles collide depends on the particle charge, cross-sectional area and their relative velocity. The actual collision outcome is determined by a detailed collision model which takes into account the aggregate morphology, trajectory, orientation, and electrostatic forces acting between charged grains. The data obtained in this research reveal the characteristics of dust populations in different environments at the end of the hit-and-stick growth, which establishes the foundation for the onset of the next growth stage where bouncing, mass transfer and fragmentation become important. For a given level of turbulence, neutral and weakly charged particles collide more frequently and grow faster than highly charged particles. However, highly charged particles grow to a larger size before reaching the bouncing barrier, and exhibit a "Runaway" growth, in which a few large particles grow quickly by accreting smaller particles while the rest of the population grows very slowly. In general, highly charged aggregates have a more compact structure and are comprised of larger monomers than neutral/weakly charged aggregates. The differences in the particle structure/composition not only affect the threshold velocities for bouncing and fragmentation, but also change the scattering and absorption opacity of dust, influencing the appearance of PPDs.

\end{abstract}

\maketitle
\section{Introduction}
\renewcommand{\theequation}{1.\arabic{equation}} \setcounter{equation}{0}
 
 Electric charging of dust grains is an important process in protoplanetary disks (PPDs). This process plays a key role in the ionization-recombination state of the plasma environment, particularly through the removal of electrons from the gas phase. The magneto-hydrodynamics of PPDs are strongly dependent on the abundance of small grains (Mori \& Okuzumi 2016; Okuzumi et al. 2019), which are efficient electron sinks. Non-zero grain charges affect the dynamical interactions between grains (Matthews et al. 2012) and, by extension, the coagulation mechanism that leads to macroscopic precursors of planetesimals. Coulomb interactions between colliding dust pairs can significantly alter the collision outcome (Matthews et al. 2012; Okuzumi 2009; Okuzumi et al. 2011a,b): growth of dust grains is strongly inhibited, or even halted, by the electrostatic repulsion between colliding aggregates.
 
An understanding of the porous structure of dust aggregates is crucial to determine their charge distribution. Although significant progress has been made in calculating the amount of charge acquired by non-porous grains (e.g., Ivlev et al. 2016), such calculations remain challenging when applied to irregular, fluffy grains (Okuzumi 2009; Matthews et al. 2012). Aggregate charging depends markedly on the open area of the aggregates: porous aggregates can acquire more charge than compact aggregates of the same mass in the same plasma environment (Matthews et al. 2012; Ma et al. 2013). Thus, improved models of the growth of charged dust in weakly-ionized PPDs need to take the evolution of dust aggregate porosity into account by, for example, calculating the orientation and rotation of colliding charged aggregates as they approach each other (Matthews et al. 2012).

Additionally, the initial size distribution of dust grains, generally assumed to be the interstellar medium distribution [a power law with index -3.5; the Mathis-Rumpl-Nordsieck (MRN) distribution; Mathis et al. 1977], is relevant to collisional outcomes. Not only do collisions occur between monomers of different sizes, but also between monomers and aggregates, as well as between aggregates of different sizes. Use of a fixed size or mass ratio between colliding pairs significantly limits the diversity of the aggregate structures, since the structure of an aggregate depends on its formation history. In this study, we examine the evolution of the porosity of charged aggregates considering an initial population of spherical monomers whose radii follow the MRN distribution (Mathis et al. 1977).  A population of aggregates is built by considering the probability of collisions between particles of all sizes present in the population. A detailed model of the collision process, which takes into account two key physical characteristics (porosity and charge), is used to resolve the collision outcome. Here we restrict the analysis to the hit-and-stick regime and identify missed collisions or collisions with great enough energy to produce bouncing.

One source of relative velocities between dust particles is turbulence. There may be different turbulence-generating mechanisms in different regions of a PPD [e.g., the convective overstability, the vertical shear instability, the zombie vortex instability (Malygin et al. 2017), or the magnetorotational instability (Simon et al. 2015)]. Regardless of the mechanism, turbulent velocities dictate the amount of energy available in a collision. In this paper, we parameterize the strength of turbulence in a PPD and analyze the influence of the turbulent strength on the electrostatic repulsion experienced by colliding aggregates.

The paper is organized as follows. Section 2 presents an overview of the methods used to determine the relative velocities and collision outcomes for grains embedded in turbulent gas flow. Section 3 describes the numerical methods for modeling the charging and particle collisions. In Section 4, we investigate the effects of particle charge and turbulence strength on the coagulation process. The resulting characteristics of the dust aggregates including the monomer size, porosity of aggregates and the time evolution of the dust population are presented. Section 5 analyzes the relationship between the properties of colliding particles and the collision outcomes. A comparison of these results to those from previous models is discussed in Section 6, and the main conclusions are summarized in Section 7.


\section{OVERVIEW OF DUST TRANSPORT AND COLLISION OUTCOMES
}
\renewcommand{\theequation}{2.\arabic{equation}} \setcounter{equation}{0}
Small particles entrained in a turbulent gas flow develop relative velocities due to the difference in their coupling times with the gas. The kinetic energy of dust particles is thus influenced by the turbulence strength and eventually determines the collision outcome. At the same time, the relative velocities between dust particles of various sizes affect the collision rate, which determines the growth rate of the dust population.

In this section, we briefly describe the type of dust motion, the possible collision outcomes and the characteristics of aggregate structure, which affect coupling with the gas.

\subsection{Relative motion of dust particles}\label{rel}

Various mechanisms impart relative velocities to solid particles in a PPD, such as Brownian motion, inward radial drift, vertical settling towards the midplane, and turbulence (Brauer et al. 2008; Weidenschilling 1977; Voelk et al. 1980; Ormel \& Cuzzi 2007; Ormel et al. 2008). The dominant source of these relative velocities depends on the disk temperature and location, as well as on the particle properties, i.e., mass and porosity. In general, Brownian motion dominates the relative velocity between the smallest particles, followed by other sources such as turbulence, radial drift, and vertical settling, in order of increasing importance with particle size (Krijt et al. 2014). In this study, we only consider contributions from turbulence and Brownian motion, as systematic relative velocities due to drift and settling are much smaller for the particle sizes relevant to our problem (Rice et al. 2004; Dullemond et al. 2004; Ormel et al. 2008). Assuming particles are well coupled to turbulent eddies, the relative speed between any two dust particles is the sum of the Brownian velocity $v_{B}$ and the turbulent velocity $v_{T}$,
\bqn
\lb{1}
v_{r}=\sqrt{v_{B}^{2}+v_{T}^{2}}.
\eqn
The Brownian velocity depends on the masses of the two colliding particles, $ m_{1}$ and $ m_{2}$,
\bqn
\lb{2}
v_{B}=\sqrt{\frac{8(m_{1}+m_{2})k_{B}T}{\pi m_{1}m_{2}}},
\eqn
where $k_{B}$ is Boltzmann's constant and the gas temperature is given by $T=280\ K/\sqrt{r}$, based on the minimum mass solar nebula (MMSN) model (Weidenschilling 1977; Hayashi 1981; Thommes et al. 2006), with $r$ the heliocentric distance (taken to be 1 AU in this study).

A closed-form analytical expression for the relative turbulent velocity $ v_{T}$ between two grains was presented by Ormel \& Cuzzi (2007) and Ormel et al. (2008), by comparing the stopping time $\tau$ of the largest particle to the turnover times of the turbulent eddies. The stopping time is the time it takes a particle to react to changes in the motion of ambient gas, and is given by (Ormel et al. 2007)
\bqn
\lb{1}
\tau=\frac{3m}{4\pi c_{g}\rho _{g}a^{2}},
\eqn
where $c_{g}$ is the gas thermal speed, $\rho_{g}$ is the gas density, and $m$ and $a$ are the mass and equivalent radius of the particle (defined in Section \ref{agg} for aggregates). For the regions of the protoplanetary disk ($r=1AU$) and the small grains considered in our study ($a \lesssim 100\ \mu m$), the stopping time is less than the turnover time of the smallest eddy, $t_{s}=Re^{-1/2}t_{L}$, where $t_{L}=1/ \Omega$ is the turnover time of the largest eddy, and the relative velocity of any two grains depends only on the difference in their stopping times (Ormel et al. 2008),

\bqn
\label{eq:vturb}
v_{T}=\frac{3}{4}(\frac{v_gRe^{1/4}\Omega}{c_{g}\rho _{g}})(\frac{m_{1}}{{a_{1}^{2}}}-\frac{m_{2}}{{a_{2}^{2}}}).
\eqn
In this expression, $v_g$ is the gas speed, and $Re$ is the Reynolds number, defined as the ratio of the turbulent viscosity, $\nu_{T}=\alpha c_{g}^{2}/\Omega$, to the molecular viscosity of gas, $\nu_{m}=c_{g}\lambda/2$ (Cuzzi et al. 1993), with $\alpha$ the turbulence strength (Shakura \& Sunyaev 1973), $\Omega$ the local Keplerian angular speed, and $\lambda$ the gas mean free path. The subscripts $1$ and $2$ refer to the two different particles. For spherical grains, with $m\propto a^{3}$, the relative velocity only depends on the difference in particle size. The relative velocity of porous aggregates, however, depends on both the mass and the effective cross section.

\subsection{Collision outcomes}

For low-velocity collisions between particles, i.e., $v < 10 ~cm~s^{-1}$, almost all collisions result in sticking at the point of contact (Ormel et al. 2008). However, collisions with energies exceeding a certain minimum threshold can result in restructuring, bouncing, erosion, fragmentation or mass transfer (Wurm et al. 2005; Kothe et al. 2010), depending on the relative velocity and the mass ratio of the colliding particles (Krijt et al. 2015).
The critical bouncing velocity between two smooth spherical grains was derived in detail by Chokshi et al. (1993) and Dominik et al. (1997) using elastic continuum theory. Upon contact, the van der Waals force accelerates the colliding particles, which elastically deform near the contact region forming a neck of material. If the kinetic energy is sufficient to break the neck, the particles separate and dissipate energy in elastic waves. Otherwise, the particles stick together. The critical bouncing velocity, corresponding to the critical kinetic energy, depends on the sizes of the grains and the material properties, and is given by (Chokshi et al. 1993)
\bqn
\lb{7}
v_{cr}\simeq 3.86\frac{\gamma ^{5/6}}{E^{1/3}a^{5/6}\rho ^{1/2}},
\eqn
with $\gamma $ and $\rho $ respectively the surface energy per unit area and the density of the grains, and $a$ the reduced radius of the two spheres, $a=a_{1}a_{2}/(a_{1}+a_{2})$. The Poisson ratios $\vartheta _{1}$, $\vartheta _{2}$ and Young's moduli $ E_{1}$, $ E_{2}$ of the two grains enter into the expression via the material constant $E=[(1-\vartheta _{1}^{2})/E_{1}+(1-\vartheta _{2}^{2})/E_{2}]^{-1}$.

For collisions between fluffy aggregates consisting of monodisperse monomers of radius $a$, assuming there is only one contact point, the local properties are the same as for two contacting spheres of radius $a$. The increased mass of the aggregates, compared to the case of monomer-monomer collisions, can be expressed as an increased density in Eq. \ref{7}, scaling the critical velocity by $N^{-1/2}$, where $N$ is the number of monomers in the two aggregates. Letting $\mu$ and $m_{0}$ be the reduced mass of the two aggregates and the mass of a monomer respectively, the critical velocity between aggregates can be expressed as (Wurm \& Blum 1998)
\bqn
\lb{708}
v_{cr}^{'}=\frac{1}{\sqrt{2}}v_{cr}\left (\frac{\mu }{m_{0}} \right ),
\eqn
where $ v_{cr}$ is the critical bouncing velocity for the monomer-monomer collision between the two contacting spheres. In our simulation with polydisperse monomers, $m_{0}$ is taken to be the average mass of all the monomers in the two aggregates.

As we are assuming that most collisions occur in the hit-and-stick regime, we merely track the relative velocity of the two aggregates at the time of contact and note whether the velocity exceeds this critical velocity. As this model does not calculate the restructuring of the aggregate grains, the simulation is terminated when more than 5\% of the previous 100 collisions result in bouncing.


\subsection{Aggregate structure}\label{agg}

The porosity of dust particles plays an important role in the collision process, as it determines the coupling of the dust particles to the motion of the ambient gas. The relative velocity developed due to the difference of dust particle coupling times affects the collision frequency (and therefore the dust growth rate), the collision outcome, and also determines whether two charged particles can overcome the electrostatic barrier and coagulate. Different quantities have been used to measure the "fluffiness" of aggregates in previous works, such as fractal dimension (Dominik \& Tielens 1997), gyration radius (Wada et al. 2008), and enlargement factor (Ormel et al. 2008). Here we adopt the compactness factor $\Phi _{\sigma}$ (Min et al. 2006). This parameter is useful in this case as it is based on an aggregate's equivalent radius $R _{\sigma}$, which defines an effective cross section for coupling with the gas and has been shown to be directly related to an aggregate's charge (Matthews et al. 2012). We use $\Phi _{\sigma}$, $R _{\sigma}$, and an aggregate's physical radius $R$ to characterize the structure of an aggregate, as described below.

\begin{figure*}[!htb]
\includegraphics[width=9cm]{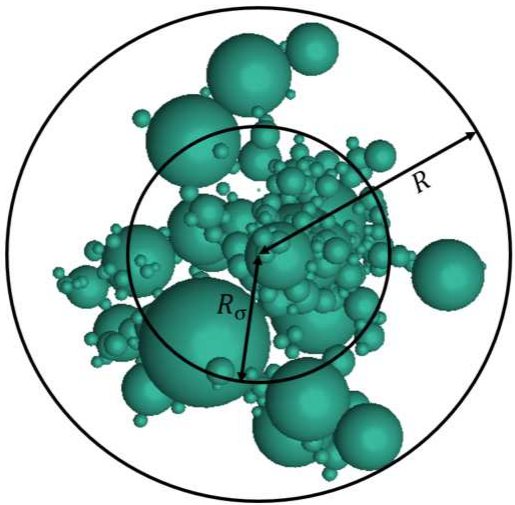}
\caption{Illustration of physical radius and equivalent radius for an aggregate. The outer circle indicates the physical radius $R$, defined as the the maximum radial extent from the center of mass. The inner circle indicates the equivalent radius $R _{\sigma}$, as defined in the text. }
\label{fig0}
\end{figure*}

The compactness factor is defined as the ratio of the volume of all the constituent monomers in an aggregate to the volume of a sphere with radius equal to $R _{\sigma}$, the radius of a circle with area equal to the aggregate's projected cross-section averaged over many orientations (Min et al. 2006). The equivalent radius of a porous aggregate is smaller than its physical radius $R$, which is defined as the maximum radial extent from the center of mass (COM), as illustrated in Figure  \ref{fig0}. 

\section{NUMERICAL TREATMENT OF DUST COAGULATION}
\renewcommand{\theequation}{3.\arabic{equation}} \setcounter{equation}{0}

\subsection{``Detailed-MC'' method}

The factors that affect the coagulation process are the probability that two particles travel towards each other (determined by their cross-sectional area and relative velocity) and the type of interaction between them, which determines the collision outcome (i.e., sticking, bouncing, etc.). We use a ``Detailed-MC'' method which combines a Monte Carlo (MC) method (Ormel et al. 2007; Gillespie 1975) and an N-Body code (Matthews et al. 2012) to model these two factors. When two solid particles are far away from each other, their relative velocity depends largely on the particle sizes, as motion is driven by coupling of the solids to the gas. The Monte Carlo algorithm is used to randomly select colliding particles, where the collision probability is a function of the particle size, as well as determine the elapsed time interval between collisions. At close approach, the detailed collision process is modeled using an N-body algorithm, Aggregate Builder (AB), to determine the collision outcome.

At the beginning of the simulation, the dust particles are grouped into 100 logarithmic bins by their radii, and the collision probability $ C_{ij}$ between the bins is initialized using the average radii and masses of particles in each bin. In each iteration, time is advanced by a random interval to the time when the next collision will occur. Then, two bins are selected based on the collision rates $ C_{ij}$, and one particle is chosen from each bin to collide, with the collision outcome modeled by AB.

\subsubsection{Monte Carlo algorithm}

The Monte Carlo algorithm is a mathematical method used to simulate the stochastic coagulation process. The fundamental postulate of this algorithm is that there exists a function $ C_{ij}(i, j) d\tau $ which represents the probability that a given pair of particles $i$ and $j$ will coagulate in the time interval $ d\tau $. Particles with larger radii and larger relative velocity have a greater chance of collision. The volume that particle $i$ sweeps out relative to particle $j$ per unit time is $\sigma _{ij}\Delta v_{ij}$, where $\sigma _{ij}$ is the effective collision cross section,
\bqn
\lb{13}
\sigma _{ij}=\left\{\begin{matrix}
 \pi(R_{\sigma i}+R_{\sigma j})^{2}\left ( 1-\frac{PE}{KE} \right )~&\text{for}&~KE>PE \\
 0~&\text{for}&~KE\leqslant PE \\
\end{matrix}\right.
\eqn
where $PE=k_{e}\frac{q_{i}q_{j}}{R_{\sigma i}+R_{\sigma j}}$, is the electrostatic energy of the particles at contact, with $q_{i}$ and $q_{j}$ the charges of the two colliding particles, and $R_{\sigma i}$ and $R_{\sigma j}$ their equivalent radii, as defined in Section \ref{agg}; $KE=\frac{1}{2} \mu \Delta v_{ij}$ is their kinetic energy, where $\mu $ is the reduced mass, $\mu =\frac{m_{i}m_{j}}{m_{i}+m_{j}}$, with $m_{i}$ and $m_{j}$ the masses of the particles, and $\Delta v_{ij}$ their relative velocity determined by Eq \ref{eq:vturb}. 

The ratio of the swept volume per unit time to the theoretical simulation volume $V$ in which the particles reside (defined by the ratio of the total dust mass to the dust density in the PPD) defines the collision rate
\bqn
\lb{10}
C_{ij}=\sigma _{ij}\Delta v_{ij}/V.
\eqn

At time $t$, the probability that the next collision will occur in time interval $ (t+\tau, t+\tau+d\tau)$ and involve particles $i$ and $j$ is
\bqn
\lb{10}
P(ij, \tau)&=(C_{tot}exp[-C_{tot} \tau])\times(C_{i}/C_{tot})\times(C_{ij}/C_{i}),
\eqn
where the partial sum $C_{i}= \sum_{j=i+1}^{N}C_{ij}$ is the probability per unit time of selecting particle $i$, and the total sum $C_{tot}=\sum_{i=1}^{N-1}C_{i}$ is the probability per unit time for any collision to occur, with $N$ the total number of dust particles (Gillespie 1975). 

The random time interval between two successive collisions, consistent with the collision probabilities $ C_{ij}$ (e.g. see Gillespie 1975), is given by 
\bqn
\lb{10}
\tau=-ln(r_{1})/C_{tot},
\eqn 
with $r_{1}$ a random number uniformly distributed between zero and one. A larger total sum indicates more frequent collisions and therefore shorter elapsed time between two successive collisions. 

The particles $i$ and $j$ are determined by finding the smallest integers $i$ and $j$ satisfying (Gillespie 1975)
\bqn
\lb{11}
\sum_{k=1}^{i-1}C_{k}<r_{2}C_{tot}
\eqn

and
\bqn
\lb{14}
\sum_{l=i+1}^{j-1}C_{il}< r_{3}C_{i},
\eqn
where $r_{2}$ and $r_{3}$ are also random numbers.

In order to reduce the computational cost in calculating the collision probabilities $ C_{ij}$, the range of equivalent radii is divided into 100 logarithmic intervals, and particles of similar size (within the same interval) are binned into the same group (Ormel et al. 2007). The average equivalent radius of each bin is used to calculate the $\widetilde{C_{ij}}$, the collision rate between particles in group $i$ and group $j$,
\bqn
\lb{13}
\widetilde{C_{ij}}=\left\{\begin{matrix}
\frac{1}{2}g_{i}g_{i}C_{ii}~&\text{for}&~i=j\\
g_{i}g_{j}C_{ij}~&\text{for}&~i\neq j
\end{matrix}\right.
\eqn
where $g_{i}$ and $g_{j}$ are the number of particles in the two groups. After particles from group $i$ and $j$ collide, $g_{i}$ and $g_{j}$ are both decreased by 1, and $g_{i+j}$ is increased by one. At the same time, a particle in the population is randomly chosen to be duplicated, and the number of particles in its group is increased by 1, so that the total number of particles is constant during the simulation. The abstract volume V is rescaled after each duplication procedure, in order to keep the dust spatial density constant.
\\

According to the power-law size distribution that we employ to model dust particles, the initial population contains a small number of particles with large radii. In order to reduce the fluctuation caused by small number statistics, instead of creating monomers randomly based on the power-law distribution at the beginning of the simulation, we create 10,000 monomers with evenly spaced radii within the range 0.5 $\mu$m $\leqslant r\leqslant$ 10 $\mu$m, and add weights to particles of different sizes according to the power-law distribution, with the total weight equal to 10,000. If particles in the $i^{th}$ group have weights $w_{1}, w_{2}, ... w_{k}$, then $g_{i}$ in Eq. \ref{13} equals $\sum_{x=1}^{x=k}w_{x}$. The physical meaning of a non-integer weight can be interpreted by expanding the space, as explained in Appendix A. When two particles are selected to collide, if they both have weights greater than 1, their weights are reduced by 1. If one of them has a weight smaller than 1, their weights are reduced by the value of the smaller one, and this value is also assigned to the weight of the resulting particle and added to that of the duplicated particle, so that the total weight of the population stays constant.

\subsubsection{Collision resolution (Aggregate Builder)}

Once the dust particles are selected, the detailed interaction is modeled using an N-body code, Aggregate Builder (Matthews et al. 2012), taking into account the morphology of the aggregates, the trajectory of the incoming particle, and the electrostatic interaction. 

\begin{figure*}[!htb]
\includegraphics[width=5cm]{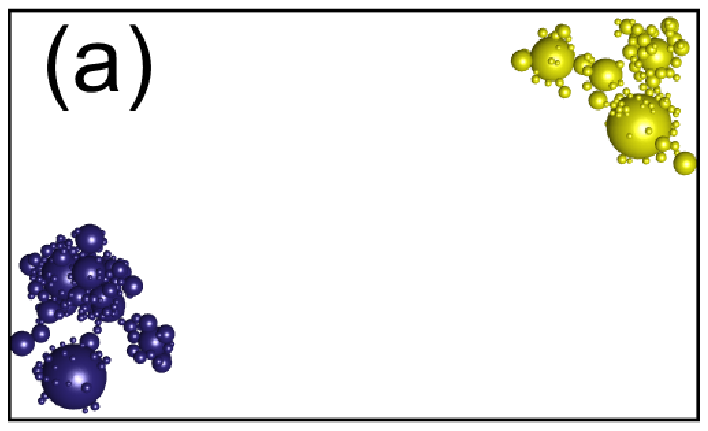}\hspace{2.3cm}\includegraphics[width=5cm]{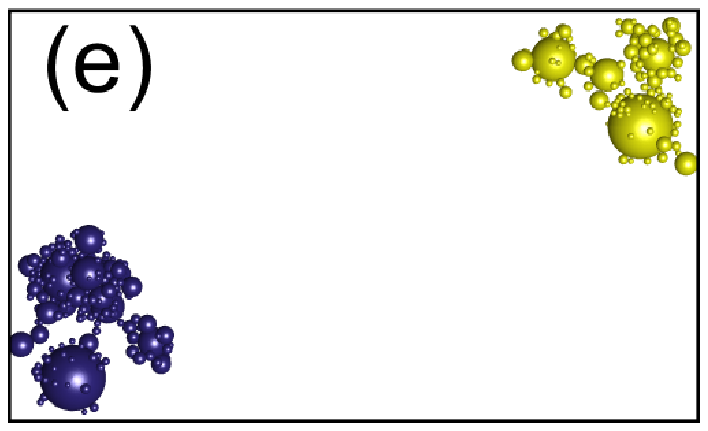}\\
\vspace{-0.182cm}
\includegraphics[width=5cm]{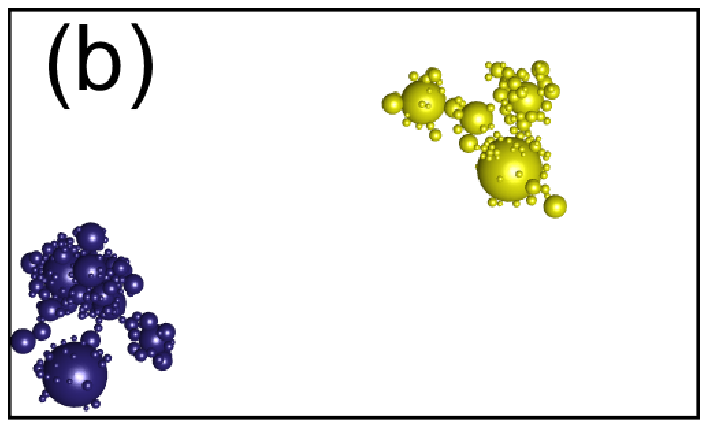}\hspace{2.3cm}\includegraphics[width=5cm]{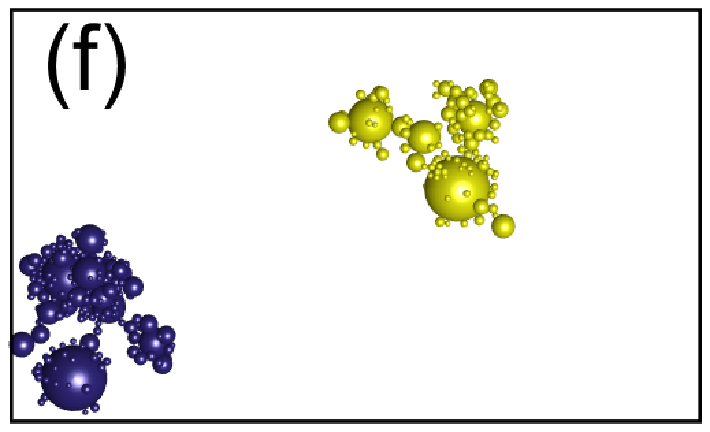}\\
\vspace{-0.182cm}
\includegraphics[width=5cm]{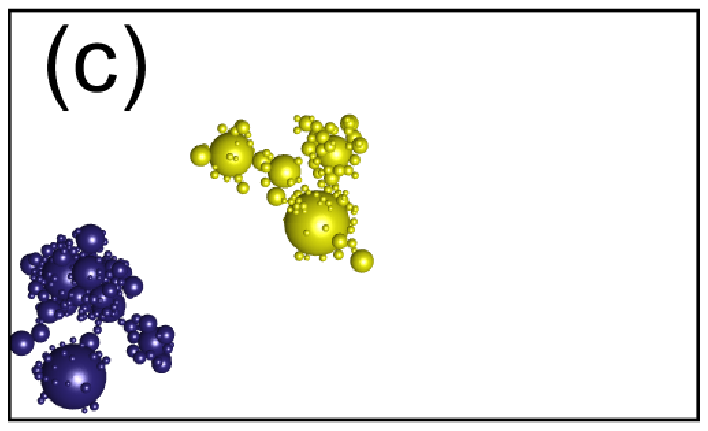}\hspace{2.3cm}\includegraphics[width=5cm]{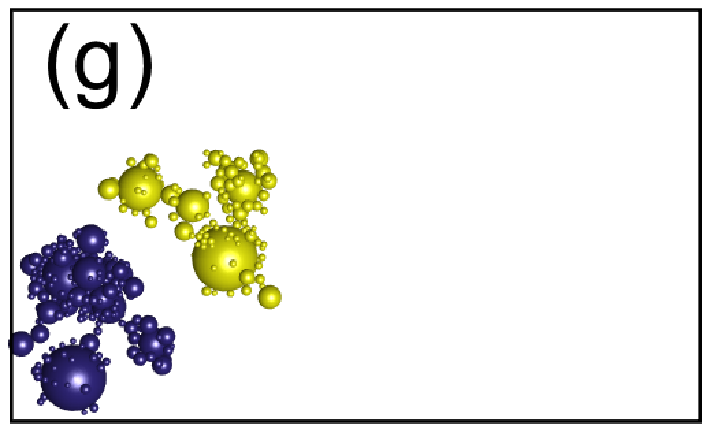}\\
\vspace{-0.182cm}
\includegraphics[width=5cm]{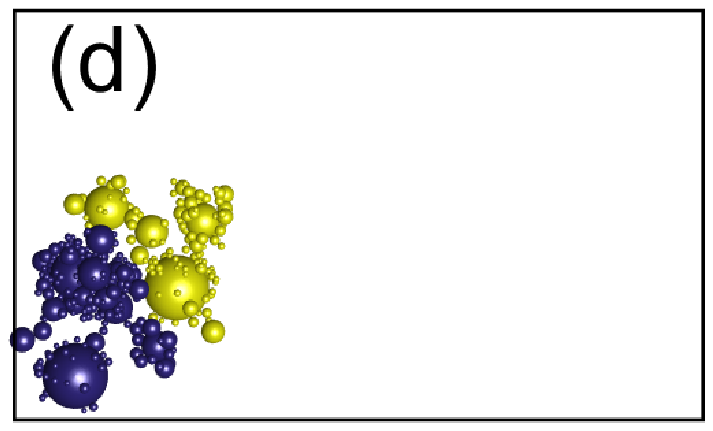}\hspace{2.3cm}\includegraphics[width=5cm]{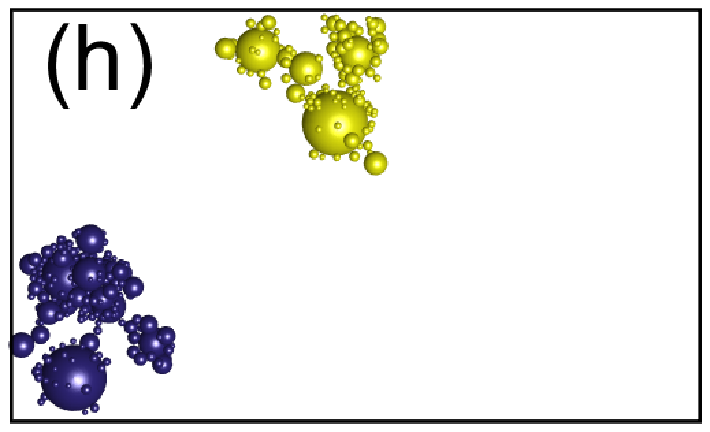}
\caption{Illustration of collision process of charged aggregates with a surface potential of -0.1 V. The blue aggregate is the target, while the yellow aggregate travels towards the target's COM plus an offset. a, b, c, d) show a hit-and-stick collision with an initial relative velocity of 0.1 m/s, and e, f, g, h) show a missed collision due to electrostatic repulsion with a initial relative velocity of 0.002 m/s. Each row shows snapshots of the two particles at different time steps (increased elapsed time from top to bottom). }
\label{fig1}
\end{figure*}

One of the particles is placed with its center of mass at the origin as the target, and an incoming particle travels towards the target's COM plus an offset up to $b=0.2\times (R_{1}+R_{2})$, where $R_{1}$ and $R_{2}$ are the maximum radii of the target and the incoming aggregates respectively (this factor is chosen for computational expediency, to reduce the number of missed interactions). The relative velocities between the two aggregates are set assuming the dust is coupled to turbulent eddies in a protoplanetary disk (Ormel et al. 2007), calculated by Eq \ref{eq:vturb}. The initial distance between two charged particles is set such that the potential energy due to the charge interactions is less than 90\% of the initial kinetic energy, while in the neutral case, it is set to be $2.5\times (R_{1}+R_{2})$. An illustration of the difference in the collision results for charged particles with different relative velocities is shown in Figure \ref{fig1}.

\begin{figure*}[!htb]
\includegraphics[width=9cm]{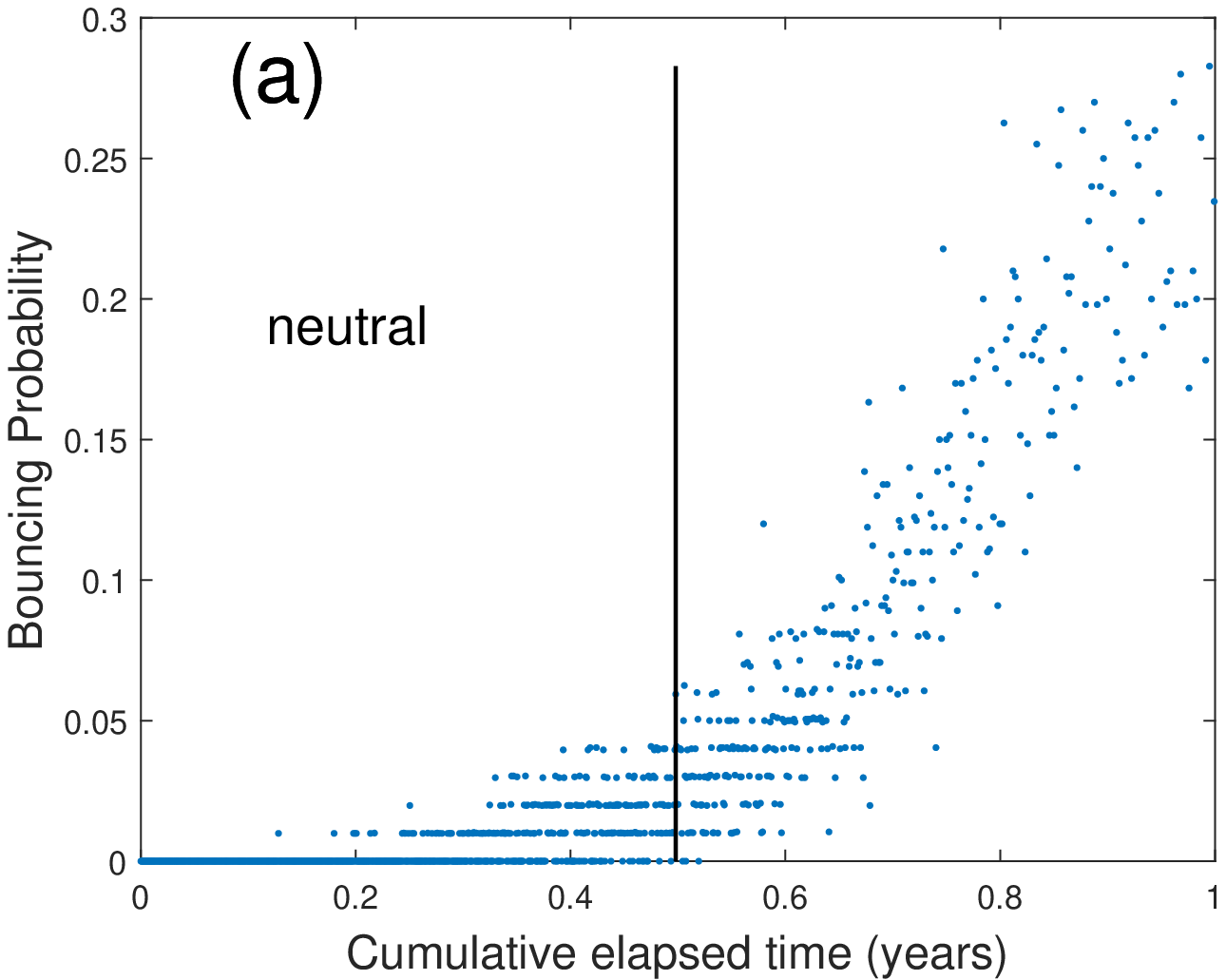}\includegraphics[width=9cm]{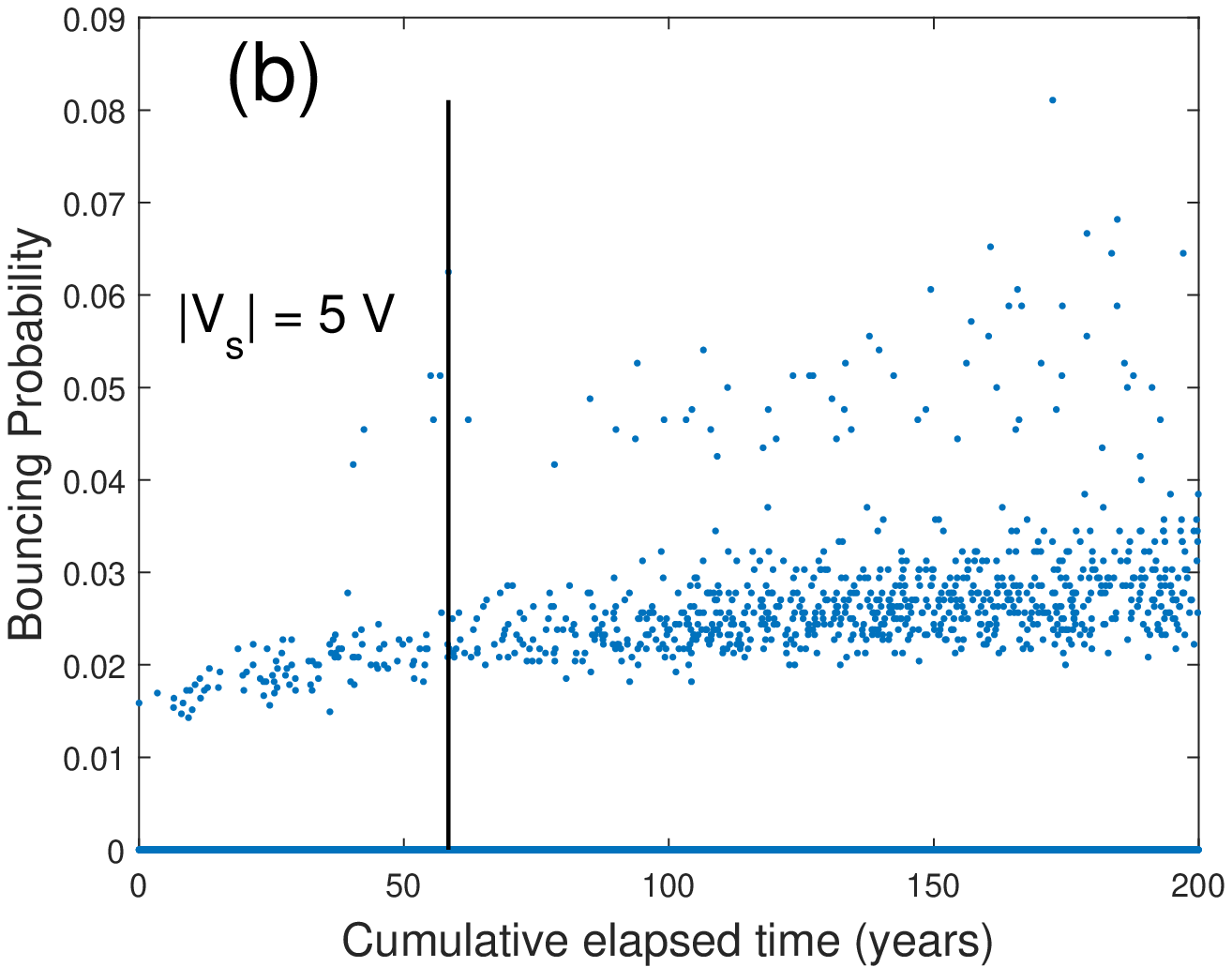} 
\caption{Bouncing probability as a function of elapsed time for a) neutral and b) charged (surface potential $\left | V_s \right |=5$ V) populations, with turbulence level $\alpha=10^{-2}$. Each point represents the ratio of the number of bouncing collisions to the total number of interactions for 100 consecutive interactions. The black vertical lines indicate the time at which the bouncing criterion is met.}
\label{f515}
\end{figure*}

We are interested in detecting true collisions, where constituent members of the two particles physically overlap. If the two particles are separated by a distance smaller than the sum of their radii, a collision check determines if any two monomers in the target and incoming aggregates overlap. A missed collision is detected when the two particles are moving away from each other. In this case, the code proceeds to the next iteration and new dust particles are selected. Upon a successful collision, the two aggregates are connected, and the total mass, charge, spin and moment of inertia are calculated for the resulting aggregate. This new aggregate replaces the incoming aggregate in the library. Subsequently, another aggregate is randomly chosen from the library (consistent with the particle weights) to replace the target aggregate, so that the total number of aggregates from the library stays constant. The collision probabilities $ C_{ij}$ are updated based on the new average equivalent radius of each bin and the change in the population of the dust particles. For each successful and missed collision, the masses, radii, equivalent radii, compactness factors, charges, relative velocities of the two particles, and the time interval between two interactions are recorded. Simulations progress until more than 5\% of 100 consecutive collisions results in bouncing. At this point, restructuring is expected to play a large role in the evolution of the population.  In general, the bouncing probability increases as the population grows. However, it develops differently for populations with different charges, due to their different growth behaviors, as shown in Figure \ref{f515}. Particles in neutral and weakly charged populations grow collectively, and the bouncing probability increases quasi-linearly after the initial stage. On the other hand, in highly charged cases, the growth is concentrated on a small proportion of the population while the majority of the population grows slowly. Therefore, a lot of hit-and-stick collisions still occur after meeting the bouncing criterion. The elapsed times and maximum particle sizes in populations when the bouncing criterion is met are listed in Table \ref{table1} and Table \ref{table2}. See section \ref{evo} for more discussion about different growth modes of dust populations in different environments.

\savebox{\tempbox}{\begin{tabular}{@{}r@{}l@{\space}}
&$\alpha$\\$\left | V_s \right |$
\end{tabular}}

\begin{table}[htb]
\caption{Maximum elapsed time (years)}
\lb{table1}
\[\begin{array}{c|cccccccccc}
\hline
\hline
\tikz[overlay]{\draw (0pt,\ht\tempbox) -- (\wd\tempbox,-\dp\tempbox);}%
\usebox{\tempbox}\hspace{\dimexpr 1pt-\tabcolsep}
&&&&$$10^{-2}$$&&&$$10^{-4}$$&&&$$10^{-6}$$\\
\hline
neutral &&&&0.5&&&3.3&&&24.2\\
0.1\ V &&&&0.4&&&3.8&&&58.3\\
0.5\ V&&&&1.0&&&52.0&&&2962.2\\
1\ V&&&&3.8&&&568.4&&&17987.0\\
5\ V&&&&58.4&&&23465.1&&&\\
\hline
\end{array}\]
\\[10pt]
{\justifying \noindent Maximum elapsed time when bouncing criterion is met (more than 5\% of previous 100 collisions result in bouncing), for different dust surface potentials ($V_s$) and turbulence levels ($\alpha$).\par}
\end{table}

\savebox{\tempbox}{\begin{tabular}{@{}r@{}l@{\space}}
&$\alpha$\\$\left | V_s \right |$
\end{tabular}}

\begin{table}[!htb]
\caption{$R_{\sigma, max}$ ($\mu m$)}
\lb{table2}

\[\begin{array}{c|cccccccccc}
\hline
\hline
\tikz[overlay]{\draw (0pt,\ht\tempbox) -- (\wd\tempbox,-\dp\tempbox);}%
\usebox{\tempbox}\hspace{\dimexpr 1pt-\tabcolsep}
&&&&$$10^{-2}$$&&&$$10^{-4}$$&&&$$10^{-6}$$\\
\hline
Neutral &&&&$$14.8\pm0.5$$&&&$$20.1\pm1.1$$&&&$$60.9\pm4.1$$\\
0.1\ V &&&&$$15.3\pm0.5$$&&&$$19.5\pm1.4$$&&&$$48.7\pm7.6$$\\
0.5\ V&&&&$$$$16.5\pm0.5$$$$&&&$$$$21.7\pm0.7$$$$&&&$$$$\textbf{81.2$\pm$0.4}$$$$\\
1\ V&&&&$$$20.0$\pm$0.4$$$&&&$$\textbf{44.8$\pm$2.0}$$&&&$$\textbf{116.2$\pm$0.2}$$\\
5\ V&&&&$$\textbf{23.1$\pm$0.8}$$&&&$$\textbf{64.5$\pm$0.4}$$&&&\\
\hline
\end{array}\]
\\[10pt]

{\justifying \noindent Average equivalent radius of the top 0.5\% largest particles when bouncing criterion is met (more than 5\% of previous 100 collisions result in bouncing), for different dust surface potentials ($V_s$) and turbulence levels ($\alpha$). The populations exhibiting runaway growth are highlighted in bold.\par}


\end{table}

\subsection{Charging of Particles}

In the case of charged particles, the electrostatic interaction exerts a force and a torque on the particles, causing deceleration (for like-charged particles) and rotation. The deceleration and deflection can alter the porosity of the resultant aggregate or result in a missed collision.

The charges on the particles are calculated using the Orbital Motion Limited Line of Sight (OML\_LOS) method (Matthews et al. 2012, 2016; Ma et al. 2013), in which dust is charged by the collection of charged particles from the surrounding plasma. Each monomer surface in the dust aggregate is divided into patches, and the charging currents due to incoming plasma species (ions and electrons) are calculated for each patch. The current density of plasma species $\alpha$ to a given surface patch depends on grains' surface potential, the speed (temperature) of the plasma particles and their number density, as well as the LOS\_factor for that patch, calculated from the directions from which a plasma particle can approach the surface of the patch, which depends on the open lines of sight from the center of the patch.  Lines of sight may be blocked by other monomers in the aggregate, so that monomers in the interior of the aggregate have small LOS\_factors, and thus collect little charge [note that the connection of two colliding particles can block the originally open lines of sight, which changes the patches' LOS\_factors, resulting in redistribution of charge on the resulting aggregate. Therefore, upon a successful collision, the LOS\_factor is recalculated for each patch based on the new monomer distribution]. The equilibrium charge distribution is obtained by summing the charge collected during a time $\Delta t$ and iterating until the charge changes by less than 0.1\%. The total charge of the aggregate is calculated by summing the charge collected on each patch at equilibrium (a detailed explanation is presented in Matthews et al., 2012).

For computational expediency, at large distances, the electrostatic force and torques are calculated using a multipole approximation up to the quadrupole terms, where the multipole moments are calculated from the charge on each patch and its position relative to the COM.  At close approach, i.e., the distance between the two particles is smaller than twice the radius of the target particle, the charge approximation using the dipole and quadrupole moments breaks down, and the electrostatic torques on the aggregates are calculated using the total charge on each monomer (see Matthews et al. 2016).  
\\

\subsection{Initial conditions}

Coagulation was modeled for conditions at the midplane of a turbulent protoplanetary disk, at a distance of 1 AU from the central young stellar object (YSO), assuming a gas temperature ($T$) of 280 K. The average molecular mass $\mu $ is taken to be 2.34 $g/mol$, and the sound speed is given by
\bqn
\lb{117}
c_{g}=\left (\gamma k_{B} T/\mu  m_{H} \right )^{1/2},
\eqn
with $\gamma$ the ratio of $C_{p}$ to $C_{v}$ for a diatomic molecule,  $k_{B}$ the Boltzmann constant, and $m_{H}$ the mass of hydrogen. The molecular viscosity of the gas is calculated as
\bqn
\lb{116}
v_{m}=\frac{\sqrt{(2/\pi)}\mu  m_{H}c_{g}}{\rho_{g} \sigma_{coll}},
\eqn
with $\rho_{g}$ the gas density and $\sigma _{coll}$ the collisional cross section of gas, set to be $\sigma _{coll}=2\times 10^{-19}~m^{2}$ (Okuzumi 2011). The ratio of the dust density to the gas density is assumed to be 0.01. 

The simulation starts with an initial population of $N$ = 10,000 spherical silicate monomers with radii $a$ evenly spaced in the range of [0.5, 10] $\mu$m and weights given by the power-law MRN distribution $n(a)da\propto a^{-3.5}da$, where $n(a)da$ is the number of particles in the size interval $(a,a+da)$ (Mathis et al. 1977). The monomers are divided into 100 logarithmic bins according to their radii for the calculation of the selection probability ($C_{ij}$). 

The precise value of the turbulence strength in protoplanetary disks is uncertain, and the values often considered range from $\sim 10^{-6}$ to 0.1 (Carballido 2011; Hartmann et al., 1998; Cuzzi 2004; Ormel 2007). In this study, we investigate turbulence strengths $\alpha=10^{-6}, 10^{-4}, 10^{-2}$.


In our simulation, the plasma environment is assumed to be hydrogen with equal electron and ion temperature, $T_{e}$ = $T_{i}$ = 280 K, as the plasma thermalizes with the gas due to collisions. In the case of low dust density, a negligible percentage of the electrons reside on the dust grains, and the number density of electrons and ions in the gas are equal. Here we use $n_{e}=n_{i}=3.5\times 10^{8}\ m^{-3}$ (Horanyi \& Goertz, 1990). The corresponding surface potential of the dust particles charged by primary plasma currents is -0.061 V. However, dust can also be charged by secondary electron emission and photoelectric emission from high energy UV and X rays. In the inner solar system, impinging electrons do not have sufficient energy to knock secondary electrons off dust particles, and thus the main charging processes in this region are the primary plasma currents and photoelectric emission. The level of UV radiation varies over locations in the disk; the inner, denser regions are less exposed to radiation, and therefore the effect of photoelectric emission is reduced. The photon flux, i.e., the product of absorption efficiency, yield and solar flux (Ma et al. 2013), ranges from $5\times 10^{12}\ m^{-2}s^{-1}$ to $1.5\times 10^{13}\ m^{-2}s^{-1}$ at 1 AU (Tobiska 1991), which can charge dust particles up to a few volts (Ma et al, 2013). A porous aggregate can collect more free electrons and absorb more UV photons due to greater surface area than can a compact aggregate of the same mass. Therefore, the efficiencies of both the plasma charging and photoelectric emission are positively related to the open area of an aggregate, and the OML\_LOS method used for plasma charging is also applicable to photoelectric emission (Ma et al. 2013). In order to explore the process of dust growth under various charging conditions, we do not specify the value of photon flux, but rather integrate the effect of photoelectric emission into the dust surface potential. We employ values of $\left | V_s \right |=0$, $0.1$, $0.5$, $1$ and $5$ V to cover the range of possible surface potentials due to charging through direct collection of plasma particles or charging through photoemission (Matthews et al. 2012; Ivlev et al. 2016). Although we simulate dust growth at 1 AU, these results can be applied to other regions of the disk with the same surface potential.


\section{RESULTS}
\renewcommand{\theequation}{4.\arabic{equation}} \setcounter{equation}{0}

In this section we compare the collision outcomes, the morphology of the resulting aggregates, as well as the evolution of dust populations for charged and neutral aggregates, in varying levels of turbulence.

\begin{figure*}[!htb]
\includegraphics[width=9cm]{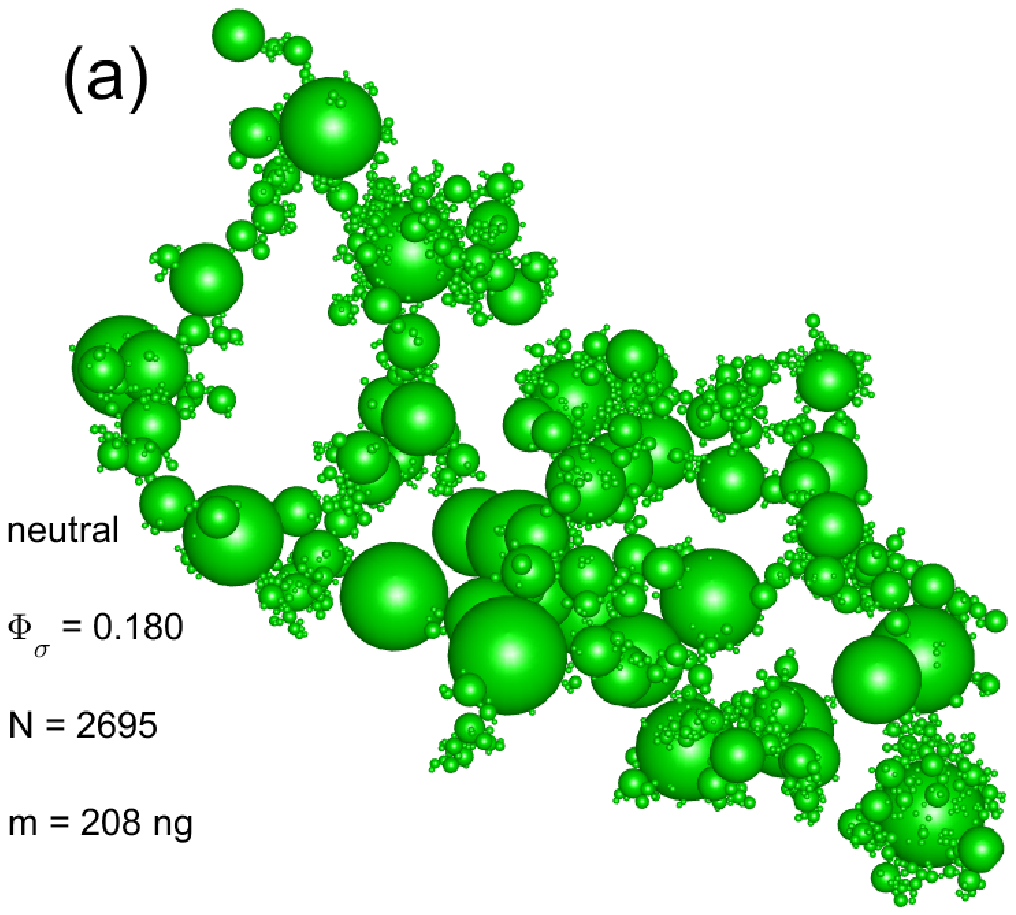}\includegraphics[width=9cm]{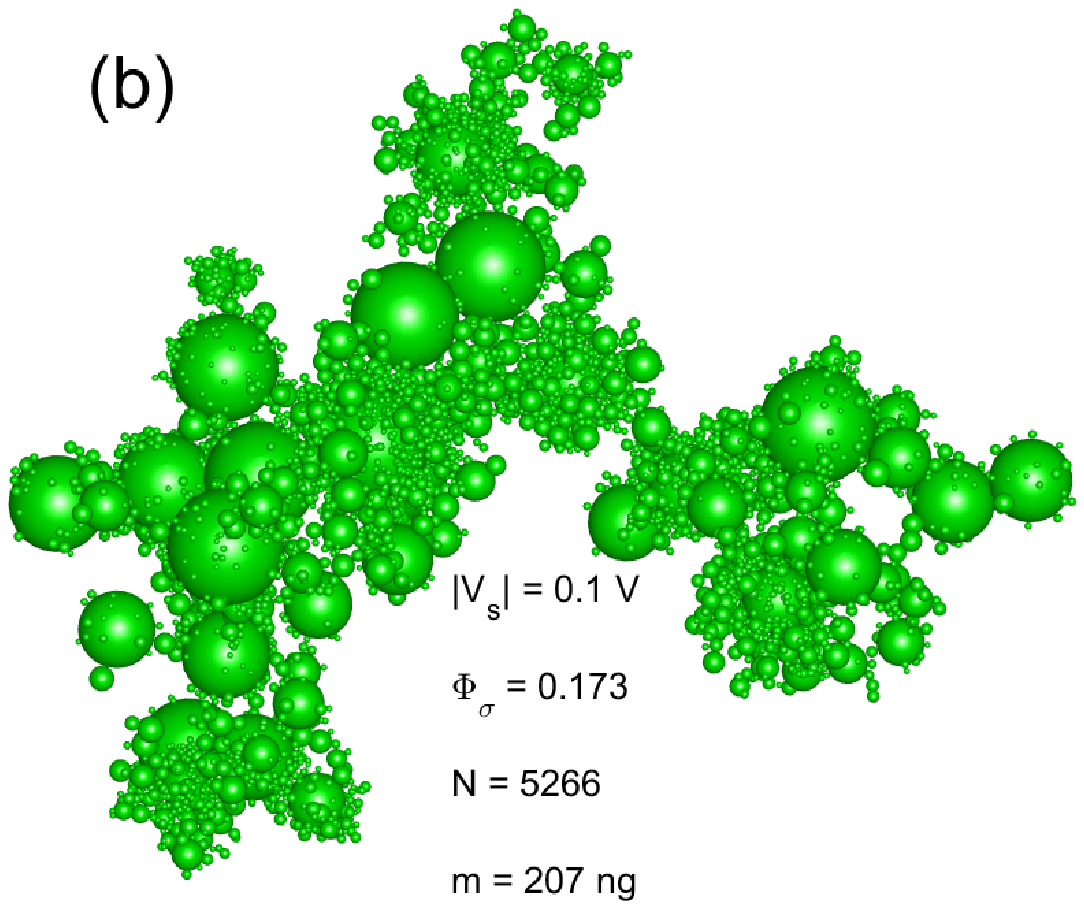}\ 
\includegraphics[width=9cm]{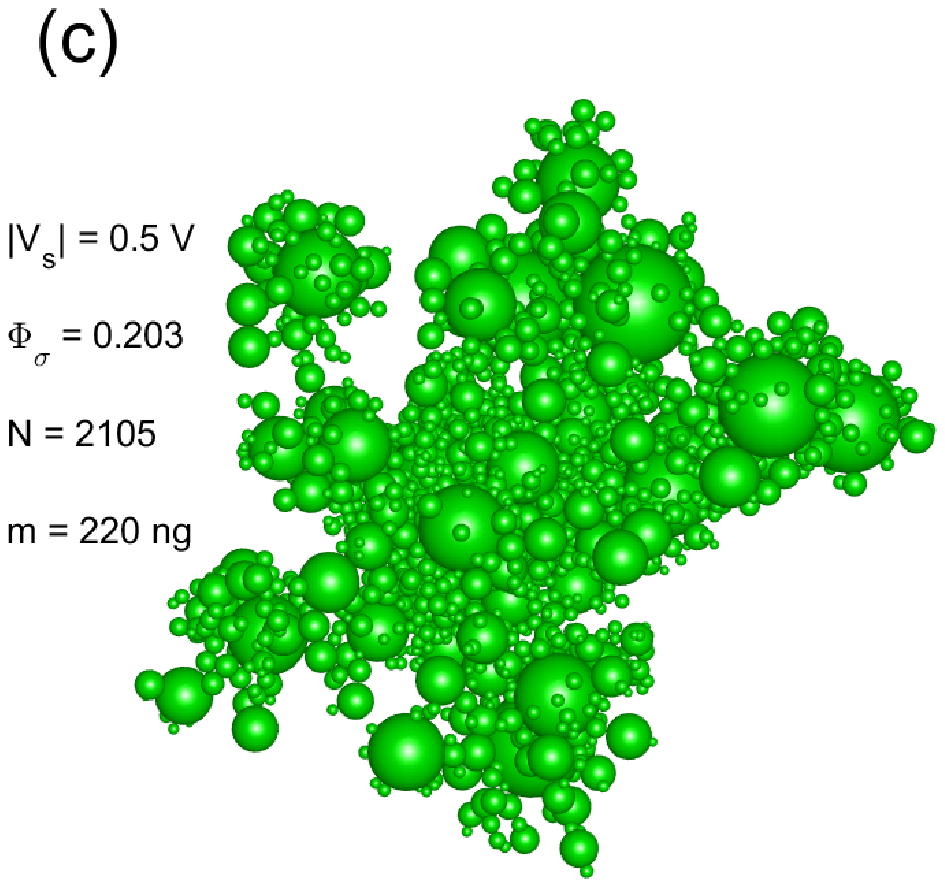}\includegraphics[width=9cm]{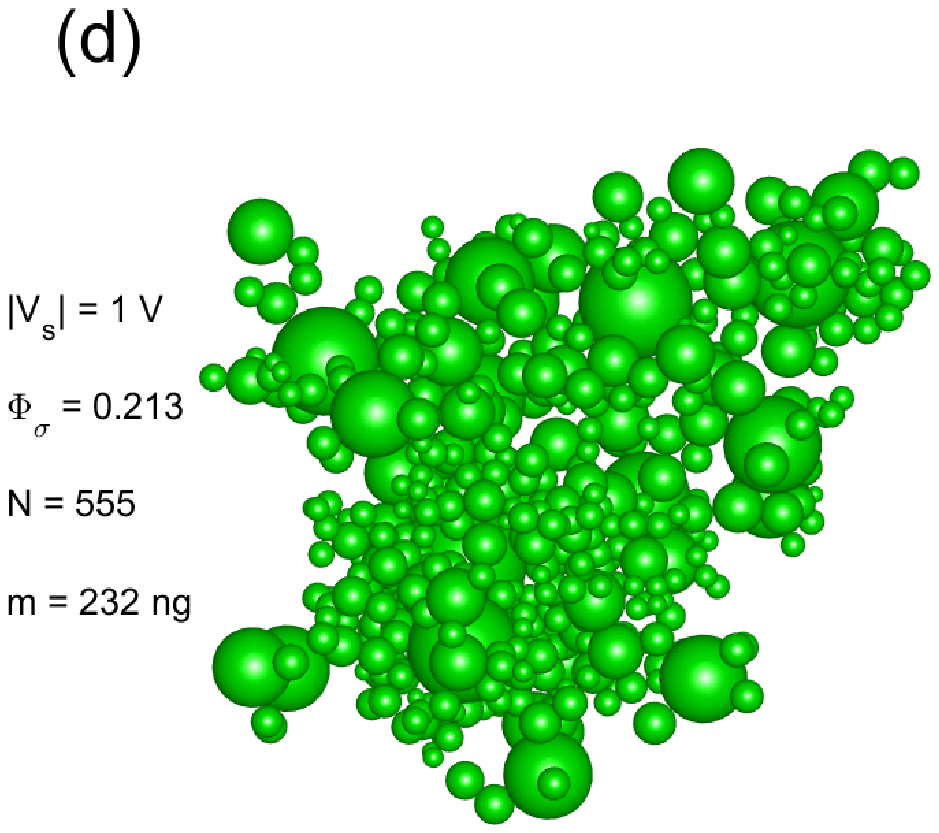}\ 
\caption{Representative aggregates with an equivalent radius of $R_{\sigma}\sim 48\ \mu m$ formed in environments with turbulence level $\alpha=10^{-6}$. The surface potential $\left | V_s \right |$ is a) neutral, b) 0.1 V, c), 0.5 V and d) 1 V. The compactness factor $\Phi _{\sigma} $, number of monomers $N$ and the mass $m$ are shown for each aggregate.}
\label{f1}
\end{figure*}

Figure \ref{f1} shows examples of aggregates formed in different plasma conditions with turbulence strength $\alpha=10^{-6}$. The selected aggregates have similar equivalent radii, but the difference between them is readily apparent: the neutral aggregate and aggregate with the smallest charge (Figs. \ref{f1}a, \ref{f1}b) are the most porous and contain large numbers of the smallest monomers ($< 1\ \mu m$). Compared to the neutral aggregate, the aggregate with the smallest charge has the greatest proportion of small monomers which tend to be attached to large monomers. As the surface potential of the particles increases, fewer small monomers are incorporated into the aggregates, the minimum size of monomers increases, and the porosity decreases (Figs. \ref{f1}c, \ref{f1}d). In addition, the highly charged aggregates are more spherical (lower aspect ratios) and symmetrical, while the neutral and weakly charged particles have more irregular shapes.

\subsection{Monomer size distribution within aggregates}

As shown in Figure \ref{f1}, highly charged aggregates contain larger monomers than do weakly charged and neutral aggregates, as small monomers have a higher charge to mass ratio and are repelled from highly charged grains. As aggregates grow larger, they develop greater relative velocity with respect to small dust grains, which enable these grains to overcome the electrostatic barrier and be incorporated into the large aggregates. As a result, for environments with low turbulence level and high surface potential, the percentages of small ($r< 1.43\ \mu m$) and medium (1.43 - 7.14 $\mu m$) monomers within aggregates increase over time, while the percentage of large monomers (7.14 - 10 $\mu m$) decreases over time, leading to a decrease in the average monomer size within aggregates [the small monomers and large monomers are defined as the bottom and top 20\% (in mass) of the initial population]. In a strongly turbulent environment, neutral and weakly charged grains incorporate a greater percentage of large monomers over time, but the average monomer size within aggregates is quasi-constant. 

\begin{figure*}[!htb]
\includegraphics[width=6cm]{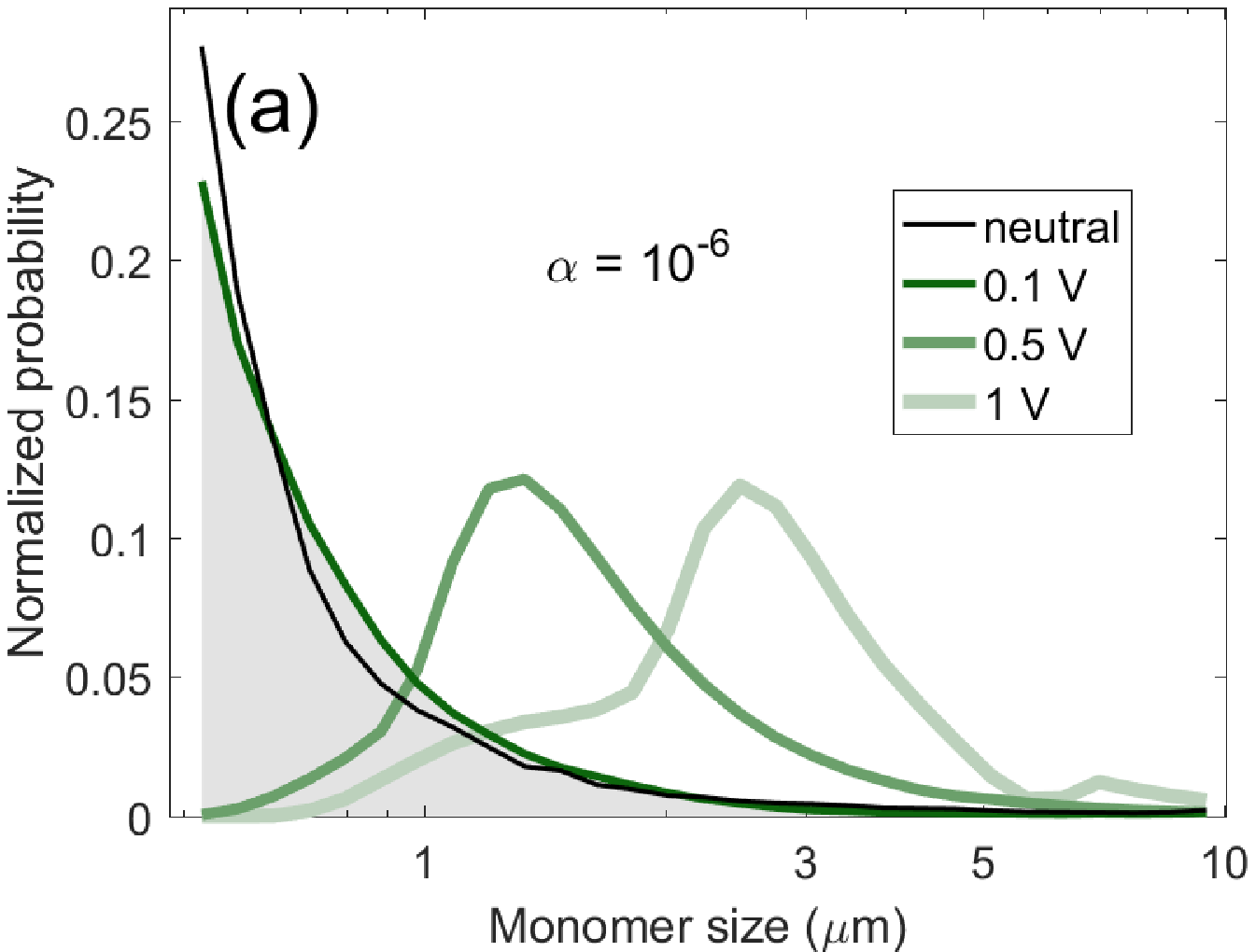}\includegraphics[width=6cm]{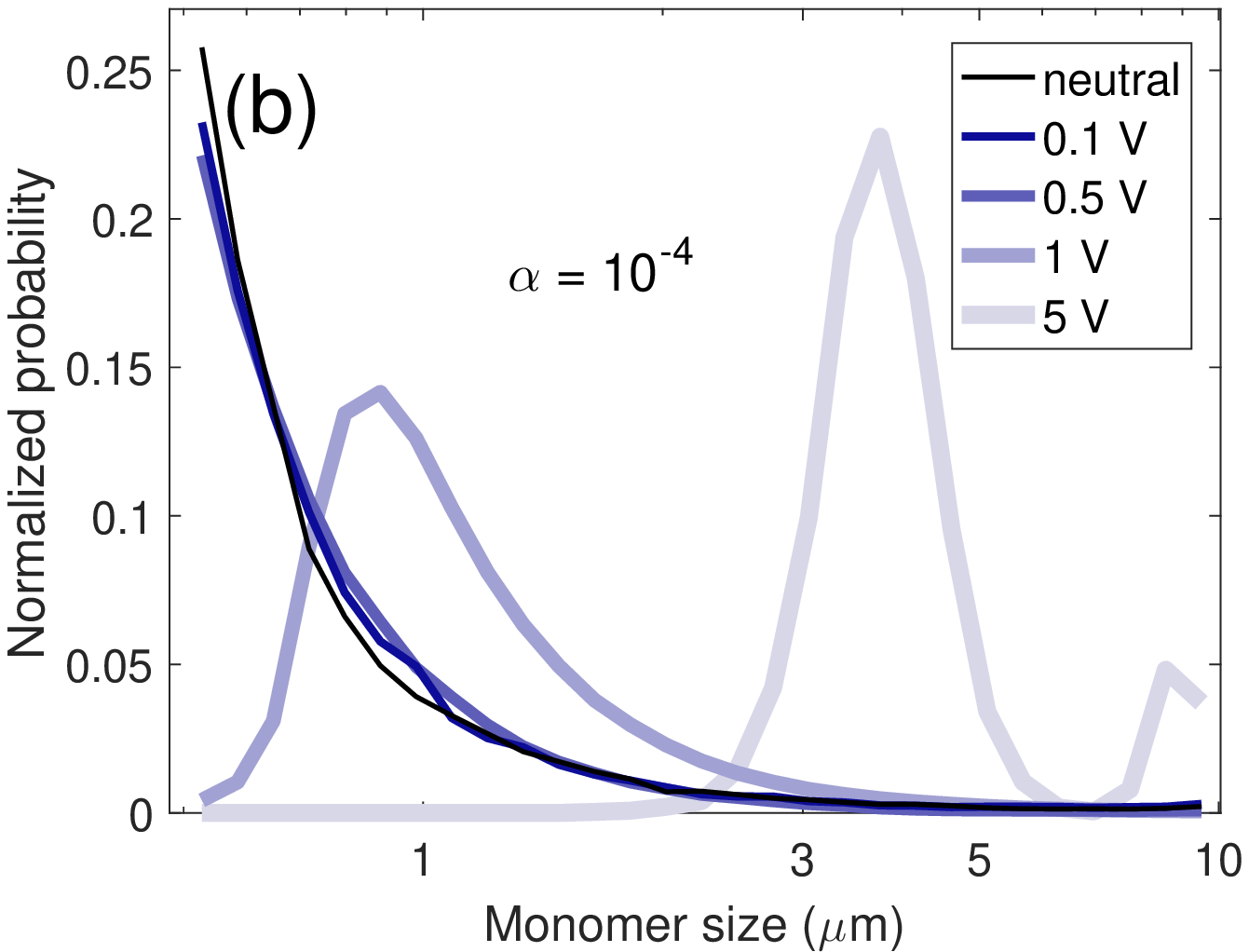}\includegraphics[width=6cm]{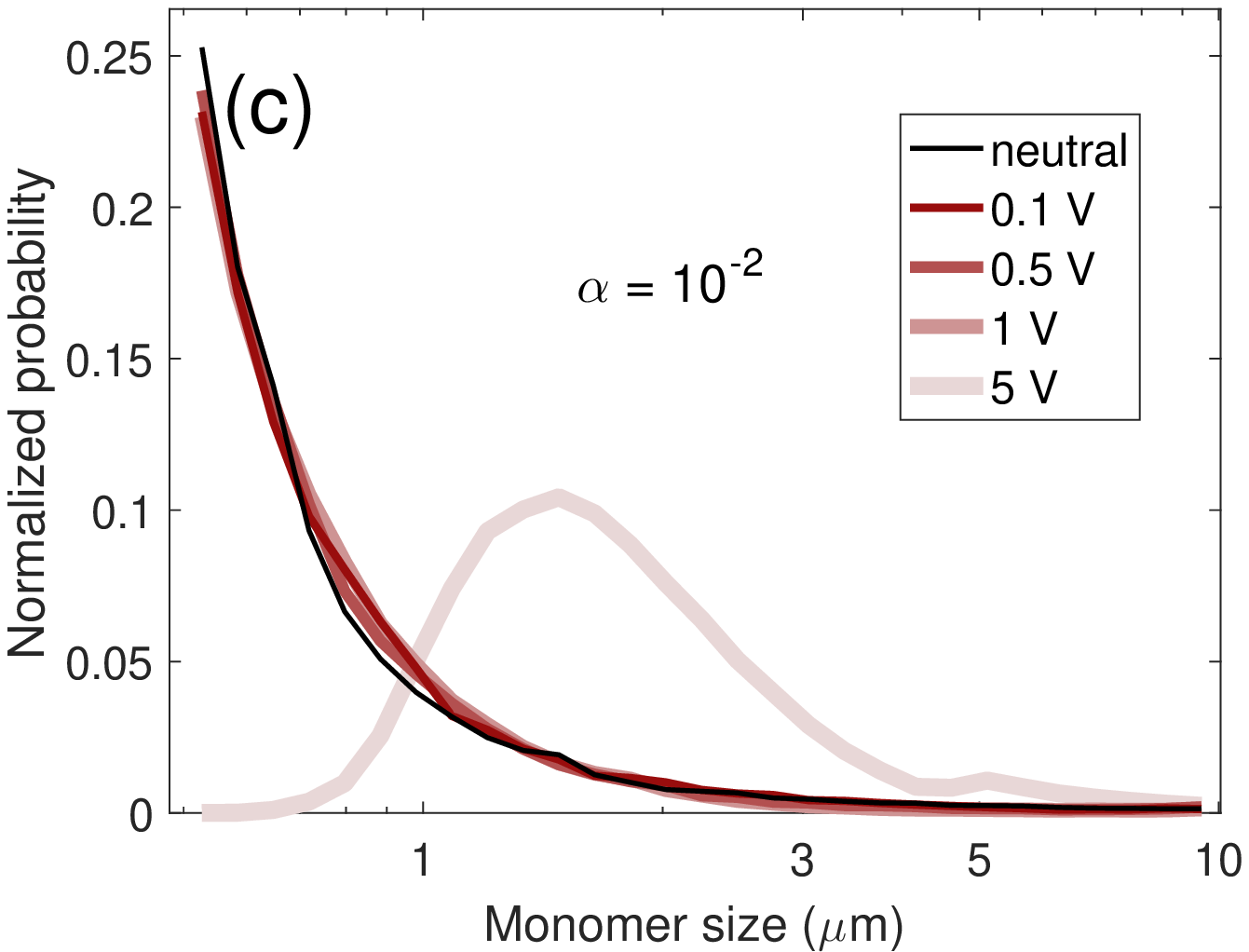} 
\caption{Monomer size distribution within aggregates (surface potential $\left | V_s \right |=0$, $0.1$, $0.5$, $1$, $5$ V; in order of decreasing color shades), at the end of the simulation (when the bouncing criterion is met). The shadowed area in a) indicates the size distribution of the initial population. Turbulence levels are $\alpha=10^{-6}$ for a), $\alpha=10^{-4}$ for b) and $\alpha=10^{-2}$ for c).}
\label{2}
\end{figure*}

A comparison of the probability distribution of the monomer sizes incorporated into the aggregates at the end of the simulation (when more than 5\% of collisions result in bouncing) is shown in Figure \ref{2}. At all turbulence levels, the distribution of monomers in the neutral and weakly charged aggregates matches the initial distribution of monomers sizes. As the surface potential increases, the monomer distribution shifts towards larger monomer sizes, and the shift increases with lower turbulence level. For environments with very low turbulence and very high charge, small monomers are not incorporated into aggregates due to repulsion. [note that for $\alpha = 10^{-4}$ and $\left | V_s \right |=5$ V, the curve is shifted beyond the monomer size of $3\ \mu m$, reflecting the fact that aggregate growth is very slow as there are so few of the large monomers which can overcome the Coulomb repulsion barrier.]

\subsection{Porosity}\label{por}

Porosity plays an important role in the collision probability because open structures couple more strongly to the gas, which reduces the relative velocity between grains. Although open, porous aggregates have a larger collisional cross section, which enhances the collision rate and aids them in out-growing the radial drift barrier (Garcia et al. 2016), small aggregates can pass through the gaps of the extended arms, making collisions less likely. In addition, the diversity of aggregate porosity can enhance collisions between particles of similar sizes by increasing the difference between their friction times. 

Charge influences the porosity of aggregates in different ways, and the porosity of the resulting aggregates is a balance between several factors. The electrostatic force causes small grains to be repelled, and the lack of small monomers filling in the gaps of aggregates increases the aggregate porosity. On the other hand, low-velocity collisions enable particles to have sufficient time to alter their path or rotate to minimize the potential energy of the configuration, producing more compact structures. This effect is most noticeable for weakly turbulent and/or highly charged environments, in which grains either have small turbulent velocities or are slowed down greatly by the electrostatic repulsion.

\begin{figure*}[!htb]
\includegraphics[width=16cm]{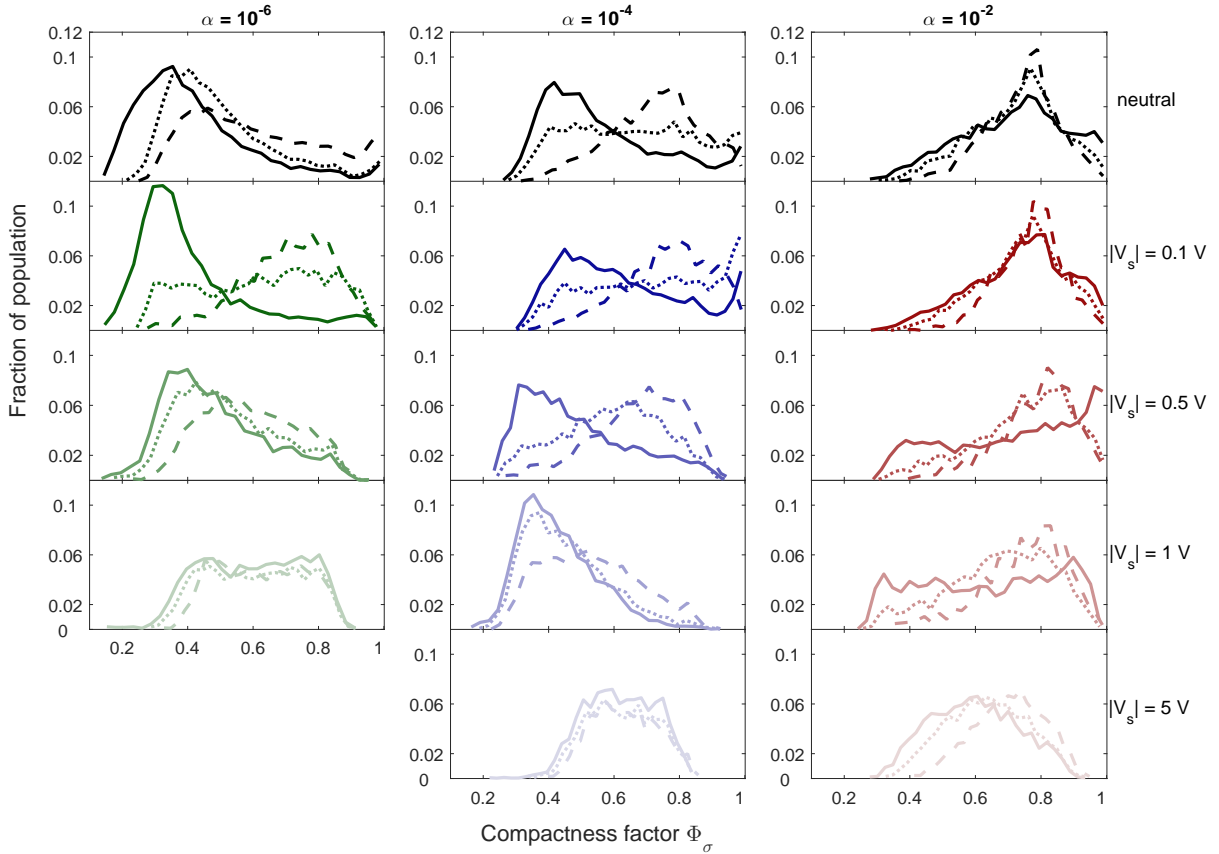}
\caption{Distribution of compactness factor in the population for aggregates with more than 10 monomers and surface potential $\left | V_s \right |=0$, $0.1$, $0.5$, $1$, $5$ V, in order of decreasing color shades (from top to bottom). The turbulence levels are $\alpha=10^{-6}$ for the left column (green), $\alpha=10^{-4}$ for the middle column (blue) and $\alpha=10^{-6}$ for the right column (red). The solid curves represent the time when 5\% of collisions result in bouncing. The dashed and dotted curves represent 1/3 and 2/3 of the total elapsed time.}
\label{f51}
\end{figure*}

The distributions of compactness factor of all aggregates within the population are compared for all charging conditions and turbulence levels, as shown in Figure \ref{f51}. The porosity of aggregates increases as they grow in size through hit-and-stick collisions. Therefore, as time progresses, a population shifts to lower compactness factors, i.e., contains a greater fraction of porous aggregates. The neutral and weakly charged populations have narrower distributions, and the peaks shift towards smaller compactness factor (greater porosity) as the populations grow and turbulence level decreases. On the other hand, the highly charged populations have broader distributions which change very little over time. 
 
\begin{figure*}[!htb]
\includegraphics[width=14cm]{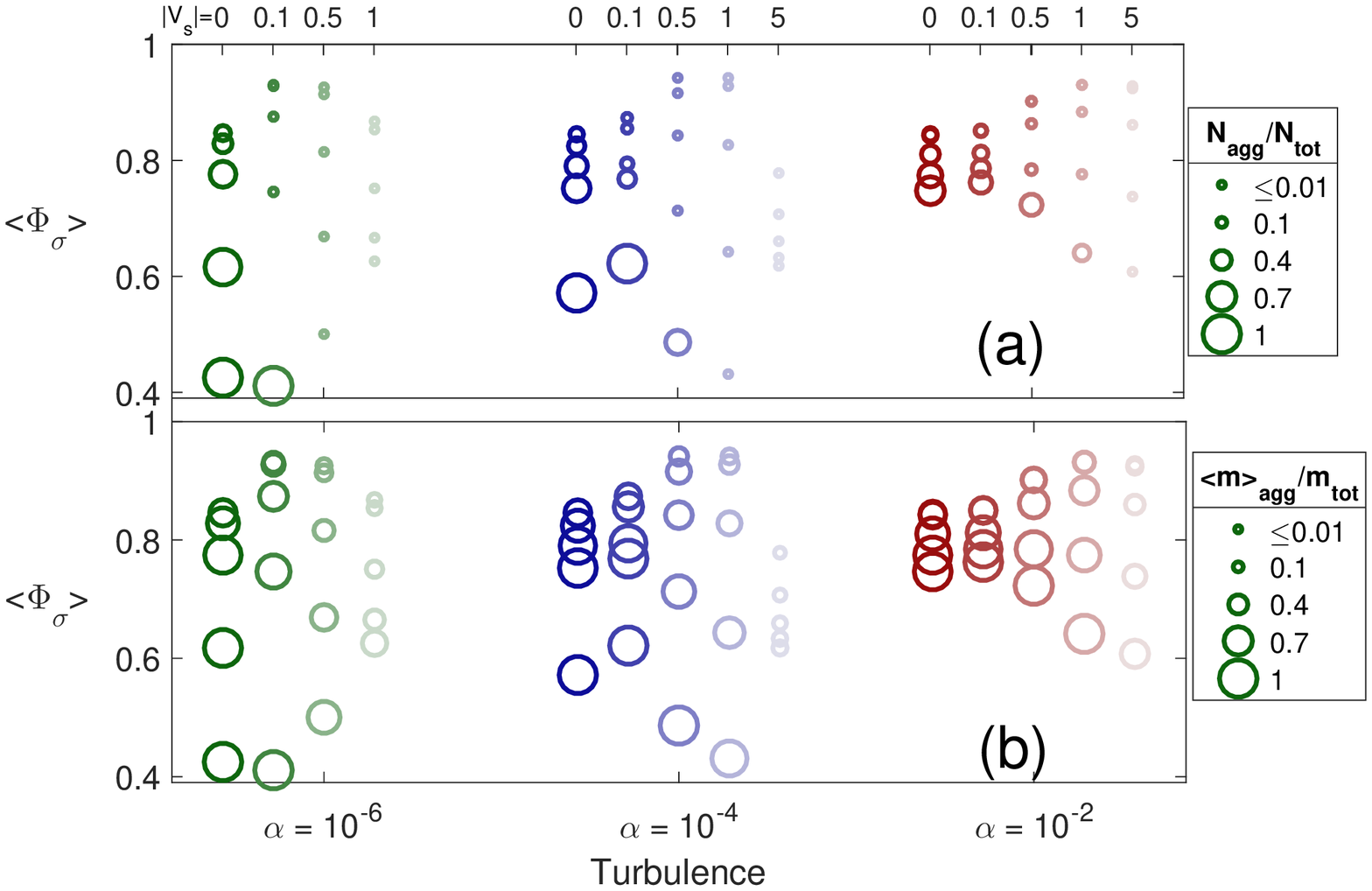}
\caption{Average compactness factor of all aggregates within the population, at different time points. In each column, the circles progress in time from top to bottom for equally spaced times until the bouncing criterion is met. The size of the circle represents the ratio of the number of aggregates to the total number of particles in the population for a), and the ratio of the mass of aggregates to the total mass of the population for b).}
\label{f50}
\end{figure*}

Figure \ref{f50} shows the mean porosity of all aggregates within the population at different time points. Each circle represents a time point which progresses from top to bottom in each column. For all environmental conditions, the mean porosity increases as the population grows. Given the same turbulence level, aggregates in highly charged populations are more compact than those with a low charge after equal elapsed time. However, at the time when the bouncing criterion is met, the highly charged aggregates are more porous in strong turbulence ($\alpha=10^{-2}$) and more compact in weak turbulence ($\alpha=10^{-4}$, $10^{-6}$). The size of the circle represents the ratio of the number of aggregates to the total number of particles in the population for Fig. \ref{f50}a, and the ratio of the mass of aggregates to the total mass of the population for Fig. \ref{f50}b. It is seen that for neutral or weakly charged dust, the number of aggregates present (Fig. \ref{f50}a) grows in time just as the fraction of the mass contained in aggregates grows (Fig. \ref{f50}b). In contrast, for highly charged dust populations ($\left | V_s \right |\geqslant 0.5$ V for $\alpha=10^{-6}$; $\left | V_s \right |\geqslant 1$ V for $\alpha=10^{-4}$; $\left | V_s \right |\geqslant 5$ V for $\alpha=10^{-2}$), the number of aggregates stays small (Fig. \ref{f50}a) while the fraction of mass within the total population contained in aggregates grows (Fig. \ref{f50}b). This indicates that in the highly charged populations the dust growth is concentrated on a small number of particles, indicating runaway growth (see next section for more discussion).

\begin{figure*}[!htb]
\includegraphics[width=6cm]{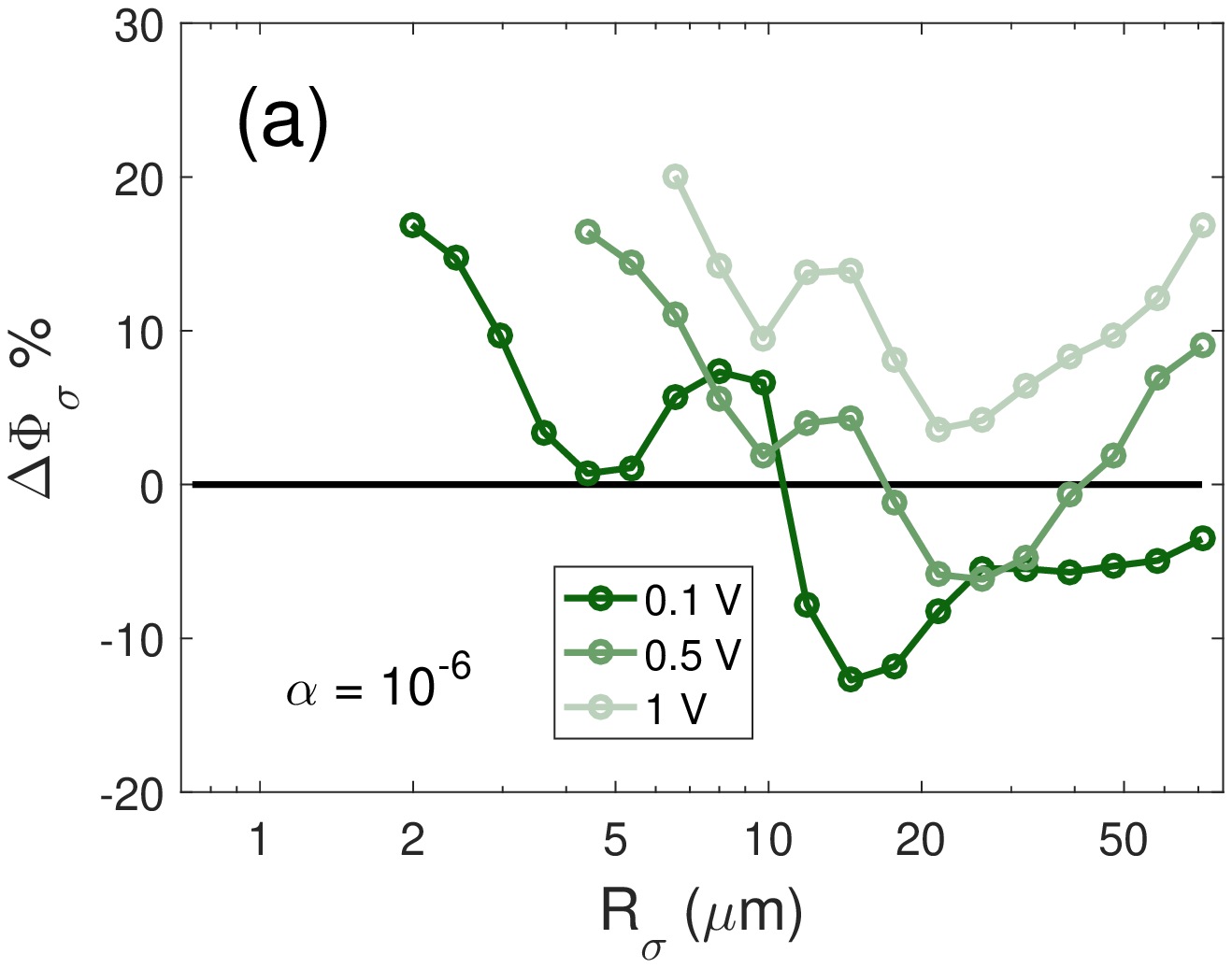}\includegraphics[width=6cm]{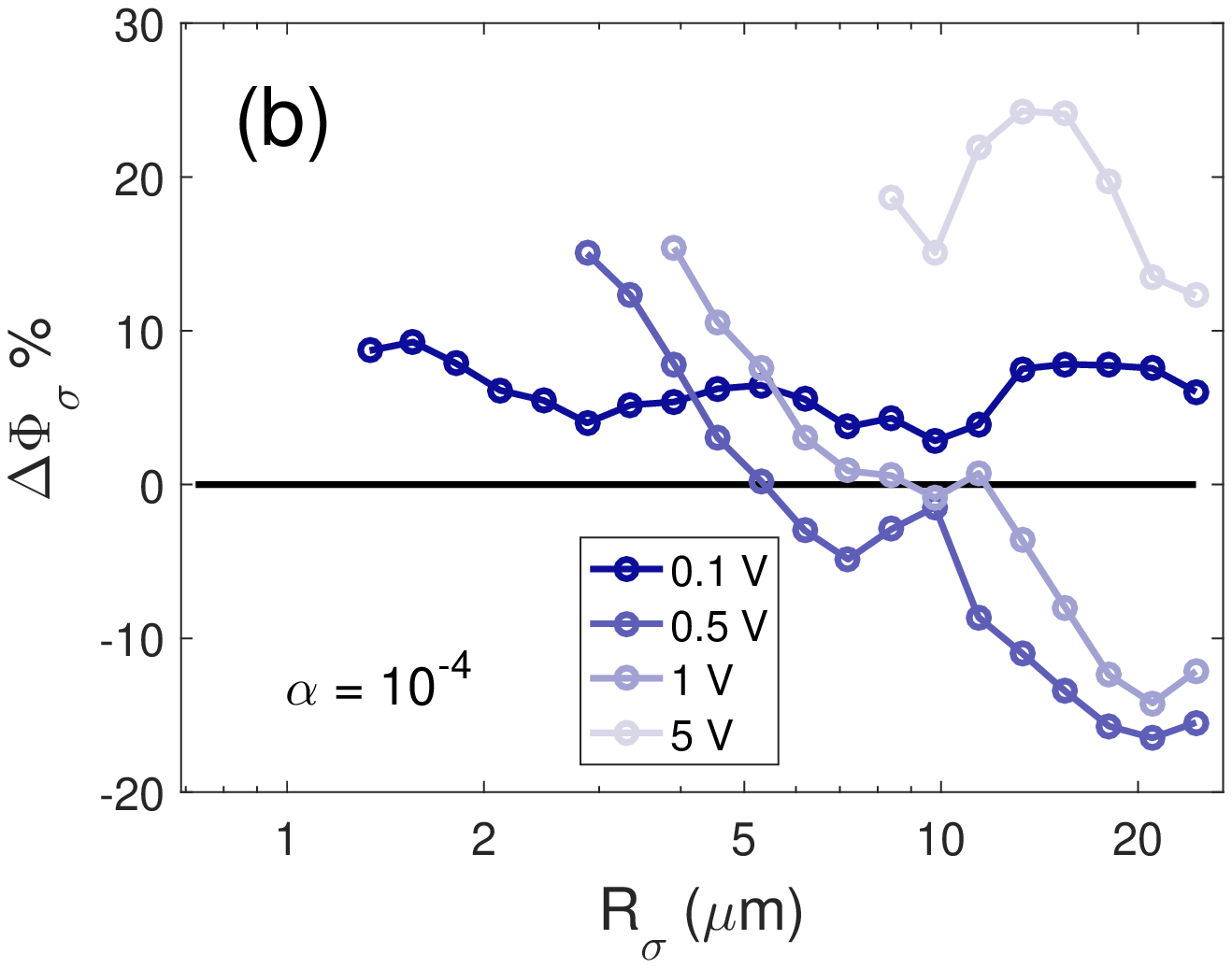}\includegraphics[width=6cm]{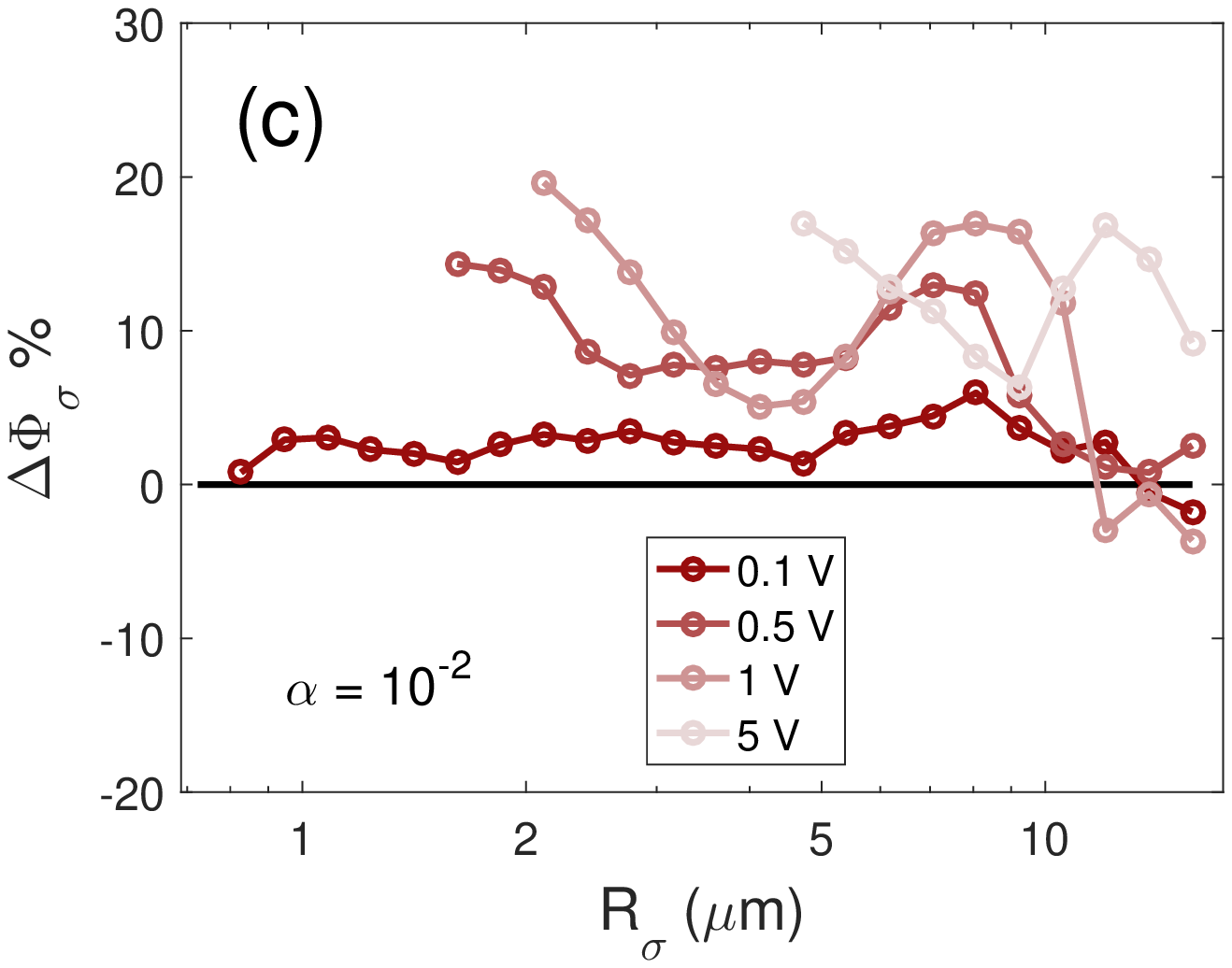} 
\caption{Percent difference of compactness factor for charged aggregates relative to neutral aggregates as a function of equivalent radius. Turbulence levels are $\alpha=10^{-6}$ for a), $\alpha=10^{-4}$ for b) and $\alpha=10^{-2}$ for c). Results shown are for all aggregates formed over the course of the simulation until the bouncing criterion is met.}
\label{f5}
\end{figure*}

Comparisons of the compactness factor for charged and neutral aggregates for different charging conditions and turbulence levels are shown in Figure \ref{f5}. Here the data is shown relative to the porosity of neutral aggregates with the same equivalent radius. For aggregates smaller than $\sim 10 \mu m$, charged particles are more compact ($\Delta \Phi _{\sigma}> 0$) than neutral particles in all cases. For large aggregates, charged particles are either more compact or more porous ($\Delta \Phi _{\sigma}< 0$) than neutral particles, depending on the environmental conditions. In general, highly charged particles tend to be more compact than those with a low charge for the same turbulence level, consistent with Figure \ref{f51}.


\subsection{Time evolution of the aggregate population}\label{evo}

\begin{figure*}[!htb]
\includegraphics[width=6cm]{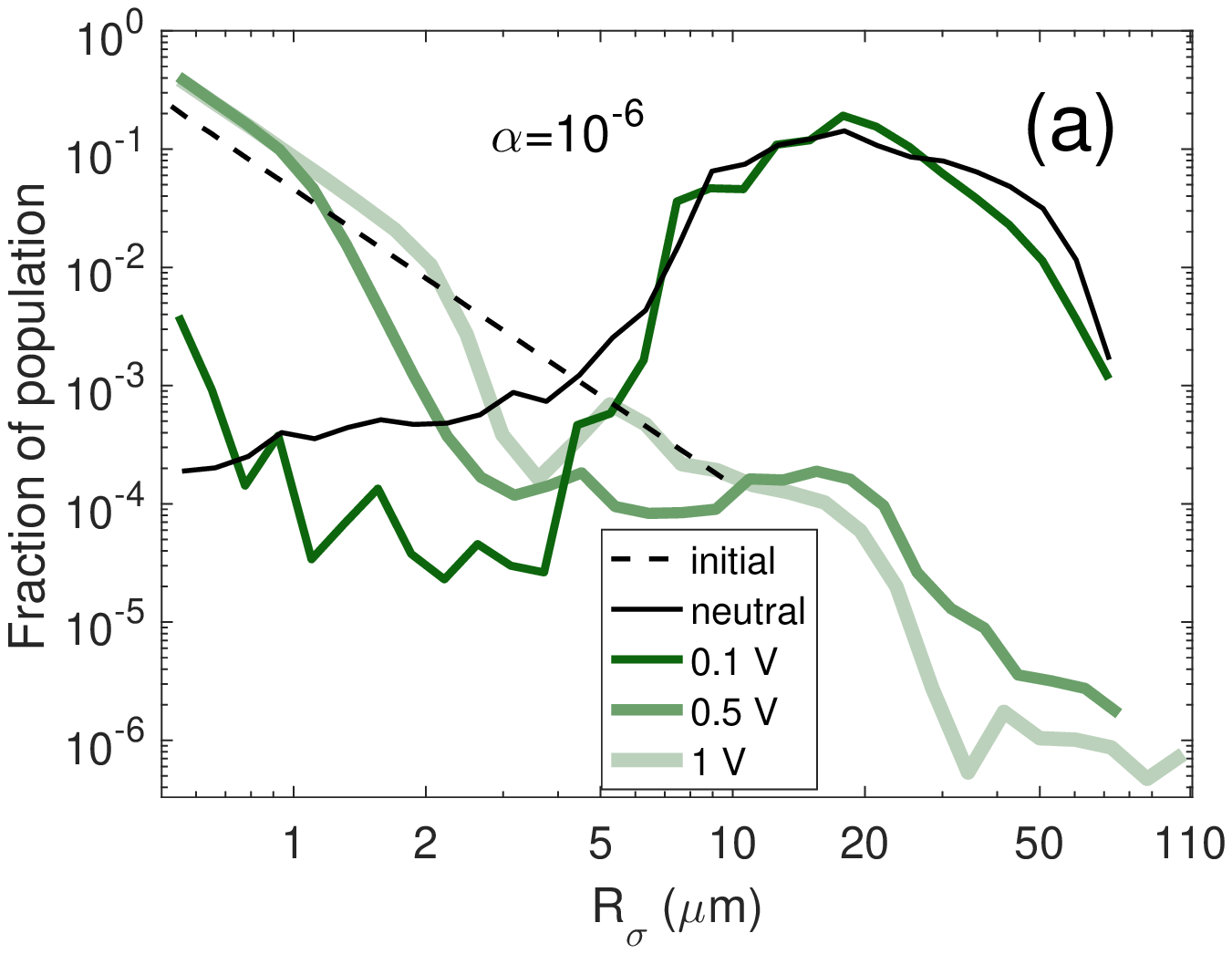}\includegraphics[width=6cm]{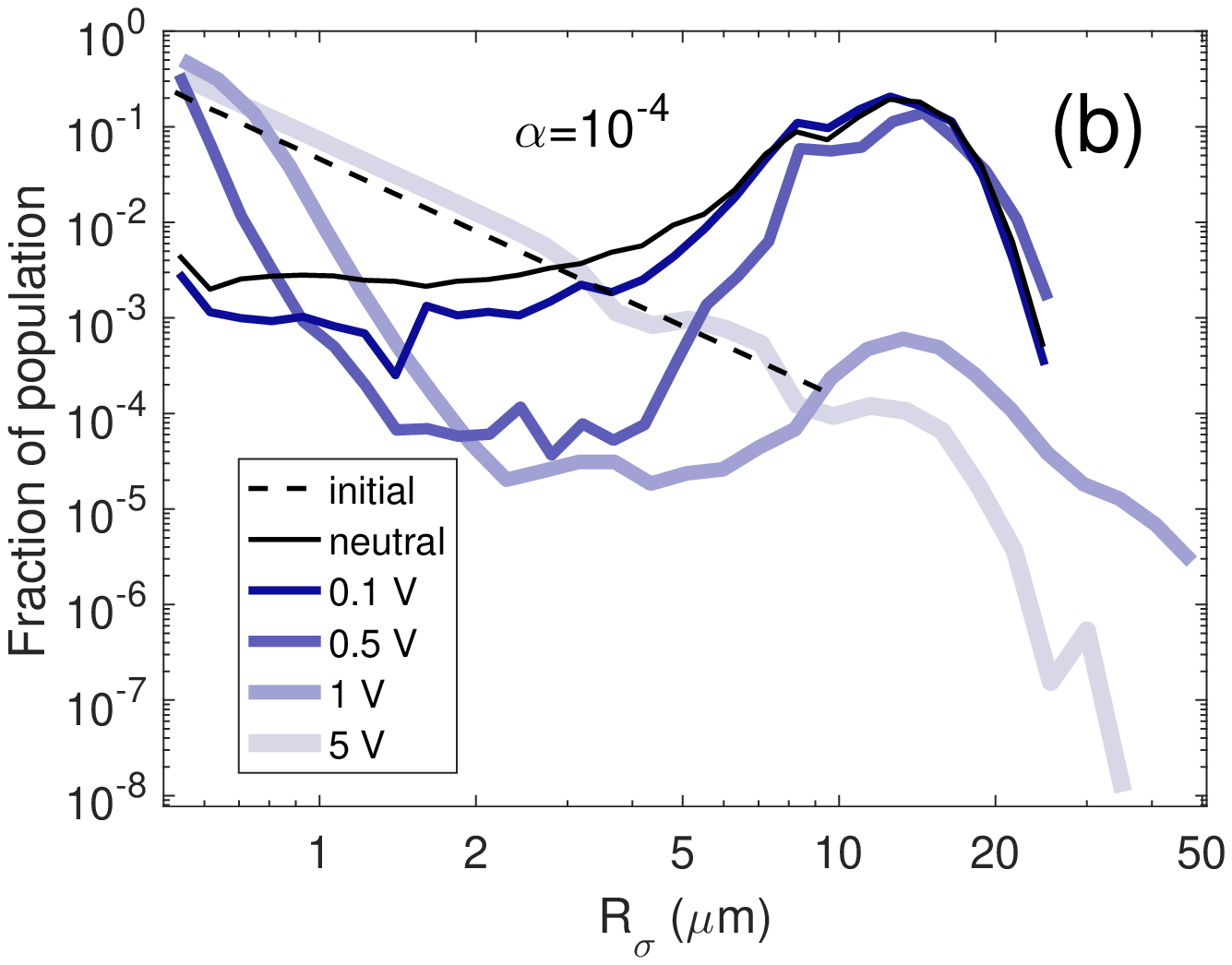}\includegraphics[width=6cm]{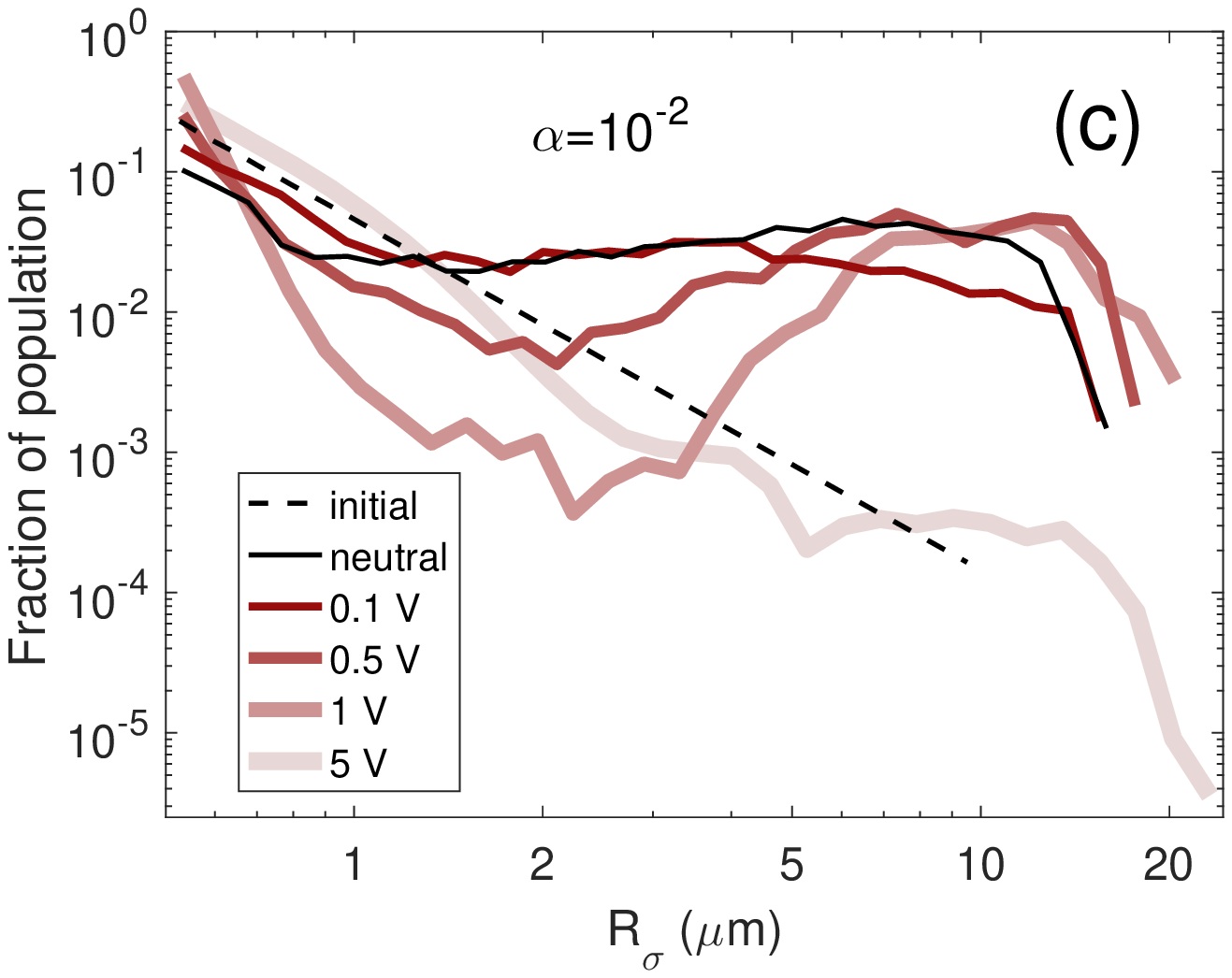} 
\caption{Size distribution of the population for charged (surface potential $\left | V_s \right |=0$, $0.1$, $0.5$, $1$, $5$ V; in order of decreasing color shades) and neutral particles (darkest shades), at the end of the simulation (more than 5\% of collisions result in bouncing). The black dashed lines indicate the distribution of initial population. Turbulence levels are $\alpha=10^{-6}$ for a), $\alpha=10^{-4}$ for b) and $\alpha=10^{-2}$ for c). }
\label{f9}
\end{figure*}

Fig. \ref{f9} compares the size distribution of the dust grains in the population for all of the turbulence strengths and charging levels when the bouncing criterion is met for each population. In highly charged cases ($\left | V_s \right |\geqslant 0.5$ V for $\alpha=10^{-6}$; $\left | V_s \right |\geqslant 1$ V for $\alpha=10^{-4}$; $\left | V_s \right |\geqslant 5$ V for $\alpha=10^{-2}$), dust particles grow to a larger size than in neutral and weakly charged cases, and the large particles represent a very small proportion of the population. Overall, the size distributions of the highly charged populations do not deviate much from the initial size distribution, meaning that a lot of monomers and small aggregates remain in the population while few large aggregates grow in size, which indicates a runaway growth. In contrast, for neutral and weakly charged populations, almost all monomers have collided and formed aggregates before the bouncing criterion is met, and the size distributions peak around the average particle size. The deviation of the population from the initial mass distribution is negatively correlated with the charge level and turbulence level.

\begin{figure*}[!htb]
\includegraphics[width=6cm]{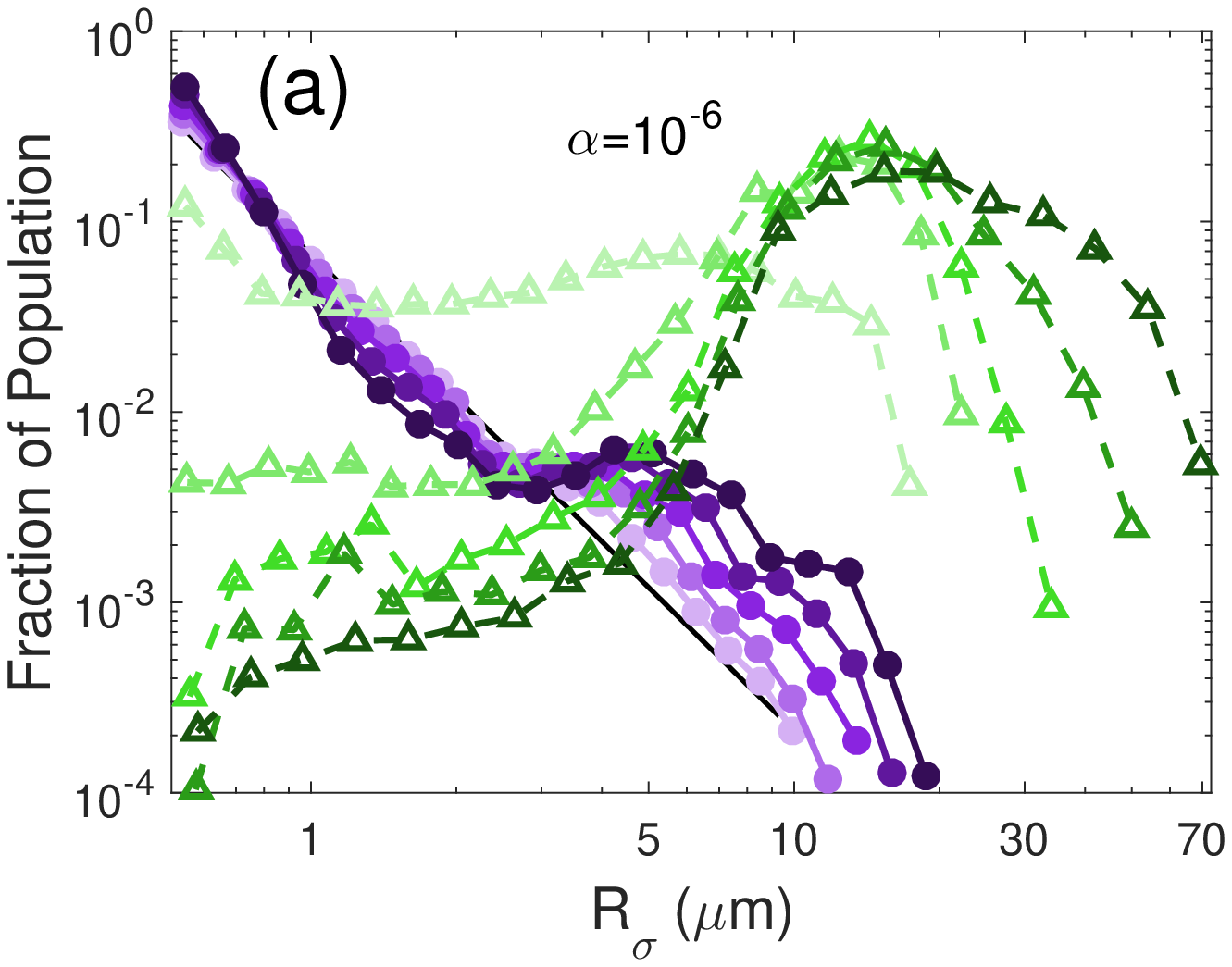}\includegraphics[width=6cm]{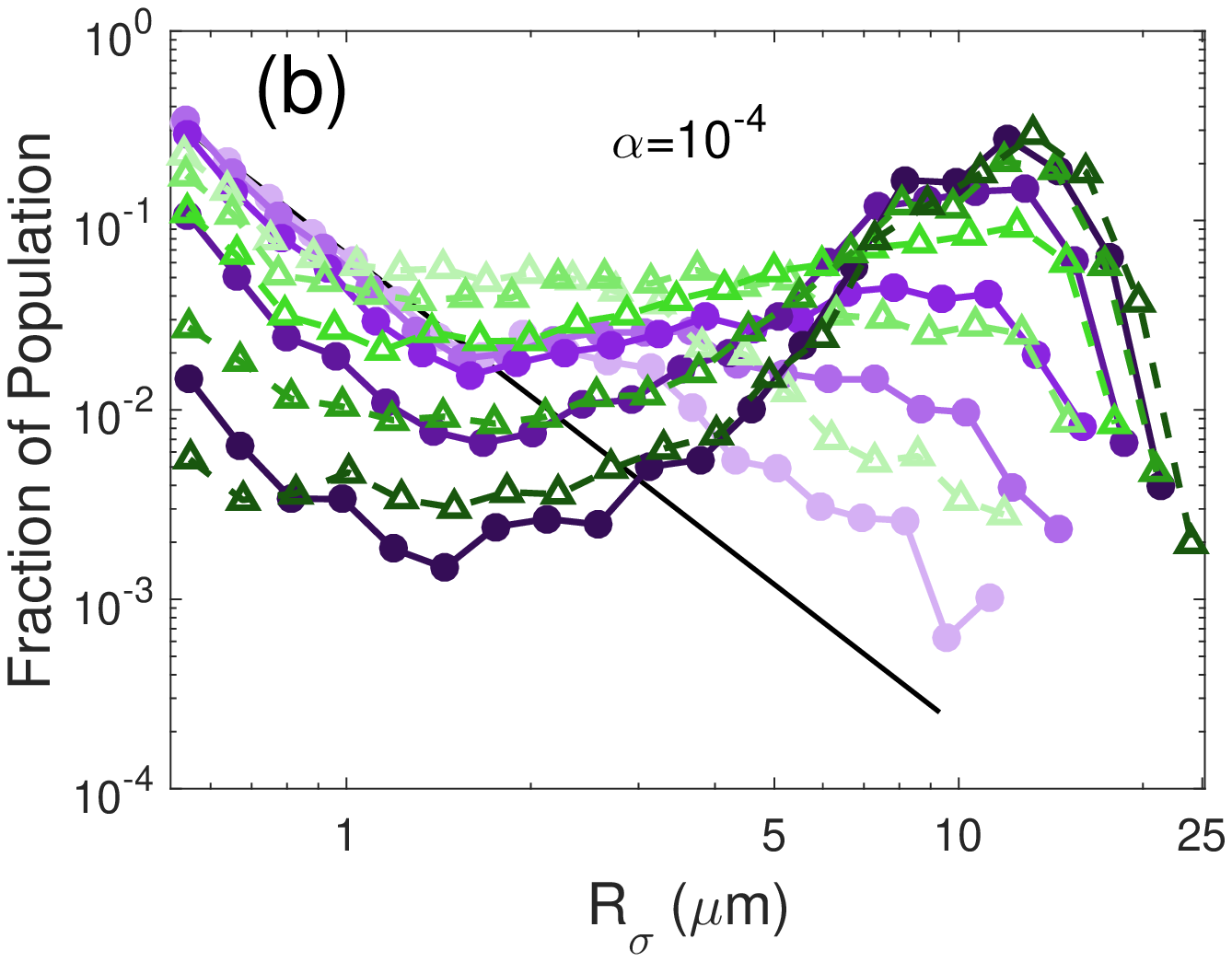}\includegraphics[width=6cm]{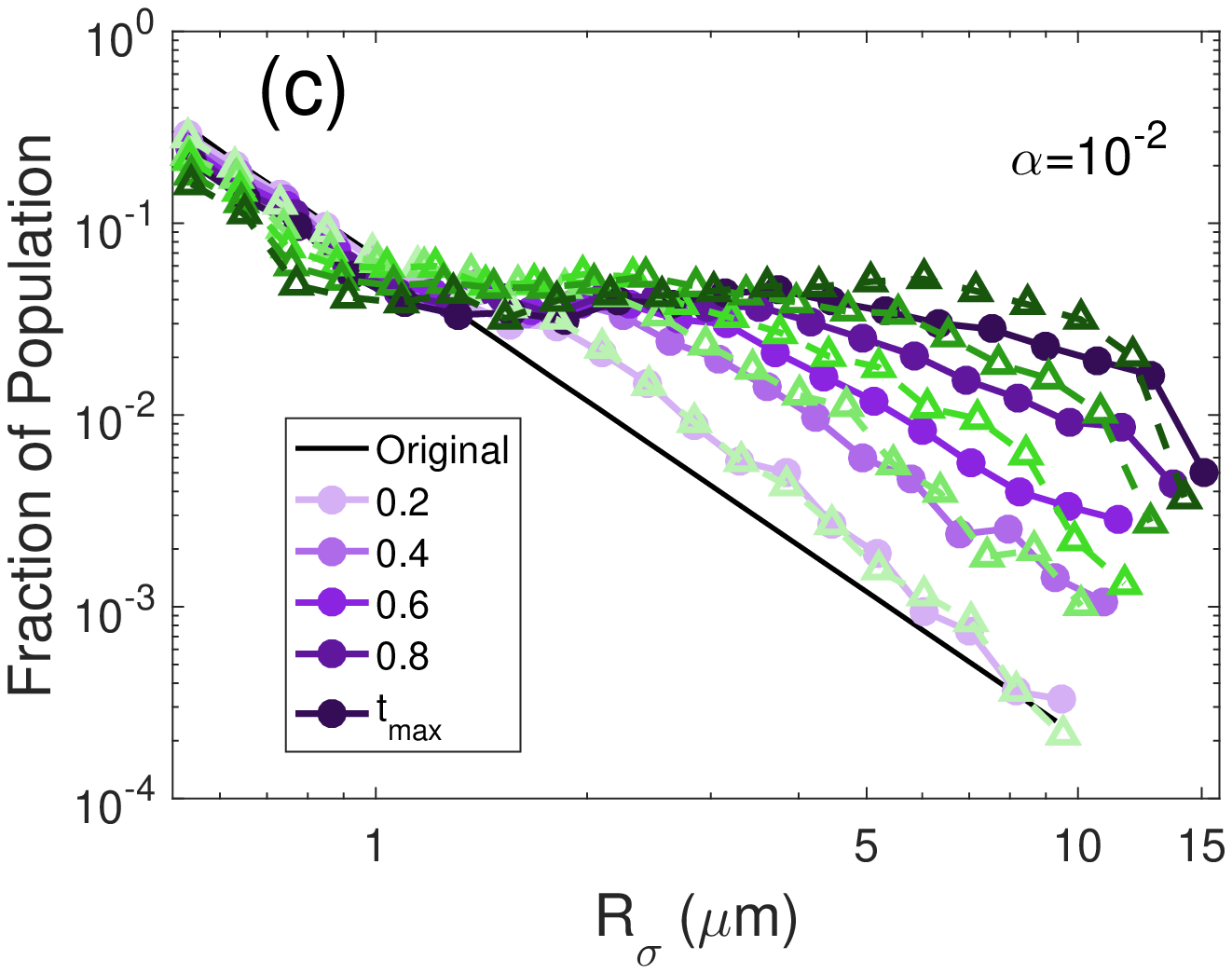} 
\caption{Evolution of distribution of particle size within total population for turbulence levels $\alpha =10^{-6}$ (a), $10^{-4}$ (b), and $10^{-2}$ (c). The purple circles represent the charged case ($\left | V_s \right |$ = 0.1 V) and green triangles represent the neutral case ($V_s$ = 0). The curves progress in time from light to dark for equally spaced times until the bouncing criterion is met. The total elapsed times are 24.1 years for a), 3.2 years for b), and 0.4 years for c).}
\label{f7}
\end{figure*}

The time evolution of the size distribution of the dust grains is shown in Figure \ref{f7}. Over time, as the dust population evolves to larger grain sizes, the smallest grains in the population tend to be depleted [at the highest turbulence level (Fig. \ref{f7}c), the small grains are depleted to a much smaller extent as aggregate growth does not progress very far before the bouncing criterion is met]. An exception occurs in the charged case with low turbulence (Fig. \ref{f7}a): the relative velocities are not great enough for the smallest monomers to overcome the Coulomb repulsion barrier, and the aggregates grow primarily through the agglomeration of the mid-sized monomers. During the early stages of coagulation, the size of the aggregates in the charged populations lags behind that of the neutral populations, as most of the small particles repel each other, resulting in missed collisions (especially in weak turbulence; Figs. \ref{f7}a, \ref{f7}b). As aggregates grow larger, the growth of weakly charged particles in the relatively strong turbulence catches up with neutral population (Figs. \ref{f7}b, \ref{f7}c), caused by higher relative velocities between charged particles, resulting from their more compact structures. At the end of the simulation when the bouncing criterion is met, the charged population even has a greater percentage of the largest particles than the neutral population, for $\alpha=10^{-2}$ (Fig. \ref{f7}c). It is also shown that the large aggregates are a small fraction of the charged population for weak turbulence ($\alpha=10^{-6}$; Fig. \ref{f7}a), and a large fraction of the population for medium turbulence level ($\alpha=10^{-4}$; Fig. \ref{f7}a). For strong turbulence ($\alpha=10^{-2}$; Fig. \ref{f7}c), the population has relatively flat distribution of particles sizes.

\begin{figure*}[!htb]
\includegraphics[width=9cm]{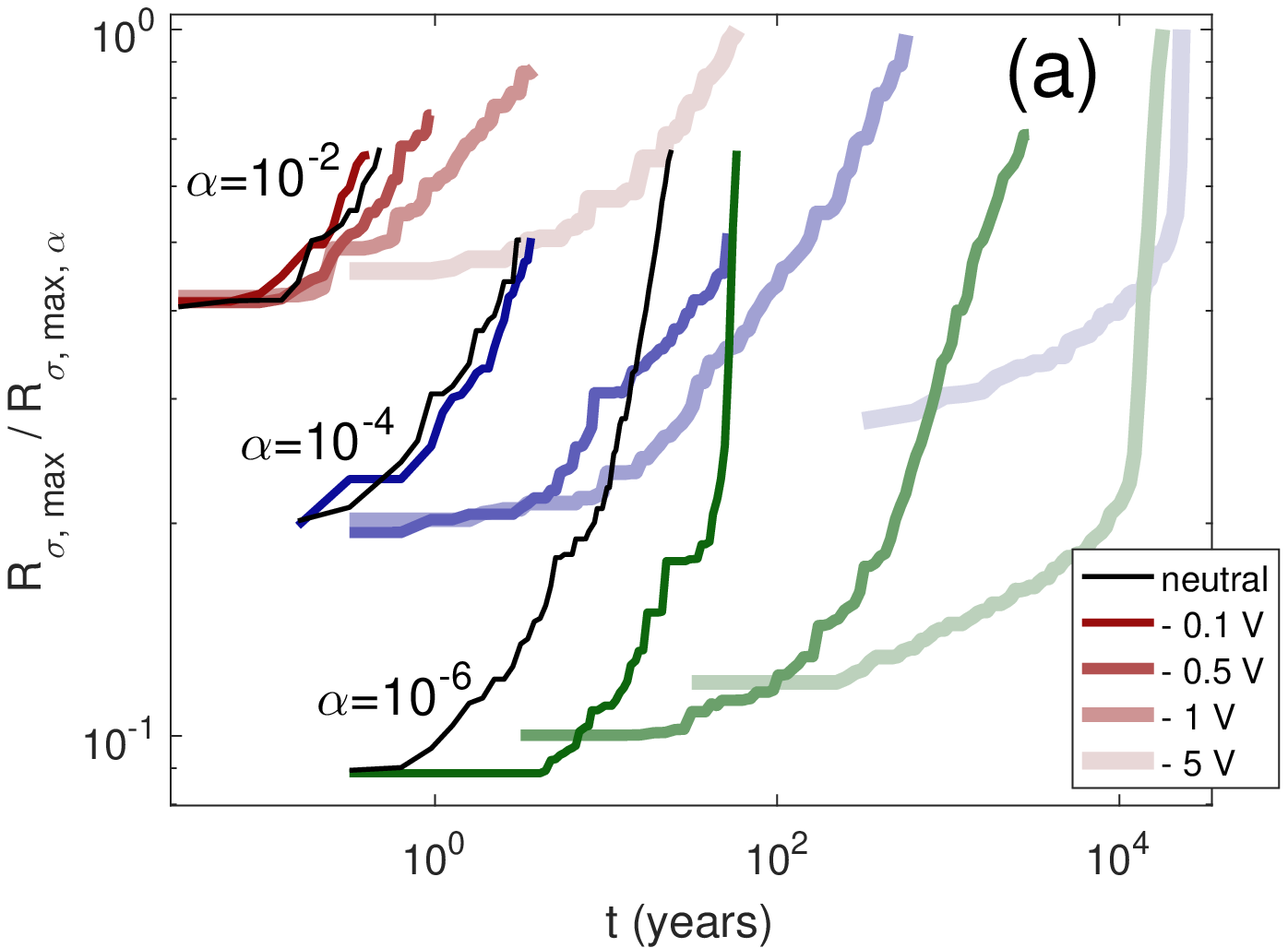}\includegraphics[width=9cm]{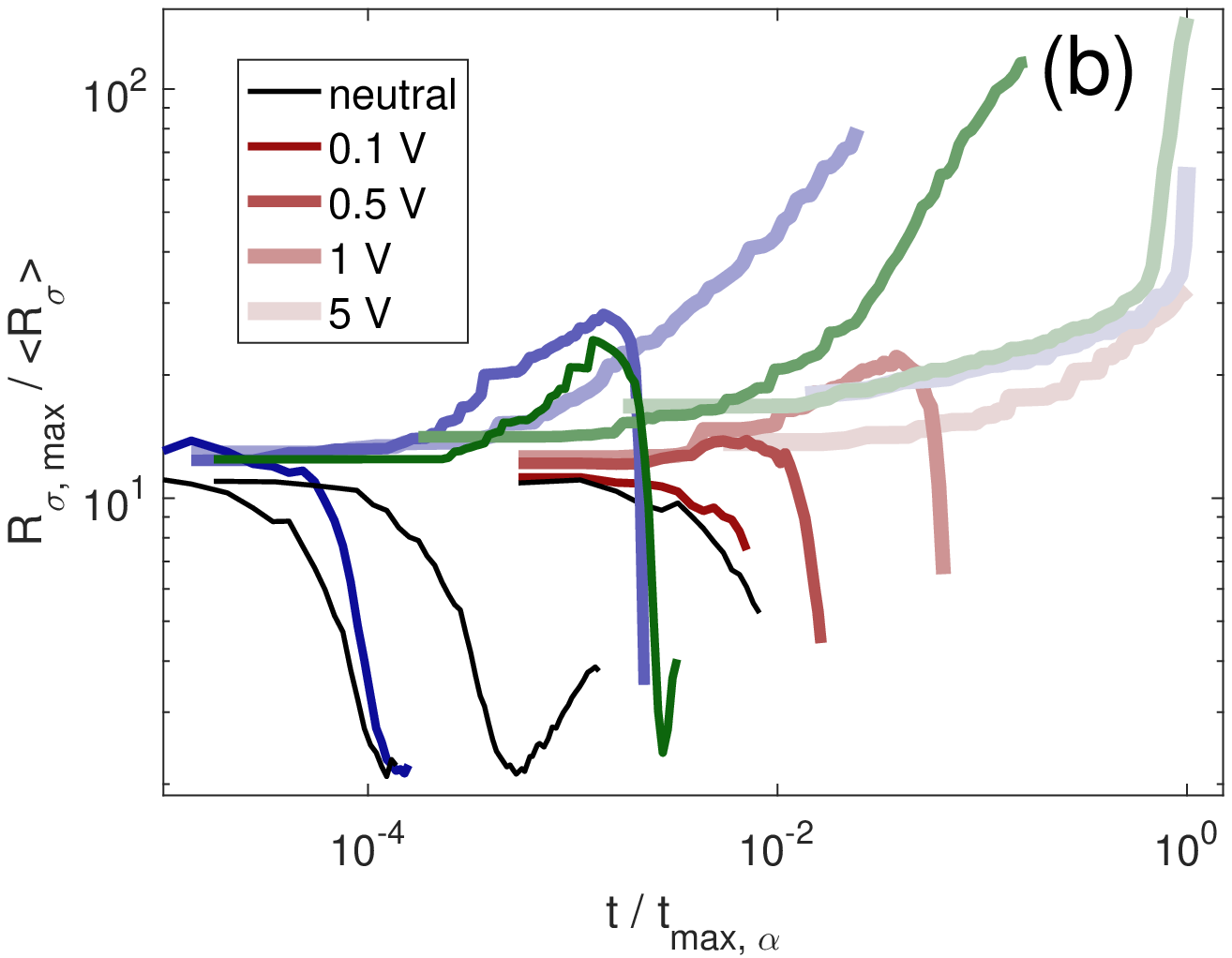}\ 
\caption{Growth rate of largest aggregate in the population. a) Normalized equivalent radius of the largest particle, and b) ratio of equivalent radius of the largest particle to the average equivalent radius of the population as a function of elapsed time for $\left | V_s \right |=0$, $0.1$, $0.5$, $1$, $5$ V, in order of decreasing color shades, with different turbulence levels (red: $\alpha=10^{-2}$; blue: $\alpha=10^{-4}$; green: $\alpha=10^{-6}$). For a), the equivalent radii are normalized by the largest value for each turbulence level: 19.5 $\mu m$ for $\alpha=10^{-2}$, 41.0 $\mu m$ for $\alpha=10^{-4}$, 115.6 $\mu m$ for $\alpha=10^{-6}$. For b), the elapsed times are normalized by the longest elapsed time for each turbulence level: 58 years for $\alpha=10^{-2}$, 23148 years for $\alpha=10^{-4}$, 17979 years for $\alpha=10^{-6}$. The right end of each curve indicates the occurrence of bouncing for more than 5\% of the particle interactions.}
\label{f8}
\end{figure*}

The different growth modes of dust populations in different charging environments are further exhibited in Fig. \ref{f8}. Fig. \ref{f8}a shows the evolution of the size of the largest aggregates in a population compared to the size of the largest aggregates for a given turbulence level at the end of the simulation. In general, the maximum particle size reached at the end of the simulation increases as the charging level increases. In highly charged, weakly turbulent environments ($\left | V_s \right |=5$ V, $\alpha=10^{-4}$; $\left | V_s \right |=1$ V, $\alpha=10^{-6}$), once a critical size is reached, the largest particle in the population grows very rapidly (runaway growth), while the rest of the population grows very slowly, which is shown by the ratio of the maximum particle size to the average particle size of the population in Fig. \ref{f8}b. In contrast, the ratio $R_{\sigma, max}/\left \langle R_{\sigma} \right \rangle$ decreases in strongly turbulent environments for $\left | V_s \right |<1$ V, meaning that the difference in particle sizes of the populations decreases over time and the particles grow collectively. For the conditions between these two extreme cases, the ratio $R_{\sigma, max}/\left \langle R_{\sigma} \right \rangle$ first increases slightly and then decreases, indicating the preferential growth of large particles in the initial stage, followed by a more even growth among the population.

The elapsed time between successive interactions increases with time, due to the reduced number density of the dust. Particles in strongly turbulent regions collide more frequently and grow faster than those in weakly turbulent regions. Given the same turbulence level, weakly charged particles collide more frequently than highly charged particles. Although populations with low charge in strong turbulence grow faster, they meet the bouncing criterion sooner than those with high charge in weak turbulence, as shown in Table \ref{table1}. The gentle collisions afforded by low relative velocities allow particles in weak turbulence to grow to larger sizes before meeting the bouncing criterion (compare results for neutral particles in Figs. \ref{f7}a, c).  In addition, particles with greater charge grow to a larger size before the bouncing criterion is met, due to reduced relative velocity by electrostatic repulsion (compare endpoints for Figs. \ref{f9}, \ref{f8}a). The maximum particle size reached before the bouncing criterion is met is positively correlated with charge for a given turbulence level, as shown in Figure \ref{f8}a and Table \ref{table2}.

Note that in addition to the relative velocity at impact, bouncing is also affected by the porosity. Compact aggregates are more likely to bounce than fluffy aggregates, because compact aggregates have much higher average coordination number, i.e., the number of contacts per monomer, than porous ones, making it more difficult to dissipate the collision energy through restructuring (Wada et al. 2011; Seizinger \& Kley 2013). In addition, compact aggregates are less coupled to the gas, resulting in greater relative velocity with other particles, which also enhances bouncing. Therefore, charge can either reduce bouncing by decreasing the relative velocity through electrostatic repulsion, or reinforce bouncing by increasing the compactness of aggregates. The first factor is more dominant for the environmental conditions considered in the simulation.


\section{ Analysis}
\renewcommand{\theequation}{5.\arabic{equation}} \setcounter{equation}{0}
The collision outcomes, i.e., the morphology of the resulting aggregates and the evolution of dust populations, depend on the properties of colliding particles, such as the relative velocity, equivalent radius and mass ratio. The particle properties also greatly determine the probability of particles being selected to collide and the probability of a successful collision, which further affect the collision outcome. In this section, we analyze the relationship between the colliding particles and the collision outcomes, and investigate how it is altered in different charging and turbulence conditions.

\subsection{Relative velocity}

\begin{figure*}[!htb]
\includegraphics[width=18cm]{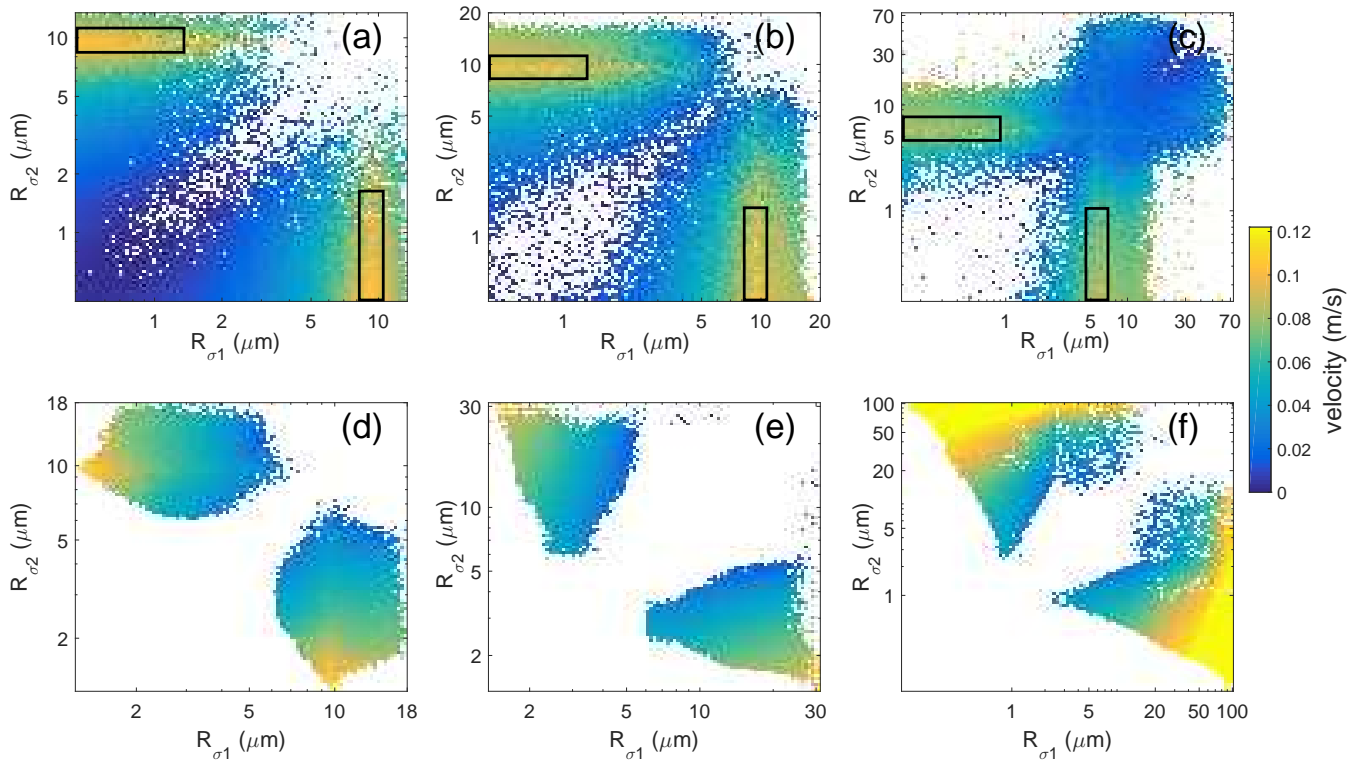} 
\caption{Relative velocity for particles at different stages of growth, a, d) 0-40,000 successful collisions, b, e) 40,000-80,000 successful collisions, and c, f) 80,000-140,000 successful collisions, comparing neutral (top row) and charged (bottom row, $\left | V_s \right |$ = 1 V) grains. The turbulence level is $\alpha=10^{-6}$.}
\label{f11}
\end{figure*}

The relative velocity of aggregate pairs plays an important role in their collision rates as it affects the volume an aggregate sweeps out per unit time, which determines how likely it is for an aggregate to encounter another. The relative velocities between particles within the population during different time periods, calculated according to Eq. \ref{eq:vturb}, are shown in Fig. \ref{f11}. Similar-sized aggregates have low relative velocities. Aggregates with greater size differences, especially colliding pairs consisting of large and small aggregates, have higher relative velocities. The blanks that can be seen on the diagonal are due to the low relative velocity for aggregate pairs in that region resulting in a low selection probability, meaning they are less likely to be selected to collide (see Section \ref{sel}). However, as more collisions take place, the difference of the relative velocity on the diagonal and the nearby regions decreases, especially for large particles (Figs. \ref{f11}b, \ref{f11}c), because the increased diversity of particle structure (porosity) enables particles of the same size to have a variety of masses and therefore different friction times, resulting in a relatively high velocity. 

Figures \ref{f11}a and \ref{f11}b show that the relative velocity drops as the equivalent radius just exceeds 10 $\mu m$, which is the maximum monomer size, because the spherically averaged density of an aggregate is much lower than it is for a spherical monomer of the same size (spherical monomers make up a large percentage of the particles at the early stage of the simulation). The decrease of the aggregate density in a PCA (particle-cluster-aggregation) collision results in a smaller difference in the friction times for an aggregate and a monomer, decreasing their relative velocity calculated from Eq. \ref{eq:vturb}. Figures \ref{f11}b and \ref{f11}c show that the regions of maximum relative velocity (indicated by black rectangles; note that they are local maxima) shift towards smaller particle size as more collisions take place. The reason is that the maximum monomer size of the population decreases over time with the depletion of large monomers, and meanwhile the increase of the ratio of small aggregates to monomers of the same size leads to an overall decreased density for small particles. The resulting stronger coupling with the gas increases their relative velocity with respect to relatively large particles (notice that this is still in the small particle regime). Therefore, particles with $5\ \mu m < R _{\sigma} < 7\ \mu m$ have higher relative velocities in Figure \ref{f11}c (indicated by the black rectangle) than in Figures \ref{f11}a and \ref{f11}b. Unlike the neutral particles, the relative velocity of charged particles increases with increasing size difference (Figs. \ref{f11}e, \ref{f11}f), except for the initial stage (Fig. \ref{f11}d).

\begin{figure*}[!htb]
\includegraphics[width=6cm]{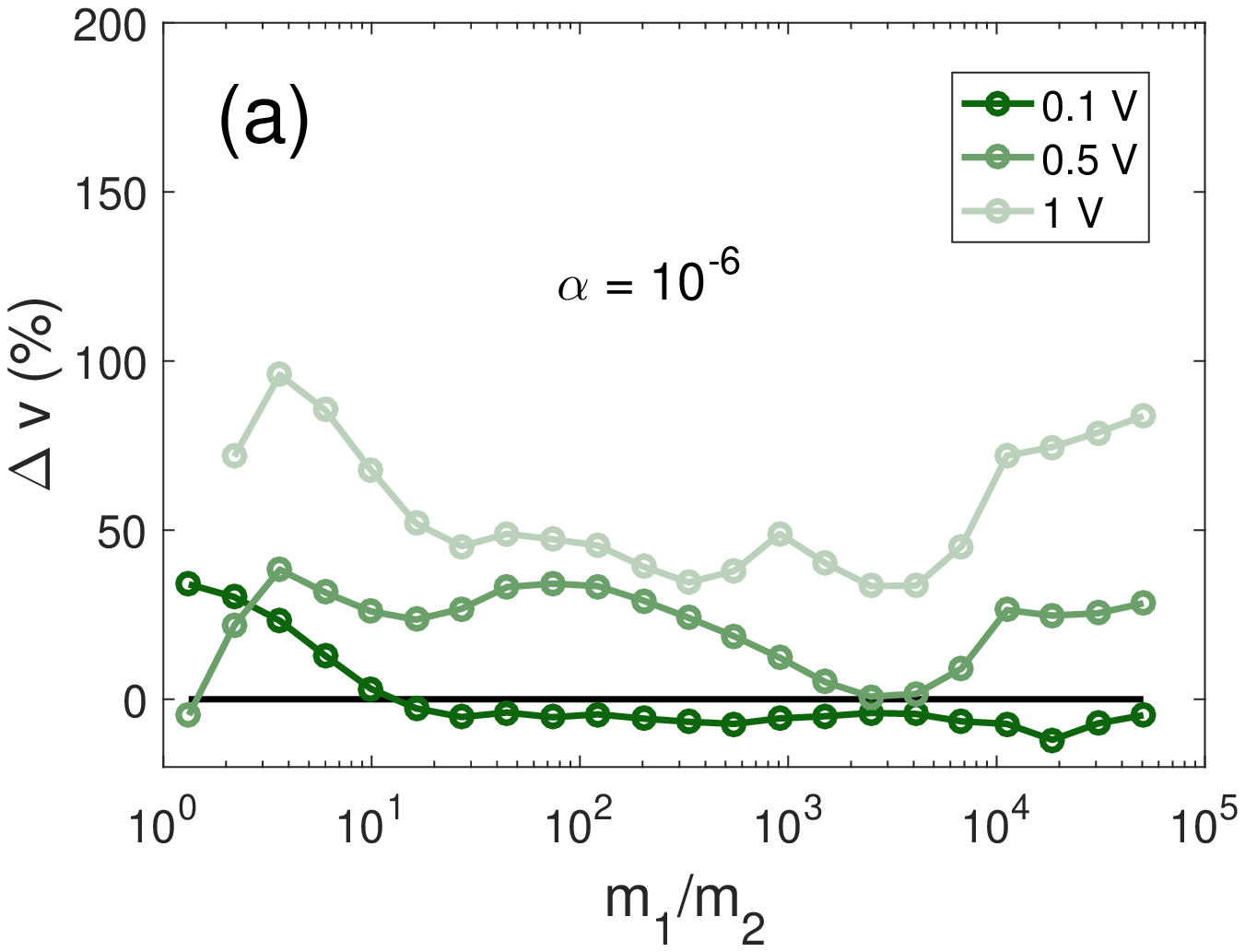}\includegraphics[width=6cm]{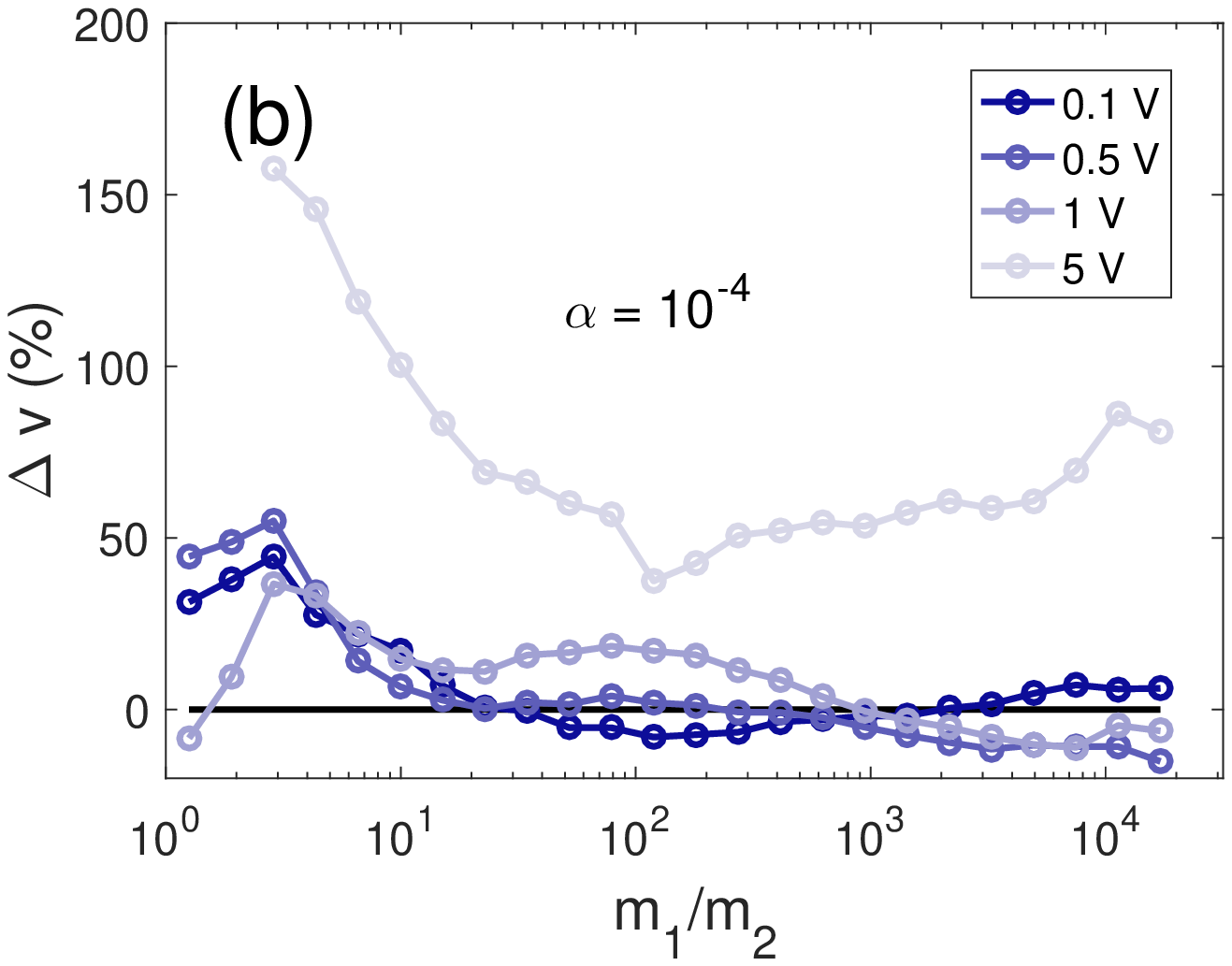}\includegraphics[width=6cm]{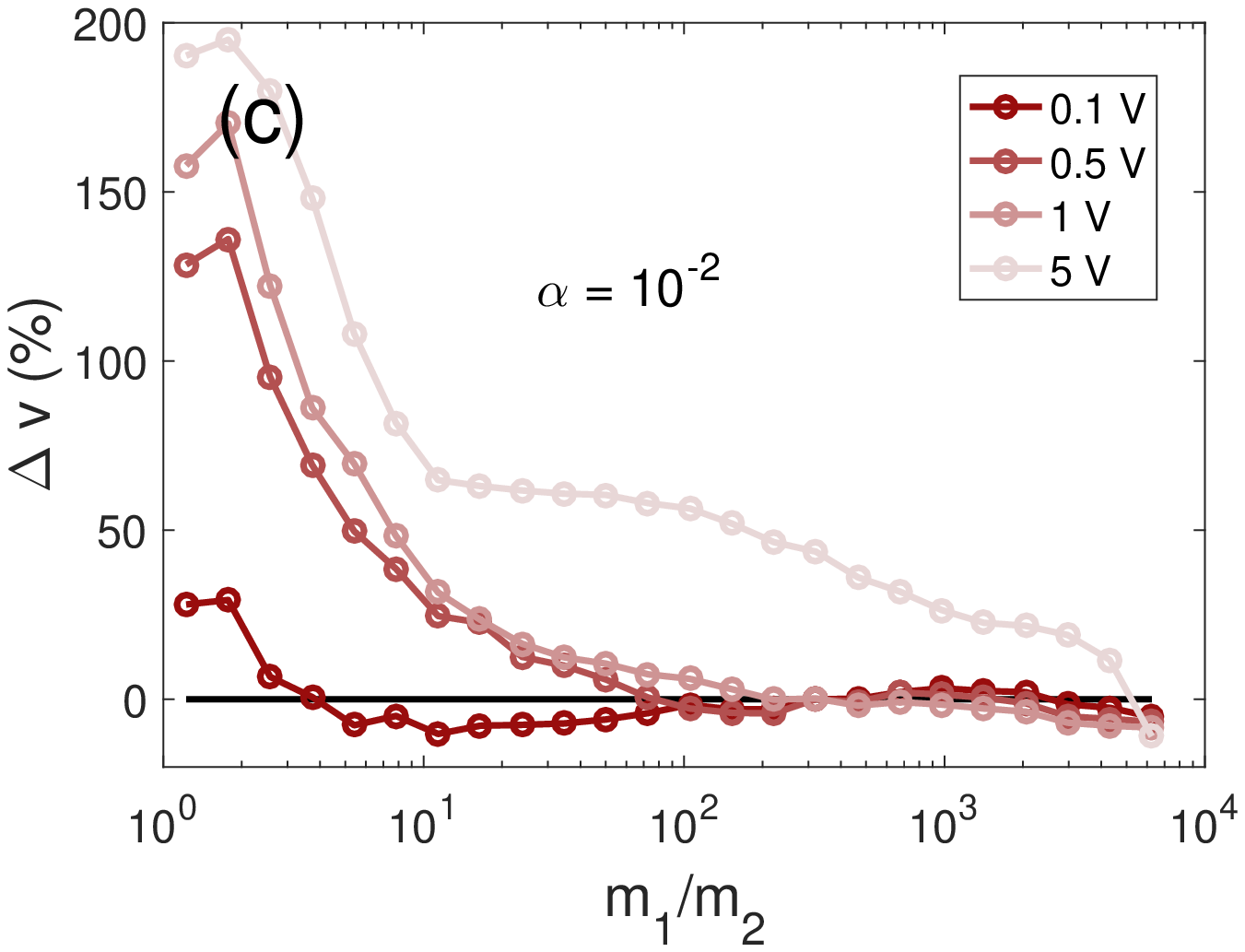} 
\caption{Percent difference of relative velocity between colliding pairs for charged aggregates relative to neutral aggregates as a function of the mass ratio. Color shades indicate dust surface potential. Turbulence levels are $\alpha=10^{-6}$ for a), $\alpha=10^{-4}$ for b) and $\alpha=10^{-2}$ for c). }
\label{f12}
\end{figure*}

As discussed in section \ref{evo}, charged particles have greater initial relative velocities than neutral particles (not considering the reduction of the velocity due to repulsion), because they are less coupled to the gas due to more compact structure. Figure \ref{f12} shows that the initial relative velocity is positively correlated with the dust surface potential for the same turbulence level. The fact that particles with the highest charge have the greatest compactness factors (Figure \ref{f5}) and largest initial relative velocities (Figure \ref{f12}) corroborates the importance of the porosity in the dust motion. 

\subsection{Selection rate}\label{sel}

\begin{figure*}[!htb]
\hspace*{-2cm}
\includegraphics[width=20cm]{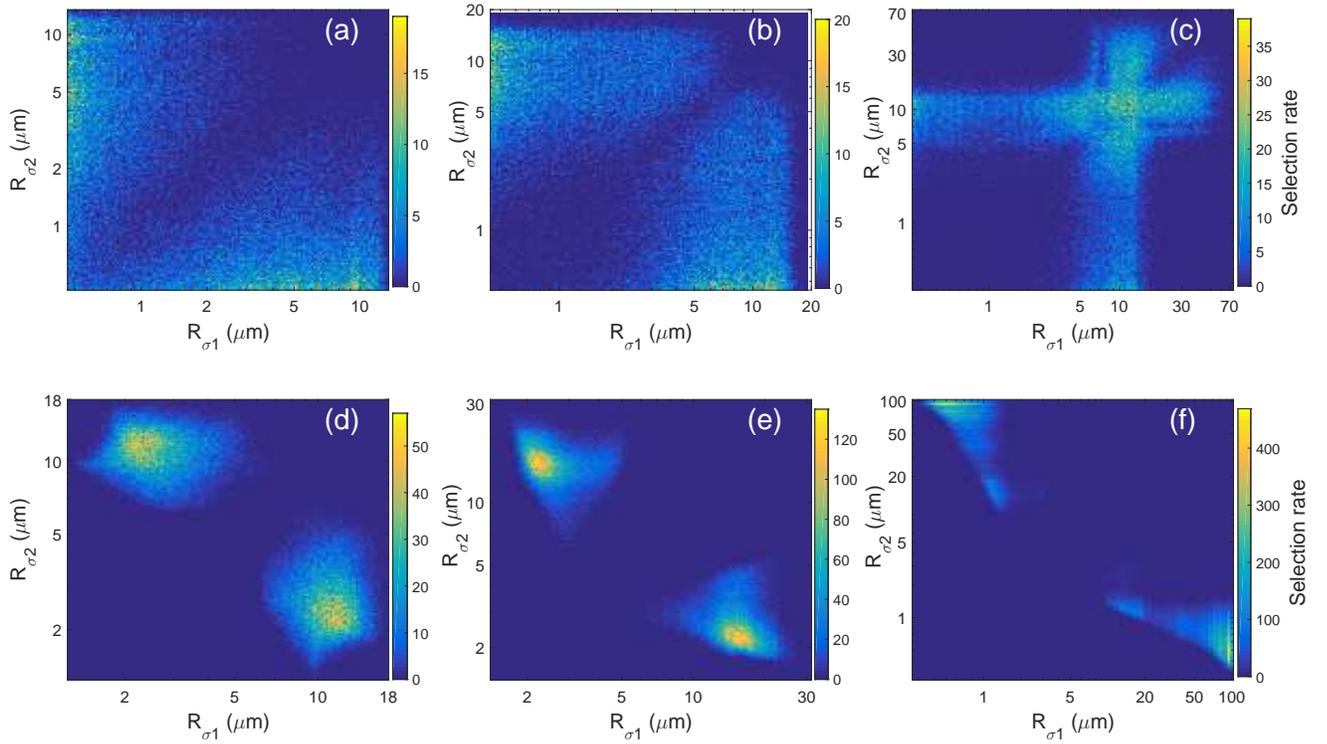} 
\caption{Selection rate for particles at different stages of growth, a, d) 0-40,000 successful collisions, b, e) 40,000-80,000 successful collisions, and c, f) 80,000-140,000 successful collisions, comparing neutral (top row) and charged (bottom row, $\left | V_s \right |$ = 1 V) grains. The turbulence level is $\alpha=10^{-6}$.}
\label{f13}
\end{figure*}

The collision kernel which determines the selection of potentially colliding pair is a function of the effective cross section, which depends on particle size and charge, and the relative velocity, which depends on size, and indirectly on charge through the porosity. As a large size difference between grains increases the relative velocity, but large particles are a small fraction of the original population, the selection rate is a function of size, charge, relative velocity and the distribution of size in population. The dominant factor varies for different size ranges and different stages of coagulation, as shown in Figure \ref{f13}, comparing the selection rates for a neutral ($\left | V_s \right |$ = 0) and highly charged environment ($\left | V_s \right |$ = 1 V) with turbulence level $\alpha=10^{-6}$. In the neutral environment (top row of Fig. \ref{f13}), the initial selection rate is broadly distributed over all particle sizes, though small monomers are most likely to be selected for collisions, reflecting the abundance of small monomers within the population (Fig. \ref{f13}a). Particles first grow by collisions between monomers of all sizes, followed by PCA and CCA (cluster-cluster-aggregation). The maximum selection rate slowly shifts towards larger particle sizes as small monomers and aggregates are gradually removed from the population, and CCA is the dominant mechanism at the late stage of the process. In contrast, in the highly charged environment (bottom row of Fig. \ref{f13}), small monomers are not selected to collide at the initial stage of the growth process (Fig. \ref{f13}d), and particles first grow by coagulation of relatively large monomers, and then by accretion of monomers onto large aggregates. The mean size of accreted monomers shifts towards smaller sizes as time goes on, as larger aggregates enable small particles to overcome the electrostatic barrier (Figs. \ref{f13}e, \ref{f13}f). The selection rate remains sharply peaked for combinations of large aggregates ($R_{\sigma}$ greater than maximum monomer size) with smaller monomers, an indication that PCA is the dominant mechanism over all time (Figs. \ref{f13}e, \ref{f13}f).  The difference in the selection rates contributes to more spherical and symmetric structures of the charged aggregates relative to the neutral aggregates (Fig. \ref{f1}), as particles growing through PCA accrete monomers more evenly than collisions between irregular aggregates.

\subsection{Size ratio of colliding partners}

\begin{figure*}[!htb]
\includegraphics[width=6cm]{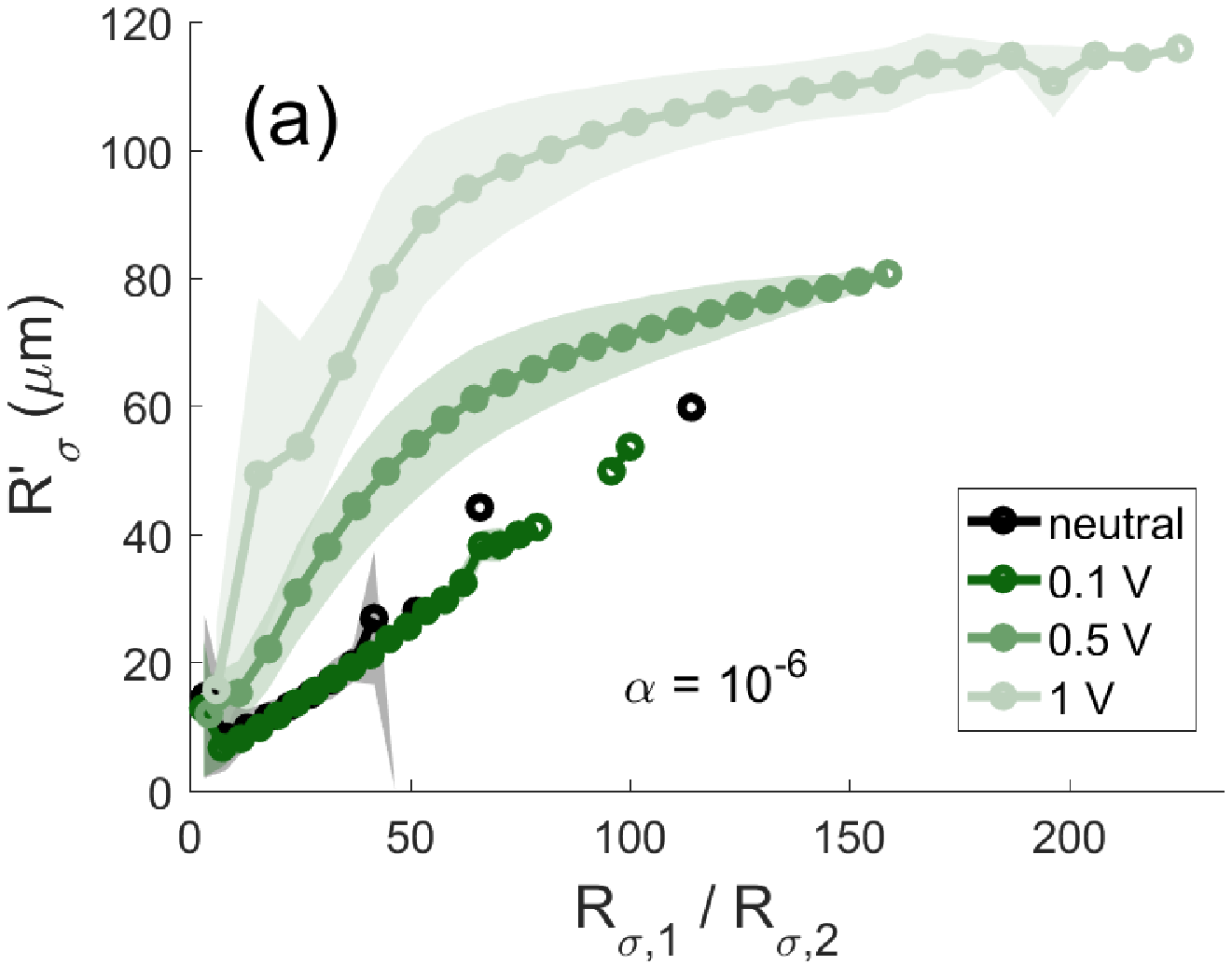}\includegraphics[width=6cm]{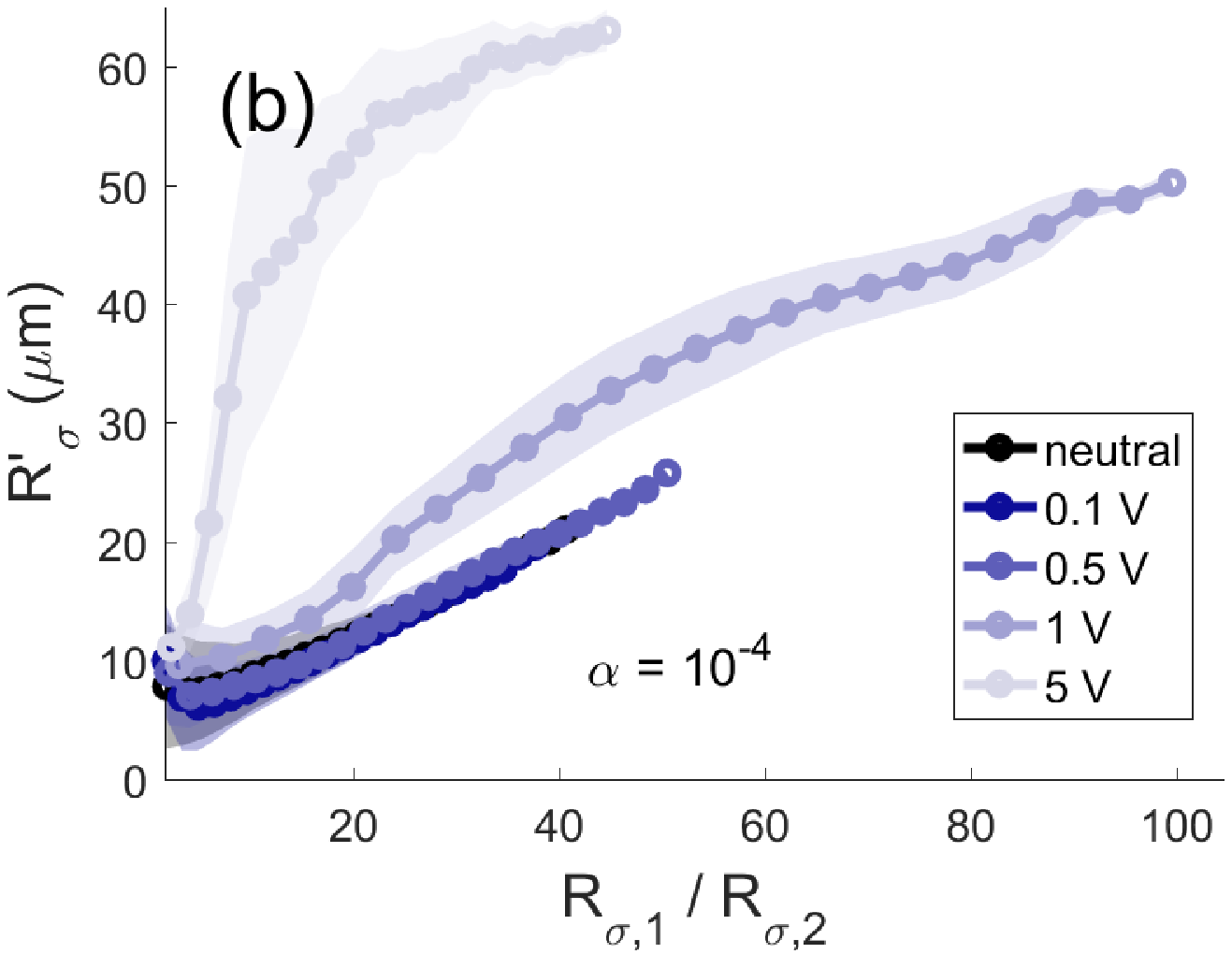}\includegraphics[width=6cm]{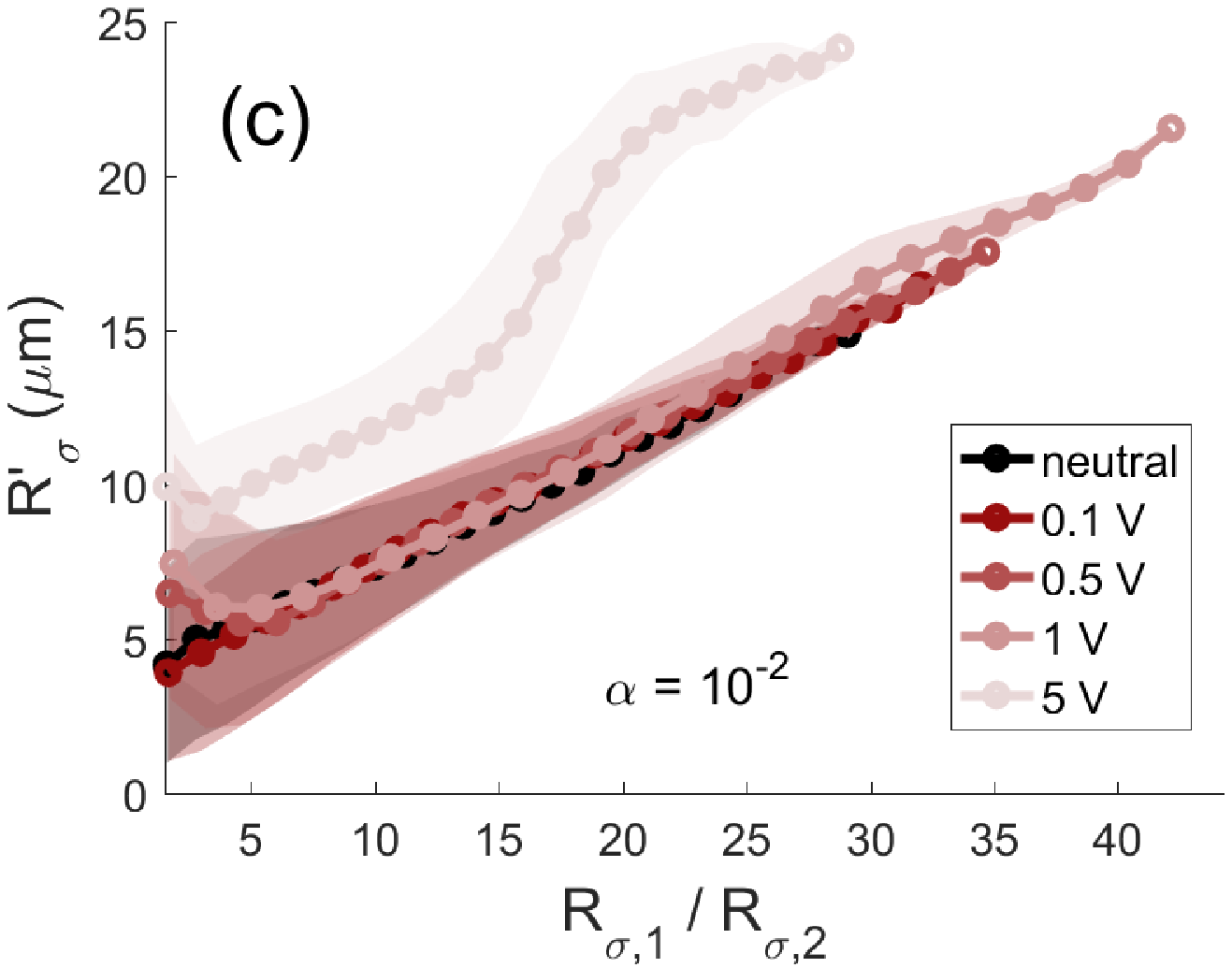} 
\caption{Average equivalent radius of the resulting particle from successful collision as a function of ratio of equivalent radii of colliding particles, for different turbulence levels and dust surface potentials. The shaded areas indicate one standard deviation of the average $R _{\sigma}$ ratio. Turbulence levels are $\alpha=10^{-6}$ for a), $\alpha=10^{-4}$ for b) and $\alpha=10^{-2}$ for c).}
\label{f14}
\end{figure*}

The size ratio of colliding particles is greatly affected by the charging condition, turbulence level and size distribution within the population. In turn, it affects the collision outcome and the structure of the resulting aggregates. In general, the size of colliding particles increases over time for neutral or weakly charged population, while the smaller particle of the colliding pair decreases over time for highly charged, weakly turbulent population, since larger particles are more capable of accreting small particles. Figure \ref{f14} shows the equivalent radii of resulting particle as a function of the size ratio of colliding particles. It is seen that the equivalent radius of the resulting particle is positively related to the dust surface potential for a given size ratio of colliding particles, because in general, the smaller particle of the colliding pair is larger in a highly charged environment, due to the repulsion of small particles. Therefore, the same size ratio corresponds to two larger colliding particles in a highly charged environment than in neutral or weakly charged environments, and results in a larger aggregate. Note that the slopes of the curves are nearly constant for neutral and weakly charged populations, but for highly charged populations the curves rise rapidly and then tend to be a constant value. The flattening of the curves is probably caused by the runaway growth, i.e., accretion of small particles onto large particles, which does not make much change to the sizes of large particles. In addition, the sizes of the small particles being accreted decrease over time, causing a rapid increase in the size ratios while the sizes of resulting aggregates are relatively constant.

\section{DISCUSSION}
\renewcommand{\theequation}{6.\arabic{equation}} \setcounter{equation}{0}
As shown, the process of dust coagulation is affected by many factors, i.e., the diversity of particle size, porosity, charge, and turbulence. These factors influence each other and come together to determine the collision process. However, most studies to date have considered only a subset of these factors, and thus do not reflect the diversity of dust particles in a real environment. The current model incorporates all of these factors to examine how the charge affects the collision kernel of particles and their resultant porosity, which further influences the collision probabilities and the growth rate of dust population. 

In this section, we compare our detailed-MC model to previous MC models for neutral particles with an initial monodisperse distribution (Section \ref{6.1}), in particular determining the effect of charged aggregates (Section \ref{6.2}).

\subsection{Comparison with previous MC models}\lb{6.1}

The key characteristic of dust particles that controls their coupling to the turbulent motion of the gas is the ratio of surface area to mass $A/m$. In the ``hit-and stick'' regime, the mass $m$ and the cross-section $A$ for growing aggregates follows a power-law relationship (Ormel et at. 2007),
\bqn
\lb{de1}
A\propto m^{\delta },~~~\frac{2}{3}\leqslant \delta \leqslant  1
\eqn
with the lower limit $\delta =\frac{2}{3}$ corresponding to compact spherical particles, and the upper limit $\delta =1$ corresponding to the aggregation of chains or linear structures. Consequently, the friction time scales with mass as
\bqn
\lb{e1}
 \tau _{f}\propto m^{1-\delta }.
\eqn

Ormel et al. (2007) defined an enlargement factor $\psi=\frac{V}{V^{\ast }}$, where $V$ is the extended volume corresponding to a sphere with a radius equal to the equivalent radius, and $V^{\ast }$ is the volume the material occupies in its compacted state [the enlargement factor is approximately equal to the inverse of the compactness factor, except that $V^{\ast }$ in the compactness factor is equal to the total volume of all the monomers in an aggregate, which is smaller than the $V^{\ast }$  in the enlargement factor]. By using the relationships $V\propto A^{\frac{3}{2}}\propto m^{\frac{3}{2}\delta }$ and $V^{\ast }\propto m$, the enlargement factor can be related to the mass as (Ormel et al. 2007)
\bqn
\lb{e2}
\psi \propto m^{\frac{3}{2}\delta-1 }.
\eqn

For compact particles ($A\propto m^{2/3}$), the friction time $\tau _{f}\propto m^{1/3}$, which increases monotonically with mass, and the enlargement factor $\psi$ is constant. On the other hand, for fluffy particles ($\delta >\frac{2}{3}$), both $\tau _{f}$ and $\psi$ increase with $m$. In the extreme case where linear aggregates are formed by collisions between equal-sized monomers, $\delta$ takes the value 1, and $\tau _{f}$ stays constant, which means the relative velocity between particles stays the same during collisional growth, where $\psi \propto m^{1/2}$. 

\begin{figure*}
\includegraphics[width=6cm]{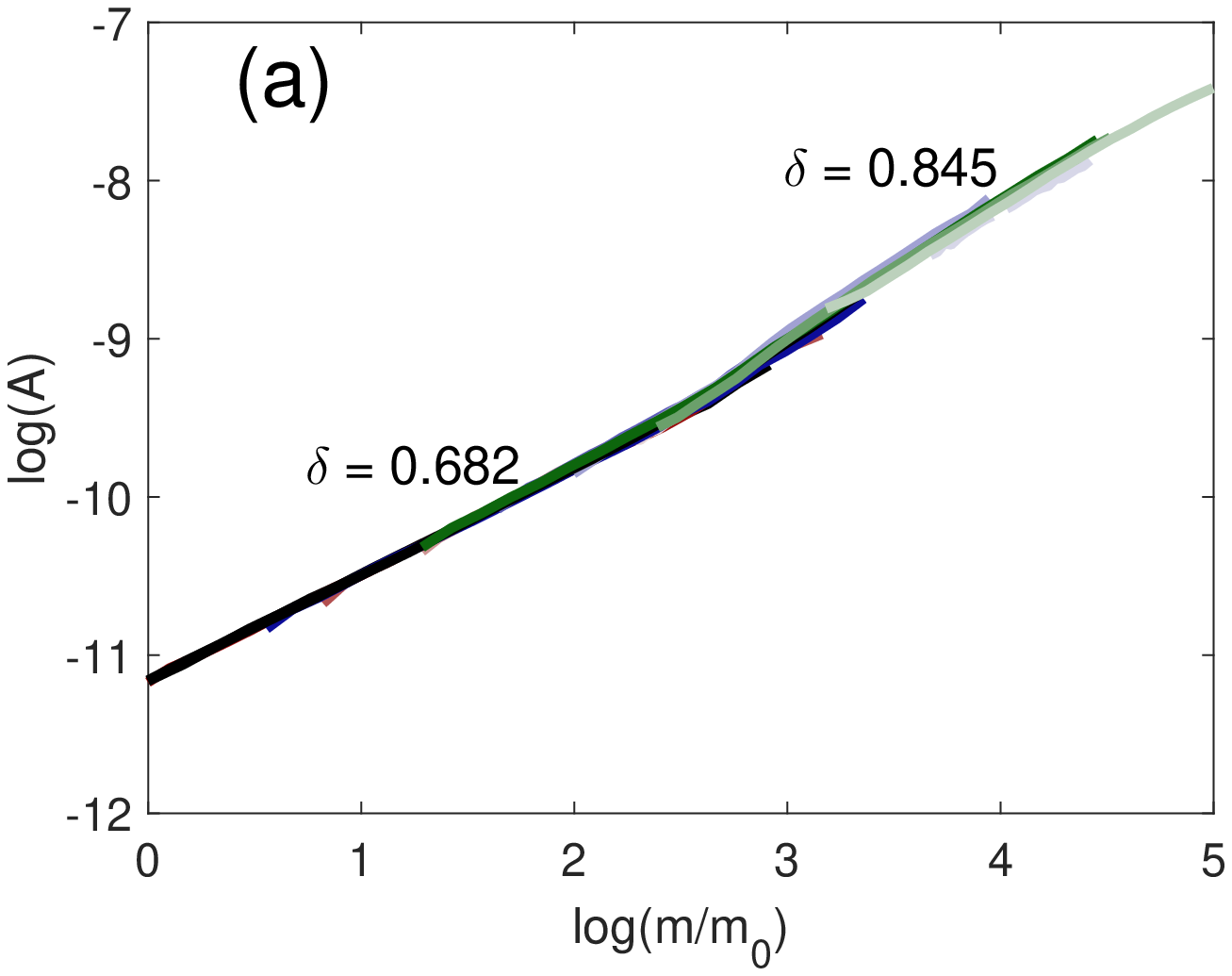}\includegraphics[width=6cm]{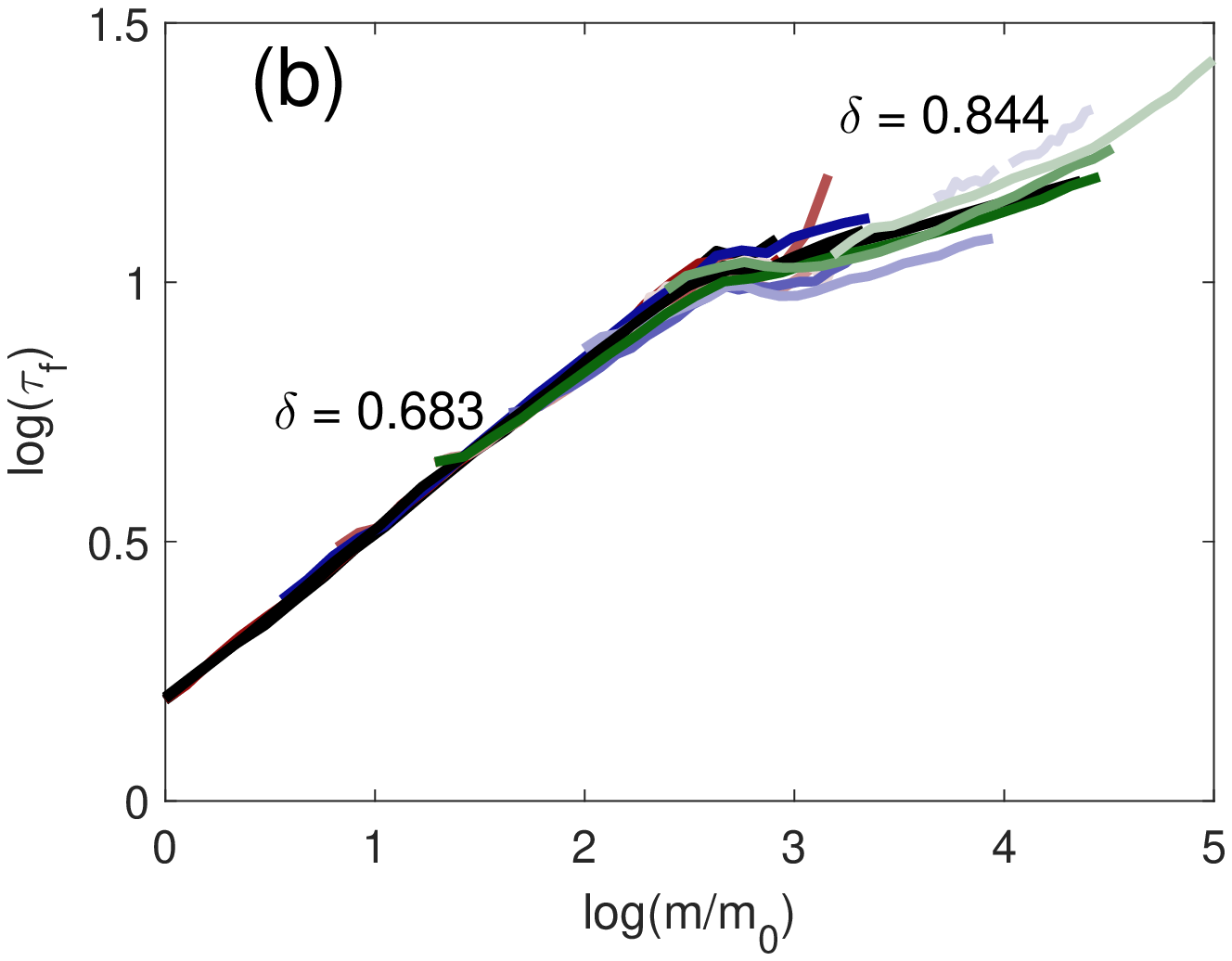}\includegraphics[width=6cm]{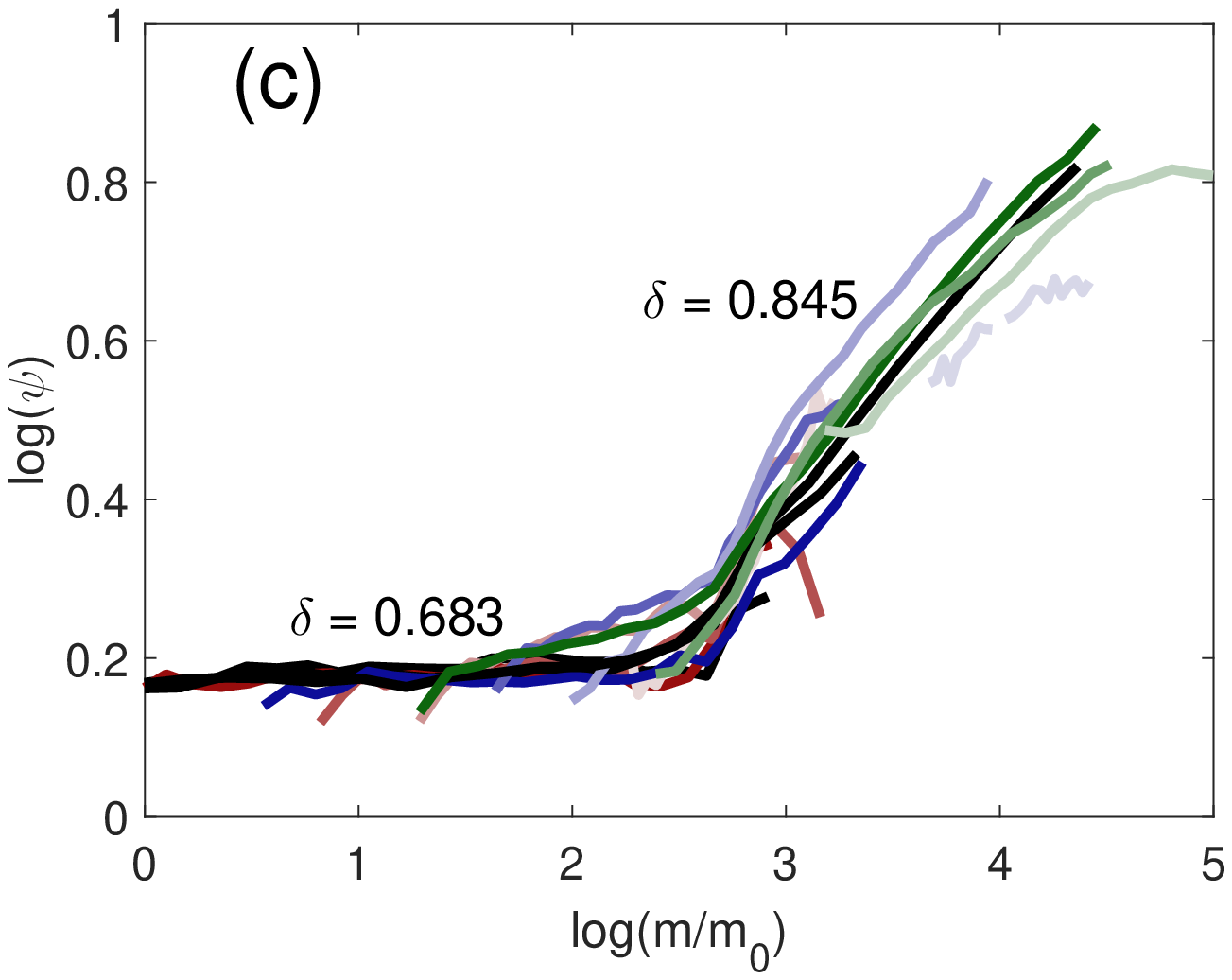} 
\caption{a) Surface area, b) friction time and c) enlargement factor as a function of mass for neutral and charged particles (surface potential $\left | V_s \right |=0$, $0.1$, $0.5$, $1$, $5$ V; in order of decreasing color shades) with more than two monomers, for different turbulence levels (red: $\alpha=10^{-2}$; blue: $\alpha=10^{-4}$; green: $\alpha=10^{-6}$). The masses are normalized by the mean mass of the initial population.}
\label{f27}
\end{figure*}

Figure \ref{f27} shows the surface area, friction time and enlargement factor as a function of mass for aggregates generated in our simulation. By comparing the slopes of the curves with Eqs. \ref{de1}, \ref{e1} and \ref{e2}, one obtains $\delta \sim 0.682$ before the turning point and $\delta \sim 0.845$ after the turning point from the relationship between $A$ and $m$ (Fig. \ref{f27}a); $\delta \sim 0.683$ before the turning point and $\delta \sim 0.844$ after the turning point from the relationship between $\tau _{f}$ and $m$ (Fig. \ref{f27}b); and $\delta \sim 0.683$ before the turning point and $\delta \sim 0.845$ after the turning point from the relationship between $\psi$ and $m$ (Fig. \ref{f27}c). The turning point corresponds to the maximum monomer mass. The increase of $\delta$ at the turning point indicates that aggregates with higher porosity have larger $\delta$.

Ormel et al. (2007) replaced the mass in Equation \ref{e2} with the volume of the particles in order to take into account the spatial extent of the collision partners, which reflects the porosity, and, by using the relationship among the volume, cross-section as well as the enlargement factor, derived a formula for the enlargement factor of the resulting aggregate for collisions between particles of all kinds of sizes:
\bqn
\lb{e3}
\psi =\left \langle \psi \right \rangle_{m}\left (1+\frac{m_{2}\psi_2 }{m_{1}\psi_1 } \right )^{\frac{3}{2}\delta _{CCA}-1}+\psi _{add}
\eqn
where $\delta _{CCA}$=0.95, and $\psi _{add}=B\frac{m_{2}}{m_{1}}\psi _{1}exp\left [ -\mu /m_{F}  \right ]$ with $B$ = 1, $m_{F}=10m_{0}$ ($m_{0}$ is the monomer size for monodisperse distribution). $\psi _{add}$ is a term added to compensate for the underestimation of the porous growth when one of the colliding particles is very small.

\begin{figure*}[!htb]
\includegraphics[width=6cm]{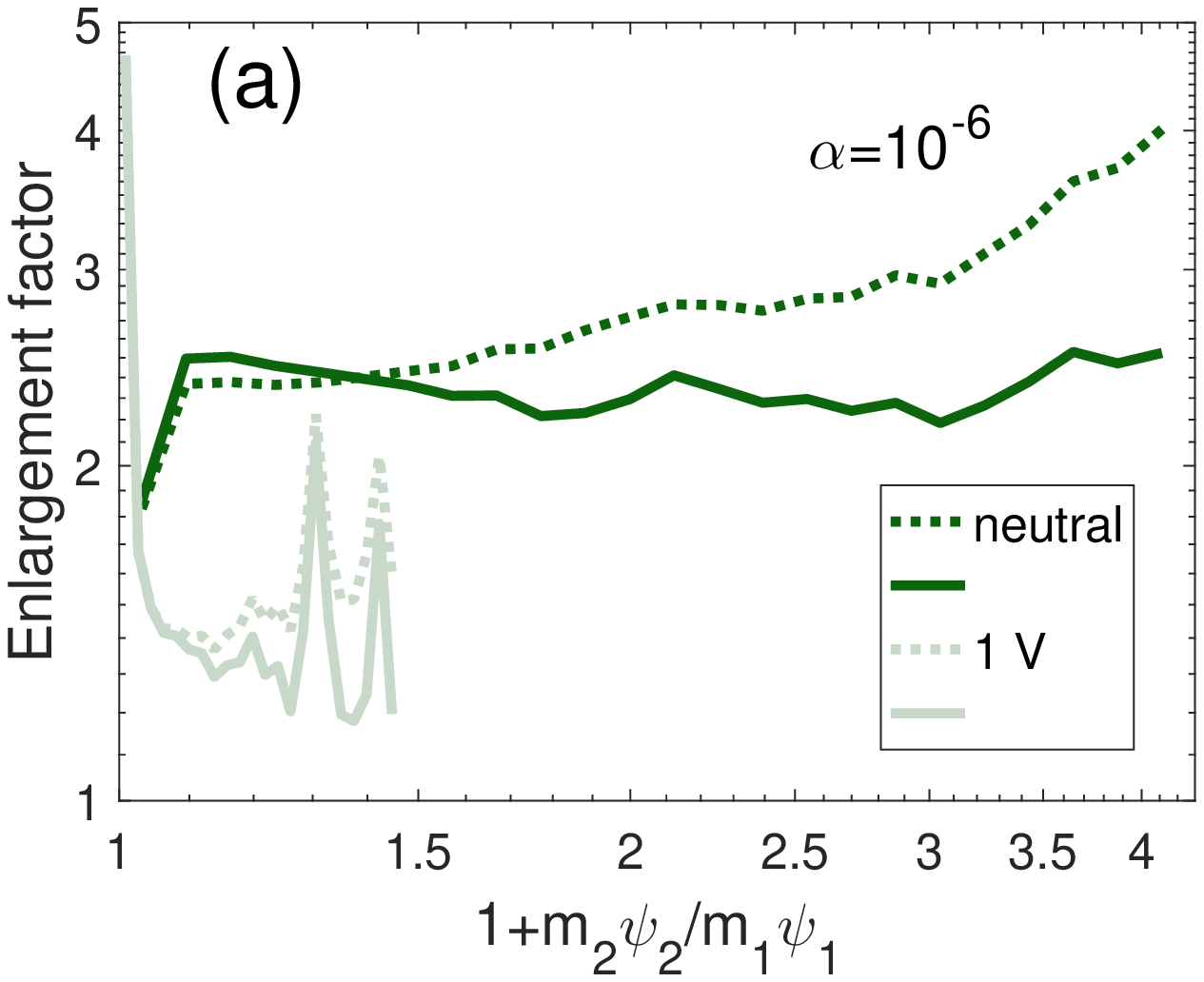}\includegraphics[width=6cm]{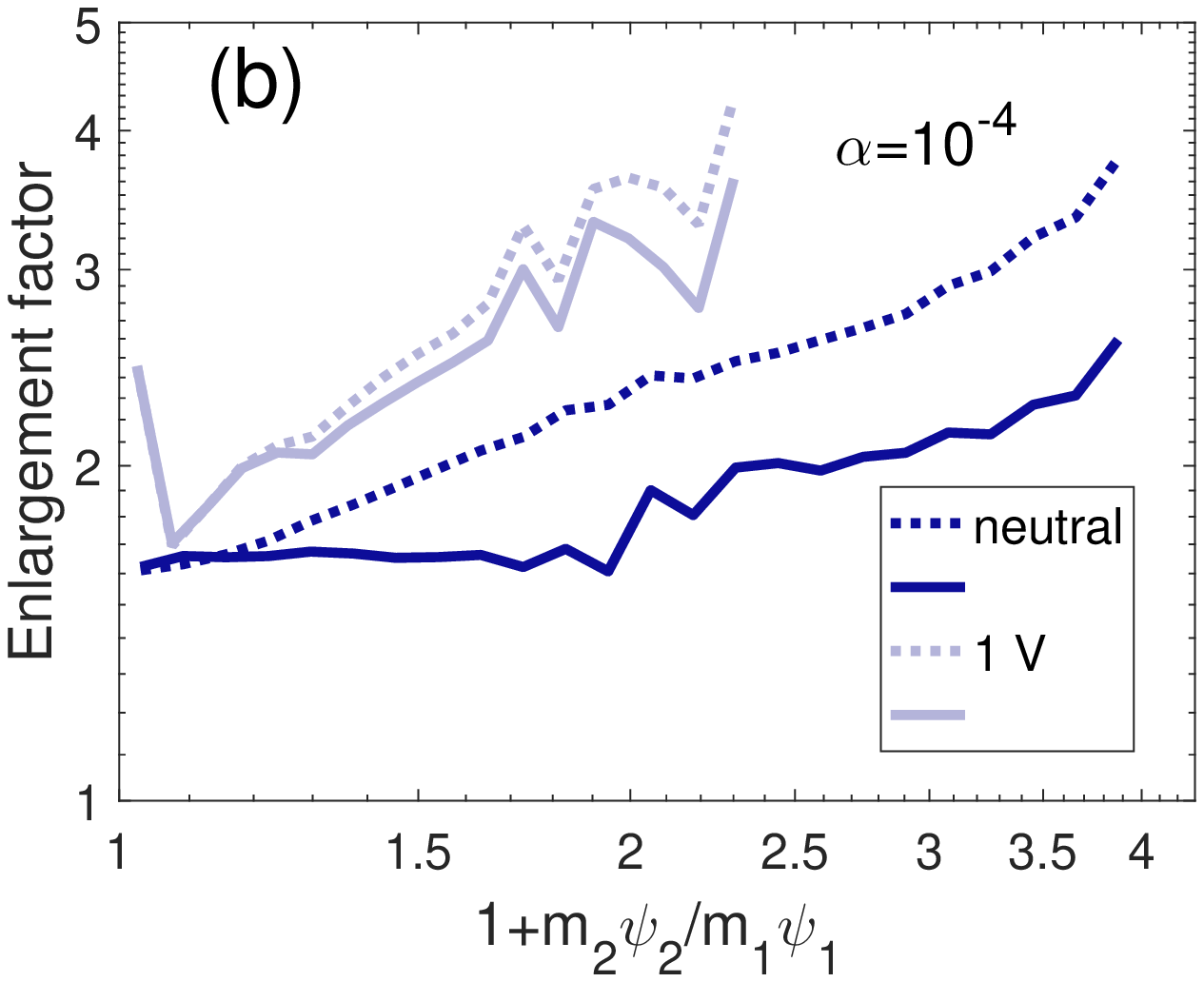}\includegraphics[width=6cm]{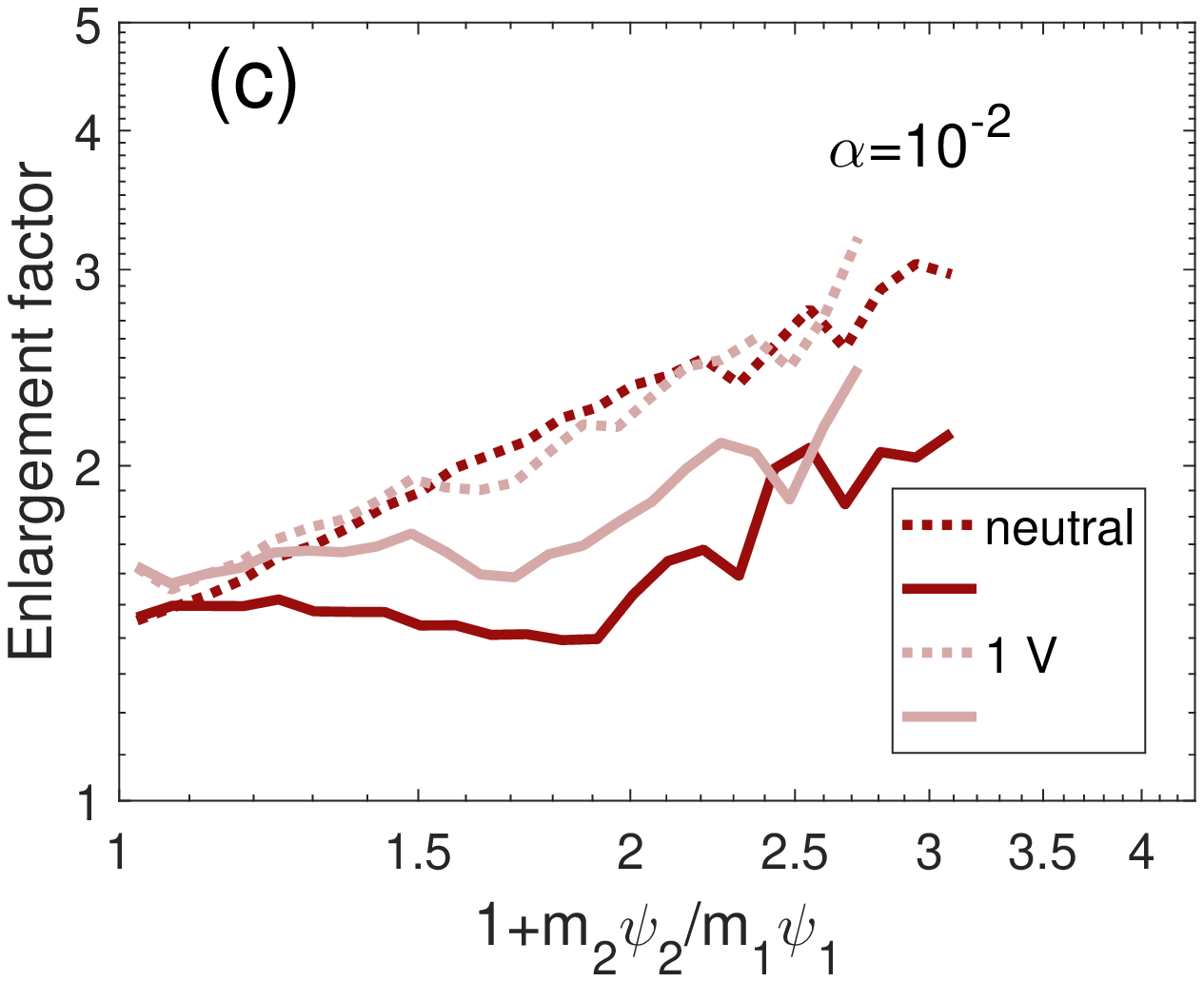} 
\caption{Comparison of the results from detailed-MC model (solid lines) to the results calculated by Ormel's MC model (dotted lines), for neutral (dark shades) and charged (light shades) particles. Turbulence levels are $\alpha=10^{-6}$ for a), $\alpha=10^{-4}$ for b) and $\alpha=10^{-2}$ for c). }
\label{dd3}
\end{figure*}

Because our detailed-MC model and Ormel's MC model have different initial populations (polydisperse vs monodisperse) and locations within the PPD, instead of comparing the results of the two models directly, it is more instructive to compare the results of the detailed-MC model to the results calculated by the formula for Ormel's MC model (Eq. \ref{e3}) using the data of the colliding particles in the detailed-MC model. Fig. \ref{dd3} shows that in low turbulence ($\alpha=10^{-6}$, $10^{-4}$; Figs. \ref{dd3}a, b), the results of the two models start off the same, and then diverge in the neutral cases, while being consistent in the charged cases. In strong turbulence ($\alpha=10^{-2}$; Fig. \ref{dd3}c), the results of the two models tend to diverge in both charged and neutral cases, with more divergence in the neutral case. Ormel's MC model predicts that charged and neutral populations have very similar enlargement factor for $\alpha=10^{-2}$, while the detailed-MC model shows a difference with larger enlargement factor for the charged case. Greater differences between charged and neutral populations occur for $\alpha=10^{-6}$, $10^{-4}$, but the effect is opposite: the enlargement factor of charged population is smaller than that of the neutral population for $\alpha=10^{-6}$, and is larger for $\alpha=10^{-4}$. For all turbulence levels, the detailed-MC model results in more compact aggregates than Ormel's MC model for equal ratio of mass-weighted enlargement factors of colliding particles, although $V^{\ast }$ in the detailed-MC model is smaller than that of MC model, which would cause a larger enlargement factor. One possible reason is that Eq. \ref{e3} is based on a monodisperse distribution, while the polydisperse distribution in the detailed-MC model leads to more efficient packing than a monodisperse distribution, i.e., the small particles fill in the gaps of aggregates. However, in the charged cases with low turbulence ($\alpha=10^{-6}$, $10^{-4}$), the two models have similar enlargement factors, because the charged population have a narrower range of monomer sizes, i.e., less diversity of monomer size, which is closer to a monodisperse distribution than the size distribution in the neutral case.

\subsection{Comparison of growth of charged aggregates}\lb{6.2}

The diversity of particle sizes not only increases the growth rate of particles (due to the higher relative velocity between grains within the population), but also reduces the porosity of aggregates (more efficient packing; Fig. \ref{dd3}), which further increases the growth rate due to the weak coupling of compact particles to the gas. For charged particles, in addition to these two effects, the size distribution also helps avoiding the freeze-out of the fractal growth. 
 
Okuzumi (2009) simulated dust coagulation with an initial monodisperse population in various plasma conditions. In contrast to Ormel et al. (2007), this simulation assumed that the dust grows into an ensemble of quasi-monodisperse aggregates with fractal dimension $D\sim 2$ and typical monomer number $N$, which increases during the fractal growth of dust. He found that the electric repulsion between charged particles strongly inhibits dust growth, which eventually ceases, for a wide range of model parameters. For example, at a radial distance of 5 AU and scale height z = H, with initial monomers size $a_{0}=0.1~\mu m$, the dust growth freezes out at $N \sim 33$ for turbulence strengths $\alpha=10^{-3}, 10^{-4}$. However, for $\alpha=10^{-2}$, the dust continues growing and reaches the subsequent growth stage involving collisional compaction. 

\begin{figure*}[!htb]
\includegraphics[width=6cm]{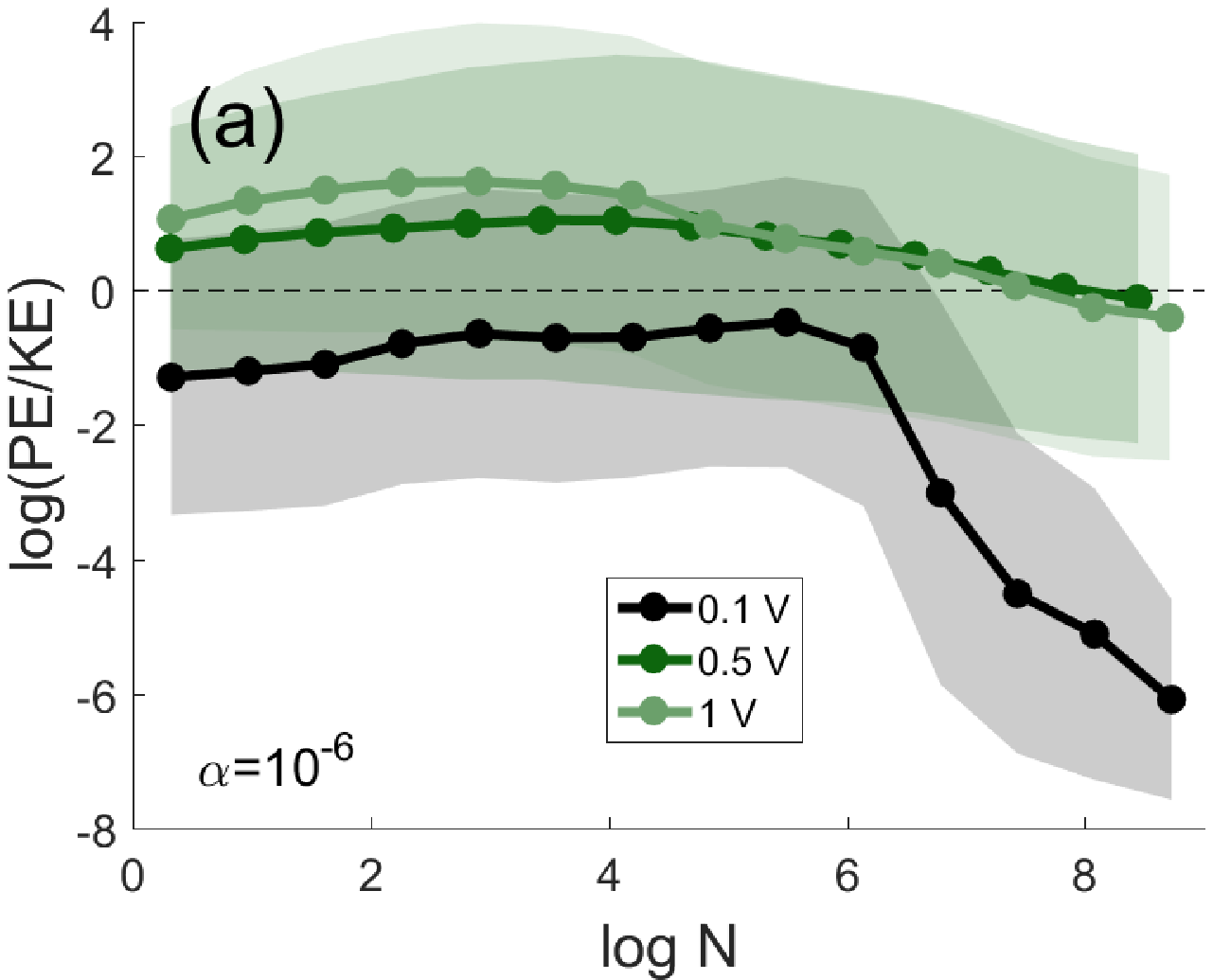}\includegraphics[width=6cm]{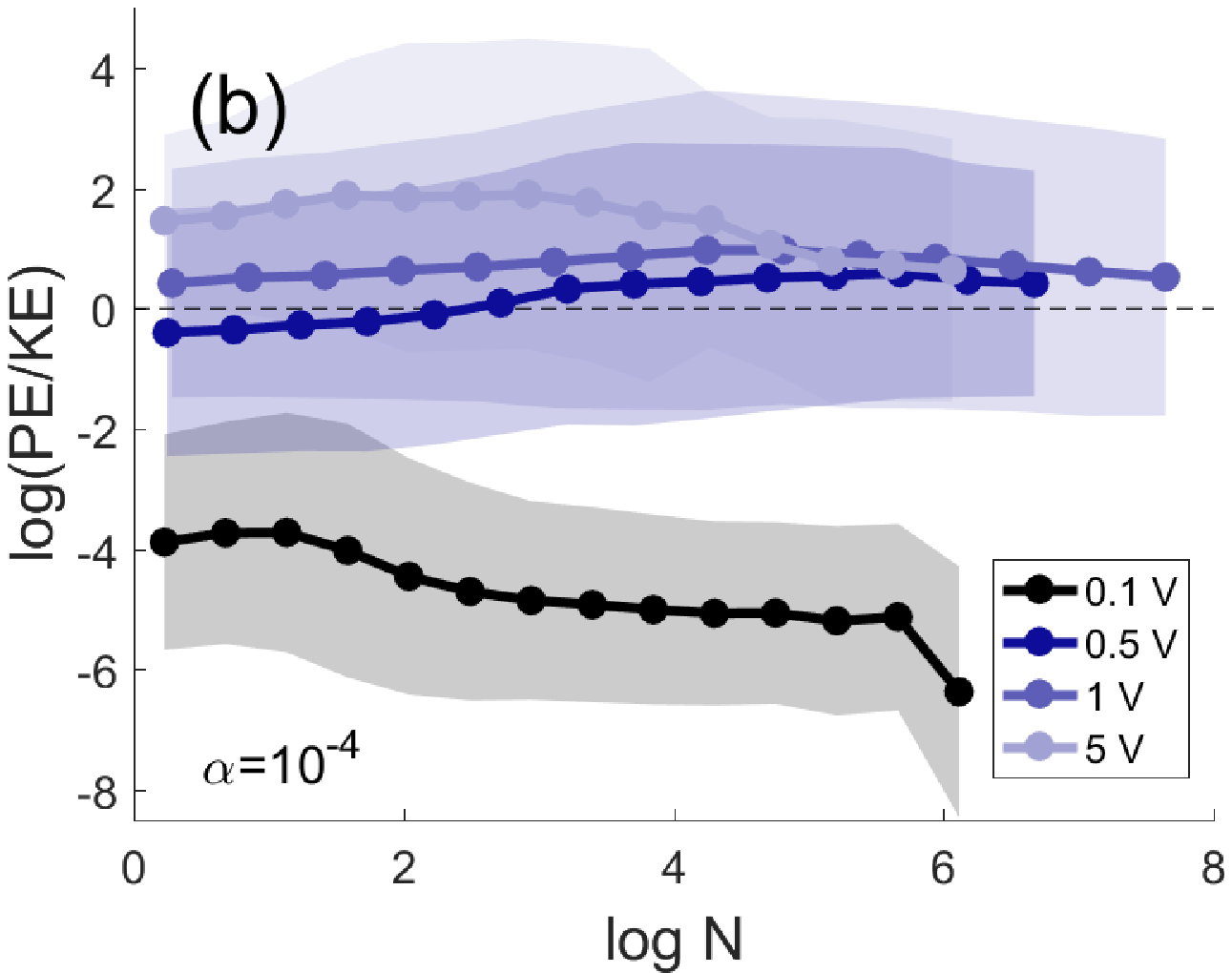}\includegraphics[width=6cm]{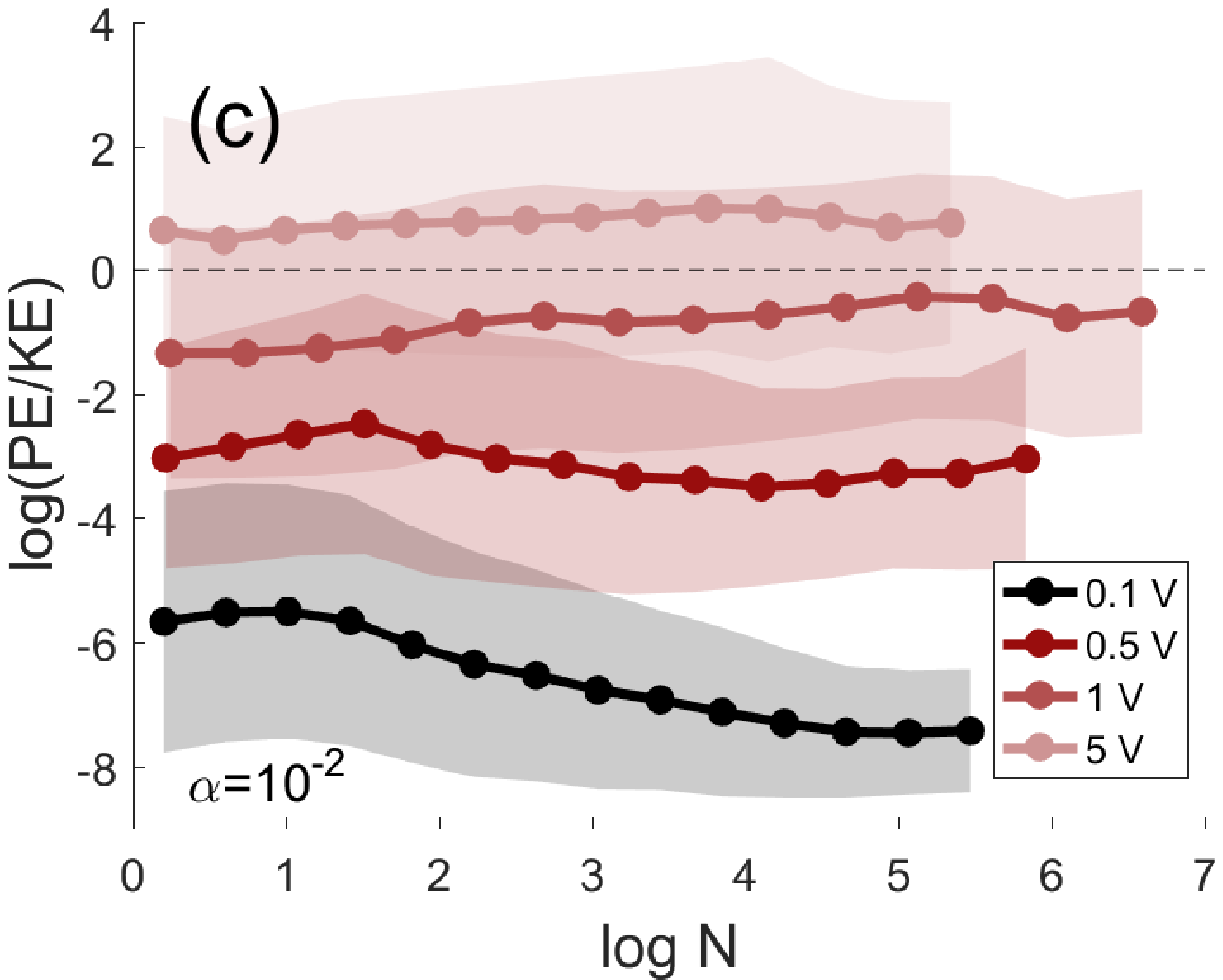} 
\caption{Ratio of electric potential energy to kinetic energy of particles as a function of the number of monomers within aggregates. Turbulence levels are $\alpha=10^{-6}$ for a), $\alpha=10^{-4}$ for b) and $\alpha=10^{-2}$ for c). The shaded areas indicate the standard deviation of the ratio.}
\label{d5}
\end{figure*}

Figure \ref{d5} shows the average ratio of electric potential energy $PE$ to kinetic energy $KE$ of particles with $N$ monomers in our simulations. Although the average ratio of potential energy to kinetic energy is larger than 1 for highly charged cases, the standard deviation of $PE/KE$ for the population shows that there is a significant fraction of the population with $PE/KE< 1$, preventing the freezing of particle growth. Okuzumi (2009) proposed a possible scenario to remove the electrostatic barrier against the fractal growth, which assumes a polydisperse size distribution where large aggregates can sweep up small aggregates. This is borne out in the results from the detailed-MC model, as the aggregates can develop sufficient relative velocity to overcome the electrostatic barrier when colliding with large particles. Okuzumi (2009) also pointed out that this can not happen if the large aggregates are as fluffy as the small aggregates (with comparable mass-to-surface-area-ratios). Therefore, the diversity of both the porosity and size of particles is important to overcoming the growth barrier.

\subsection{Application of the results to other regions in the disk}\lb{d3}

\begin{figure*}[!htb]
\includegraphics[width=9cm]{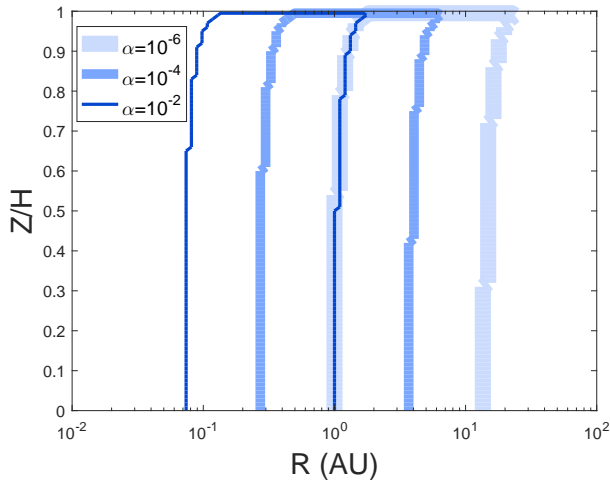}
\caption{Contour lines indicating the regions in a PPD, where the relative velocity between a 0.5-$\mu m$-radius particle and a 10-$\mu m$-radius particle is within the limits of [0.12 1.21] m/s, for turbulence strengths $\alpha=10^{-6}$, $10^{-4}$, and $10^{-2}$.}
\label{f22}
\end{figure*}

The environmental parameters that determine the relative velocity between particles are the radial distance, scale height ($H=c_{g}/\Omega$), and the turbulence strength. For a given particle pair, varying the turbulence level at a fixed disk location can result in the same range of relative velocity as changing the disk location with a fixed turbulence level. For example, the relative velocity between two particles, with radii of 0.5 $\mu m$ and 10 $\mu m$ respectively, at 1 AU (midplane) is 0.12 m/s for $\alpha=10^{-6}$ and 1.21 m/s for $\alpha=10^{-2}$. The contour lines in Figure \ref{f22} indicate the regions in the disk where such a particle pair has a relative velocity within these two limits, for various turbulence strengths. Therefore, although we model the dust coagulation at a specific location in the disk (midplane at 1 AU) and adopt specific turbulence levels ($\alpha=10^{-6}, 10^{-4}, 10^{-2}$), the results can be applied to other regions of the disk by adjusting the turbulence strength.

%




\section{Conclusions}
\renewcommand{\theequation}{6.\arabic{equation}} \setcounter{equation}{0}
We have presented a model of grain growth that incorporates the detailed physical characteristics of aggregates during the collision process employing a MC algorithm to model the collisional evolution of a population of dust particles in a protoplanetary disk. We defined two quantities, the compactness factor and the equivalent radius, to describe the porosity of an aggregate and quantify the effect of the collision process on the structure of the colliding aggregates. We compared charged and neutral aggregates for different levels of turbulence, which drives collisions. Our main findings are:

1. Highly charged aggregates contain fewer small monomers, resulting in a size distribution shifted towards larger monomer sizes and greater average monomer size; this shift increases with larger surface potential and lower turbulence (Fig. \ref{2}). The average monomer size within aggregates does not change much over time in strongly turbulent environments for $\left | V_s \right |<1$ V, and decreases over time in weakly turbulent environments with $\left | V_s \right |>0.5$ V.

2. In general, charged aggregates are more compact than are neutral aggregates, and highly charged aggregates tend to be more compact than those having lower charge for the same turbulence level (Fig. \ref{f5}). One reason is that charged particles tend to avoid the prominent region of the colliding partners due to the high local potential, and are more likely to stick between the extended arms. This effect is most noticeable for weakly turbulent and/or highly charged environments, in which the particles have sufficient time to deviate/rotate to minimize the potential energy of the configuration due to the slow movement. A second reason is that aggregates grow mainly through PCA in highly charged, weakly turbulent environments, i.e., by accretion of monomers onto large aggregates. It is easier for monomers to pass through the pores and fill in the gaps of aggregates, reducing the porosity.

3. Aggregate growth in the charged population first lags behind neutral populations due to electrostatic repulsion. As particles grow larger, the growth of weakly charged particles in relatively strong turbulence catches up with neutral particles (Fig. \ref{f7}), due to the greater number density (more particles remain in the population due to repulsion) and higher relative velocities between charged particles (resulting from their more compact structures). 

4. Particles in strongly turbulent regions collide more frequently than those in weakly turbulent regions. Given the same turbulence level, weakly charged particles grow faster than highly charged particles (Table \ref{table1}). However, highly charged particles grow to larger size before reaching the bouncing barrier, due to reduced relative velocity by electrostatic repulsion. The maximum particle size reached before the bouncing criterion is met is positively correlated with charge for a given turbulence level (Fig. \ref{f8}a and Table \ref{table2}).

5. For a highly charged environment with low turbulence, once a critical size is reached, the largest particles in the population grow very rapidly (runaway growth), while the rest of the population grows very slowly (Fig. \ref{f8}). The particles formed by runaway growth are a small proportion of the population, resulting in a population with a few large aggregates and a lot of remaining monomers and small aggregates. For populations without runaway growth, particles grow collectively, and almost all monomers have collided and formed aggregates before the bouncing criterion is met (Fig. \ref{f9}).

6. Charge has a greater impact on the porosity, monomer size, collision probability and the growth rate of dust particles in weak turbulence than in strong turbulence.

7. Diversity of particle size not only increases the growth rate of the particles (due to the higher relative velocity), but also reduces the porosity of the aggregates (more efficient packing; Fig. \ref{dd3}), which further increases the growth rate due to the weak coupling of compact particles to the gas. For charged particles, the diversity of both the porosity and size are important to overcoming the growth barrier (Fig. \ref{d5}).

In conclusion, it has been shown that charge and porosity play an important role in the evolution of a dust population. Charge decreases the growth rate of dust particles, due to missed collisions and smaller capture cross section (more compact structure). The longer growth timescale causes particles to be more subject to the radial drift barrier, and particles may have been accreted to the central star before growing to large sizes (Birnstiel 2016). In addition, dust particles also encounter bouncing and fragmentation barriers during their growth. The threshold bouncing and fragmentation velocities depend on impact energy, material, monomer size and porosity. Compact aggregates are more likely to bounce/fragment than fluffy aggregates, and aggregates comprised of small monomers are more resistant to fragmentation/compaction. Particles in charged environments are overall more compact and are comprised of larger monomers, which makes them more susceptible to bouncing and fragmentation. On the other hand, although charge increases the initial relative velocity between particles, due to the weaker coupling to the gas, the velocity is reduced during the electrostatic interaction, which decreases bouncing/fragmentation. On top of these two factors, the collision in charged cases tends to occur between a large particle and a small particle, which is unlikely to cause catastrophic fragmentation. Instead, mass transfer is more likely: when a small particle impacts a large aggregate, part of its mass is transferred to the target, leading to further growth of large particles. Therefore, charge may assist the population to overcome the fragmentation barrier. In addition to the impact on the growth barriers, the charge also greatly affects the optical properties of dust, such as the scattering and absorption opacity, by altering particle porosity/composition and the abundance of remaining small particles, which influences the temperature distribution, spectral energy distribution and appearance of PPDs (Kirchschlager \& Wolf 2014; Krijt 2015).

A linear regression and principal component analysis were used to determine that the charge, compactness factor, equivalent radius and relative velocity are the greatest contributors to the collision rate and the properties of the resulting aggregates in the hit-and-stick regime. The next step of this research is to include other types of collision outcomes in the simulation, such as bouncing, fragmentation, erosion and mass transfer, and develop a heuristic model for the collision rate as well as the physical characteristics of the resulting aggregate based on the data recorded from actual collisions. This new kernel will be used to simulate the evolution of a dust population over long time periods relevant to protoplanetary disk evolution.
\\


\textbf{Acknowledgments:} Support from the National Science Foundation grant PHY-1707215 is gratefully acknowledged.

\renewcommand{\theequation}{6.\arabic{equation}} \setcounter{equation}{0}

\appendix
\section{ Introduction about why fractional weights are used}
Suppose the smaller weight of the colliding particle is 0.2, the collision can be imagined to take place in a space which is 5 times as large as the original one, so that there is one such particle in the enlarged volume and the weight of each species become 5 times as large. Therefore, the probability that particle $i$ and particle $j$ will collide is: $C_{ij}(new)=5w_{i}(old)\times 5w_{j}(old)\sigma _{ij}\Delta v_{ij}/5V=5C_{ij}(old)$, where 'old' refers to the original space and 'new' refers to the enlarged space. Since the chance that the collision takes place in the original space is 1/5 (the ratio of the volume of the original space to that of the enlarged space), the probability that particle $i$ and particle $j$ will collide in the original space is $C_{ij}(new)/5$, which equals to $C_{ij}(old)$.

\def\bibindent{1em}

\end{document}